\RequirePackage{fix-cm}
\RequirePackage{fixltx2e} 
\documentclass[aps,pre,twocolumn,superscriptaddress,10pt,nofootinbib]{revtex4-1}

\usepackage{subfigure}
\usepackage{amsfonts, amssymb, amsmath}
\usepackage{bm}
\usepackage{graphics}
\usepackage{graphicx}
\usepackage{wasysym}
\usepackage{natbib}
\usepackage{mathtools}

\usepackage[utf8x]{inputenc}
\usepackage[T1]{fontenc}
\usepackage{paralist}

\usepackage[capbesideposition={right,bottom}]{floatrow}

\usepackage[usenames,dvipsnames]{xcolor}

\usepackage{color}
\definecolor{darkblue}{rgb}{0,0,0.6}
\definecolor{darkred}{rgb}{0.6,0,0}

\colorlet{lightred1}{orange!50}
\colorlet{lightred2}{orange!30}
\colorlet{lightred3}{orange!10}
\colorlet{lightred4}{orange!5}

\usepackage{tikz}
\usetikzlibrary{decorations.pathreplacing}

\usepackage{hyperref}

\hypersetup{
bookmarksopen=true,
pdftitle= Kibble-Zurek mechanism from different angles: The transverse XY model and subleading scalings , 
pdfauthor= B. Ladewig et al., 
pdftoolbar=false, 
pdfstartview={FitH},		
pdfmenubar=true,			
pdfhighlight=/O,			
colorlinks=true,			
urlcolor=darkblue,
citecolor=darkblue,		
linkcolor=MidnightBlue,	
}
\usepackage[normalem]{ulem}
\usepackage{comment}

\usepackage{pifont}

\newcommand{\Fig}[2]{Fig.~\ref{#1}\textcolor{MidnightBlue}{#2}}

\newcommand{\Eq}[1]{Eq.~(\ref{#1})}
\newcommand{\eq}[1]{(\ref{#1})}
\newcommand{\Eqs}[1]{Eqs.~(\ref{#1})}
\newcommand{\Tab}[1]{Tab.~\ref{#1}}

\newcommand{\adtime}{t}
\newcommand{\subexponent}{l}
\newcommand{\adparameter}{\hat{v}_k}
\newcommand{\adparameterstar}{\hat{v}_{k^*}}
\newcommand{\adparameterstarj}{\hat{v}_{k_j^*}}

\newcommand{\adiabaticbasisA}{a_k}
\newcommand{\adiabaticbasisB}{b_k}

\usepackage{abbrevs}
\usepackage{etoolbox} 
\newabbrev\RG{Renormalization Group (RG)}[RG]
\newabbrev\KZ{Kibble-Zurek (KZ)}[KZ]
\newabbrev\KZM{Kibble\texorpdfstring{{\textendash}}{--}Zurek Mechanism (KZM)}[KZM]
\newabbrev\DDP{Dykhne-Davis-Pechukas (DDP)}[DDP]

\robustify{\RG}
\robustify{\KZ}
\robustify{\KZM}
\robustify{\DDP}

\makeatletter 
\renewcommand\maybe@space@{%
  \maybe@ictrue 
  \expandafter   \@tfor
    \expandafter \reserved@a
    \expandafter :%
    \expandafter =%
                 \nospacelist
                 \do \t@st@ic
  \ifmaybe@ic 
    \space
  \fi
}
\makeatother

\begin{document}

\title{Kibble\texorpdfstring{{\textendash}}{--}Zurek mechanism from different angles: The transverse XY model and subleading scalings}

\author{Björn Ladewig}
\email[]{bladewig@thp.uni-koeln.de}
\affiliation{Institut f\"ur Theoretische Physik, Universit\"at zu K\"oln, 50937 Cologne, Germany}

\author{Steven Mathey}
\email[]{smathey@thp.uni-koeln.de}
\affiliation{Institut f\"ur Theoretische Physik, Universit\"at zu K\"oln, 50937 Cologne, Germany}

\author{Sebastian Diehl}
\affiliation{Institut f\"ur Theoretische Physik, Universit\"at zu K\"oln, 50937 Cologne, Germany}

\date{\today}

\begin{abstract}

The Kibble{\textendash}Zurek mechanism describes the saturation of critical scaling upon dynamically approaching a phase transition. This is a consequence of the breaking of adiabaticity due to the scale set by the slow drive. By driving the gap parameter, this can be used to determine the leading critical exponents. But this is just the `tip of the iceberg': Driving more general couplings allows one to activate the entire universal spectrum of critical exponents. Here we establish this phenomenon and its observable phenomenology for the quantum phase transitions in an analytically solvable minimal model and the experimentally relevant transverse XY model. The excitation density is shown to host the sequence of exponents including the subleading ones in the asymptotic scaling behavior by a proper design of the geometry of the driving protocol in the phase diagram. The case of a parallel drive relative to the phase boundary can still lead to the breaking of adiabaticity, and exposes the subleading exponents in the clearest way.  Complementarily to disclosing universal information, we extract the restrictions due to the non-universal content of the models onto the extent of the subleading scalings regimes.

\end{abstract}


\maketitle





\section{Introduction \label{Sec:Introduction}}

The \KZM \cite{Kibble1976,Zurek1985,Zurek1996} is a beautiful instance of the interplay of universality at an equilibrium critical point and a slow (non-equilibrium) drive of the coupling parameters. The mechanism roots in the breaking of adiabaticity and the creation of measurable excitations, applying to finite temperature as well as quantum phase transitions (see, e.g., Refs.~\cite{Dziarmaga2010,DelCampo2014})

The physical setup is as simple as paradigmatic: Consider the slow drive of a coupling, say $g_0$ relative to the critical point $g_0^*$: $g_0(t)-g_0^*= v_{0,n}t^n$ ($v_{0,n}$: generalized `velocity' for a drive of order $n$), starting far away from the phase transition. The concept of the \KZM can then be understood from different viewpoints. 

The first perspective puts the observable phenomenology center stage: Starting from the disordered side and approaching the phase transition, the state of the system is not globally symmetry-broken, but hosts spatial fluctuations of the order parameter on a scale given by the correlation length $\xi$. Therefore, domains of an average size of $\xi$ reside in one of the symmetry-broken states. 

Once adiabaticity is broken, the state is essentially frozen, and the correlation length saturates to $\xi^*$. The `frozen' domain structures are separated or punctured by (topological) defects \cite{Zurek1993,Zurek1996}. The density of these defects, $n_E$, is again related to the length scale $n_E \sim \xi^*{}^{-(d-p)}$ with $d$ the dimension and $p$ the dimensionality of the defects \cite{Chandran2012}. Furthermore, both quantities scale algebraically with the velocity of the drive.

To complement this observation and the emerging power-law dependence on the drive velocity, consider a second, scaling perspective: A system is initially prepared in its ground state far away from the critical region in the disordered phase. Early on, for a slow drive, the time-evolution will be adiabatic, as the characteristic time scale $\tau(t) \sim 1/\Delta (t)$ ($\Delta$ the energy gap, for a quantum phase transition) as a function of the time $t$ is small compared to the rate of change in the coupling: $|(g_0(t)-g^*_0)/\dot{g}_0|\propto t$. Nevertheless, close to the critical region near the transition, the correlation length $\xi$ as well as the characteristic time scale start to diverge, with a degree of divergence governed by the critical exponents $(z,\nu,\dots )$ determined by the universality class \cite{Sachdev2011}:
\begin{align}
\begin{aligned}
&\xi \sim |g_0-g_0^*|^{-\nu}, \\
&\tau \sim \xi^z .
\end{aligned}
\end{align}
Once $\tau(t)$ becomes of the order of the change of the coupling, at the time $t^*$ defined by $\tau(t^*)\sim t^*$, adiabaticity gets broken. The system is essentially `frozen' (impulse regime, see Sec.~\ref{Sec:BasicsAI}) with a finite length scale, which cannot diverge anymore. It gives a direct estimate of the saturated length scale $\xi^*$ with a power-law scaling in the velocity \cite{Polkovnikov2005,DeGrandi2010b,Barankov2008,Sen2008}
\begin{align}
\xi^*=\xi_0 \sim v_{0,n}^{-\frac{1}{nz+1/\nu}},
\label{Eq:IntroKZMScaling}
\end{align}
supporting the observed scaling.\footnote{
A third perspective is given by the sonic horizon: To refine the `freeze-out' scenario also the spreading of the defects/quasiparticles after breaking adiabaticity should be taken into account. The system is not completely frozen afterwards as there is still a finite velocity scale set by $v \approx \xi^*/t^*$ \cite{Francuz2016,Sadhukhan2019}, which in the quantum case is nothing but the (maximal) speed of the excited quasi-particles. It leads to a continued finite growth of the correlated regions \cite{Francuz2016,Sadhukhan2019}. Nevertheless even taking this important aspect into account will still lead to the same scaling of the correlation length with the velocity of the drive \Eq{Eq:IntroKZMScaling} (but with a modified prefactor). Apart from that, see also, e.g., Ref.~\cite{Liu2018} for a numerical analysis of the entire time-resolved process.}

Both the observable based as well as the scaling perspective have been investigated and verified in a broad spectrum of experiments in systems like superfluid ${}^{3}$He \cite{Bauerle1996,Ruutu1996}, liquid crystals \cite{Bowick1994,Chuang1991}, finite temperature as well as quantum phase transitions in ultracold gases \cite{Weiler2008,Lamporesi2013,Corman2014,Chomaz2015,Donadello2014,Navon2015,Yukalov2015,Beugnon2017,Sadler2006,Clark2016,Anquez2016,Chen2011,Braun2015,Meldgin2016}, trapped ions \cite{Ulm2013,Pyka2013,Mielenz2013}, ferroelectrics (multiferroic crystals) \cite{Chae2012,Lin2014,Griffin2012}, superconducting systems/Josephson tunnel junctions \cite{Monaco2002,Maniv2003,Monaco2009,Golubchik2010}, colloidal particles (in two dimensions) \cite{Deutschlander2015}, hydrodynamic systems \cite{Casado2006}, qubits \cite{Gong2016,Zhang2017,Cui2016,Cui2020}, Dicke models \cite{Baumann2011,Klinder2015}, and a Rydberg simulator \cite{Keesling2019}.

\begin{figure*}[t]
\centering

\begin{tikzpicture}
\node[anchor=south west,inner sep=0] at (0,0) {\includegraphics[width=\textwidth]{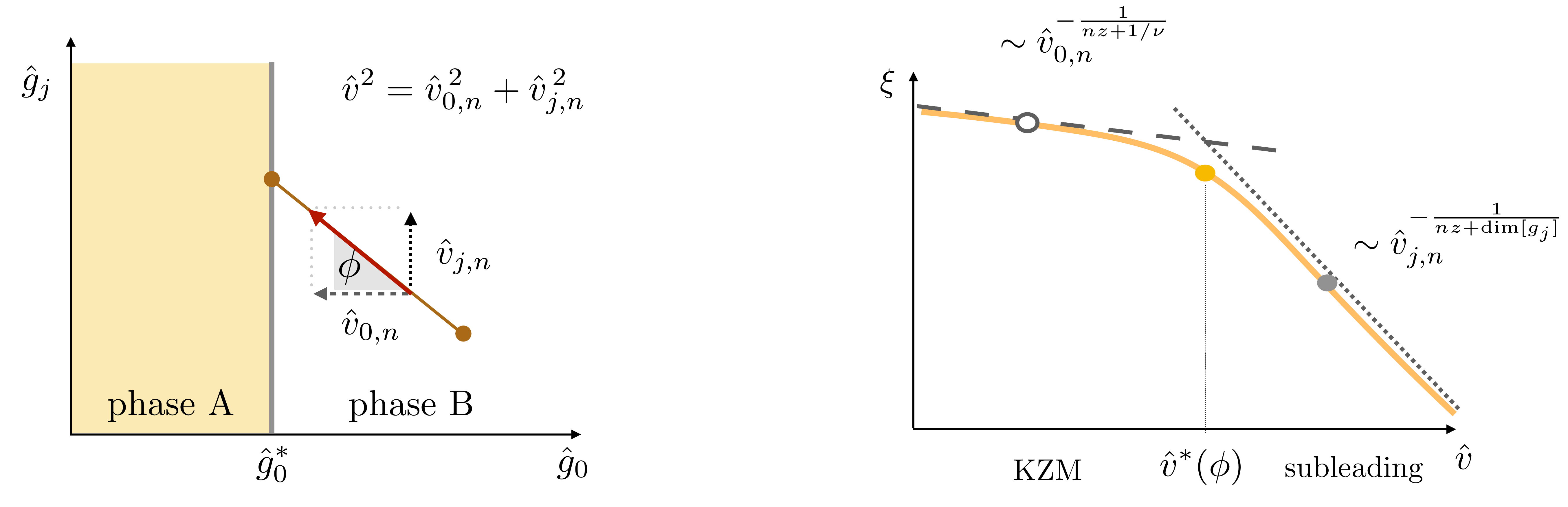}};
\node at (1,6.0) {\textbf{(a)}};
\node at (17.2,6.0) {\textbf{(b)}};
\end{tikzpicture}
\caption{Summary (generalized \KZM): a generalized drive of order $n$ of a dimensionless equilibrium relevant coupling $\hat{g}_0$ and an irrelevant coupling $\hat{g}_j$ in the phase diagram \textbf{(a)} allows one to extract both scaling dimensions from the finite length scale $\xi^*$ [orange curve in the log-log plot in \textbf{(b)}] due to adiabaticity breaking. The two different scalings regimes in (b) can be explained most directly for a drive of order $n=1$ by looking at the dimensionless velocity $\adparameter$ and the $k$-resolved excitation density $p_k=\exp(-\pi \adparameter^{-1})$, measuring adiabaticity breaking (see also Sec.~\ref{Chap:MinimalModel}). Once $\adparameter \gtrsim 1$ adiabaticity is broken [see again \Eq{Eq:KeyResultIdea} and \Eq{Eq:KeyResultAdBreaking}]. If $\hat{v} \gg \hat{v}^*(\phi)$ we observe the subleading scaling [e.g., full circle in (b)], otherwise the \KZM scaling.}
\label{Fig:Summary}
\end{figure*}

All these perspectives give valuable insights into the \KZM. In a recent work, the scaling perspective was picked up and formalized into an adiabatic \RG framework. As in the scaling approach, the key ingredient is to formulate the breaking of adiabaticity in an \RG language. In this approach, the \KZM was identified as the `tip of the iceberg' \cite{Mathey2020}: A \emph{generalized KZM} scenario can be established. It allows one not only to access the leading critical exponents as known previously, but in fact the whole spectrum of universal critical exponents underlying a second order phase transition. In particular, also equilibrium irrelevant couplings/operators can lead to an observable length scale, or differently put: \emph{Irrelevant} couplings at equilibrium can be made \emph{relevant} by a proper drive, leading to diverging length scales in the slow drive limit. For \emph{any} critical exponent $\text{dim}[g_j]$, such a length scale takes a form fully analogous to \Eq{Eq:IntroKZMScaling},
\begin{align}
\xi_{j} \sim v_{j,n}^{-\frac{1}{nz+\text{dim}[g_j]}},
\label{Eq:IntroSubScaling}
\end{align}
where $v_{j,n}$ is the `velocity' used to drive the coupling $g_j$. Here $\text{dim}[g_j]>0$ corresponds to an equilibrium relevant coupling (in particular $\text{dim}[g_0]=1/\nu$) and $\text{dim}[g_j]<0$ to an irrelevant one. The direct consequence for driving multiple couplings, say $g_0$ and $g_j$ [see \Fig{Fig:Summary}{a}], is that there are two \emph{competing} scales, $\xi_0$ and $\xi_j$. The observable scale is the smaller one, setting the largest possible scale of correlations:
\begin{align}
\xi^* \sim \text{Min}[ \xi_0,\xi_j].
\label{Eq:IntroTwoScales}
\end{align}

In this work, we make use of and combine both perspectives: From the \RG perspective we identify the critical exponent spectrum for explicit models and how a proper drive can be constructed to access this hierarchy. Completing the concept of the generalized \KZM, we then consider quantitative measures of adiabaticity breaking, here the excitation density $n_E$. The context of this work is briefly summarized in \Tab{Tab:WorkinContext}.

\begin{table}[t]
\begin{tabular}{c | c | c}
\toprule 

KZM  & leading coupling & subleading couplings  \\ \colrule

 observable & \ding{51} & this work  \\
scaling & \ding{51}  &  \cite{Mathey2020} \\ \botrule
 \end{tabular}
 \caption{(Generalized) \KZM framework from the observable and scaling perspective for driven leading and subleading couplings with scalings according to \Eqs{Eq:IntroKZMScaling},\eq{Eq:IntroSubScaling}.}
 \label{Tab:WorkinContext}
 \end{table}
 
This allows us to connect the more formal \RG predictions like \Eq{Eq:IntroSubScaling} with observables, which can be well approximated or even calculated exactly. This includes in particular non-universal scales, like the crossover velocities separating different scaling regimes from \Eq{Eq:IntroTwoScales}, which are not accessible from the \RG analysis. Furthermore, deep in the paramagnetic or ferromagnetic phases of spin models, the density is directly related to the density of defects, like spin flips or domain walls \cite{Dziarmaga2005}, which underlie the \KZM as outlined above. In particular, the excitation density and the scale $\xi^*$ are directly related according to (in one dimension) \cite{Dziarmaga2010}
\begin{align}
\xi^* \sim n_E^{-1} .
\label{Eq:IntroExvsLength}
\end{align}

A valuable platform to test both the traditional and -- as demonstrated here -- new aspects of the \KZM is the Ising quantum phase transition between a ferromagnetic and paramagnetic phase in the transverse Ising/XY model, which was already extensively studied, see, e.g., Refs.~\cite{Zurek2005,Dziarmaga2005,Polkovnikov2005,Damski2006,DeGrandi2010,Hodsagi2020,Biaonczyk2018,Divakaran2008,Divakaran2010,Dutta2015,Rams2019,Francuz2016,Barankov2008,Damski2006,Sen2008,Deng2009,Deng2009b,Rams2011,Mukherjee2010,Mondal2009}. 
The transverse Ising model (or the corresponding universality class), for instance, was realized experimentally in highly controllable quantum systems or simulators (e.g. the Rydberg simulator \cite{Keesling2019} or trapped ions \cite{Cui2016,Cui2020}), where the quantum version of the \KZM was verified.

\subsection{Key results \label{Sec:KeyResults}}

\emph{Mechanism and Observability}: We analyze the generalized \KZM for the transverse XY model, as well as an exactly solvable minimal model with $z=3$. In both cases, different drives as shown in \Fig{Fig:Summary}{a} are considered, interpolating between a transversal drive into the relevant $g_0$ direction and a parallel drive in the irrelevant $g_j$ direction. Such drives are parameterized by an angle $\phi$ and a velocity modulus $\hat{v}$ (at the level of dimensionless velocities $\hat{v}_{j,n}$). The main reason for choosing such a drive is to reveal the two \emph{different} scaling regimes of $\xi^*$, as shown in \Fig{Fig:Summary}{b}.

As anticipated above, the \KZM is based on adiabaticity breaking close to a critical point. Even an initially slow drive becomes fast compared to the other scales involved, and in particular, to the gap. We make use of this idea by introducing a rescaled, dimensionless `velocity' $\adparameter$, which depends on the momentum $k$ of a specific mode under consideration. This becomes possible as the different $k$-sectors in the time evolution decouple for both models. More precisely, for each velocity $v_{j,n}$, the rescaled velocity takes the form 
\begin{align}
\adparameter^{(j)} \sim v_{j,n} k^{-(nz+\text{dim}[g_j])},
\label{Eq:ScaleDependentVelocity}
\end{align}
which already has similarities to \Eq{Eq:IntroSubScaling}. This rescaled velocity has two advantages: Its scaling with $k$ already encodes the information of the critical exponents in \Eqs{Eq:IntroSubScaling} and \eq{Eq:IntroKZMScaling}. Furthermore, it has a rather direct relation to the excitation density. To establish the connection to the excitation density $n_E$, we first remark that also $n_E$ can be decomposed into the $k$-resolved densities $p_k$ ($N$: number of lattice sites)
\begin{align}
n_E= \frac{1}{N} \sum_k p_k .
\label{Eq:DefDensity}
\end{align}
To give a simplified picture, the qualitative relation between $p_k$ and the velocity $\adparameter^{(j)}$ is 
\begin{align}
\begin{aligned}
&\adparameter^{(j)} \gg 1 &&\Leftrightarrow && p_k \sim \mathcal{O}(1),\\
&\adparameter^{(j)} \ll 1 && \Leftrightarrow && p_k \ll 1,
\label{Eq:KeyResultIdea}
\end{aligned}
\end{align}
which gives meaning to the statement that a fast drive breaks adiabaticity (see Sec.~\ref{Chap:MinimalModel} for more details). In turn, we can identify a momentum scale $k_j^*$ separating the two regimes:
\begin{align}
&\adparameterstarj^{(j)} \approx 1, && k_j^* \sim v_{j,n}^{\frac{1}{nz+ \text{dim}[g_j]}}.
\label{Eq:KeyResultAdBreaking}
\end{align}
This onset of adiabaticity breaking also appears at the level of the excitation density: Combining \Eqs{Eq:DefDensity},\eq{Eq:KeyResultIdea} and \Eq{Eq:KeyResultAdBreaking}, we roughly get
\begin{align}
n_E=\frac{1}{N} \sum_k p_k \sim k_j^* \sim (\xi_j^{*})^{ -1},
\end{align}
(see Sec.~\ref{Chap:MinimalModel} for a more detailed discussion). In the case of driving $g_0$ and $g_j$ this leads to two length scales $\xi_0$ and $\xi_j$ and therefore the competition in \Eq{Eq:IntroTwoScales}. To make use of this competition, we consider the dimensionless velocities $\hat{v}_{0,n}$ and $\hat{v}_{j,n}$ and parametrize the drive by an angle $\phi$ and velocity $\hat{v}$. Now consider \Fig{Fig:Summary}{b}: shown are $\xi^*$ interpolating between the smaller of $\xi_0$ (dashed) and $\xi_j$ (dotted) for some fixed $\phi$. The two regimes are separated by $\hat{v}^*(\phi)$. By tuning $\hat{v}$, we can observe either the subleading scaling for $\hat{v} \gg \hat{v}^*(\phi)$ (e.g. filled circle) or the \KZM scaling otherwise (e.g., empty circle).

\emph{Microscopic vs. effective couplings}: When we consider drives in, e.g., an Ising model, we control the microscopic couplings like the transversal field or the ferromagnetic coupling, dragging the system through the phase diagram. Nevertheless, from the \RG point of view, the scaling of $\xi^*$ due to the (generalized) \KZM results from the \emph{effective} (renormalized) couplings of the long-wavelength theory in the critical region. It is possible that the relation of these couplings is non-trivial, so that e.g. a microscopic coupling is rather connected to a series of relevant and irrelevant effective couplings. In such a case, even though we approach the phase boundary orthogonally in terms of our microscopic `knobs', we are actually driving multiple effective couplings, an example is given in \Fig{Fig:SpinVsFermion}{}. Since also driven irrelevant couplings can lead to a scaling according to \Eq{Eq:IntroSubScaling}, this has the potential to obtain a `misleading' scaling regime, similar to \Fig{Fig:Summary}{b} for larger velocities, and places a need for caution in the interpretation of experiments on the \KZM. In an RG approach to generic interacting models, the relation between microscopic and effective couplings is complicated and not particularly transparent. Here we demonstrate this effect very explicitly: It not only surfaces in renormalization group transformations but also in the diagonalizing transformation of the microscopic spin model to a set of fermionic momentum modes, see Sec.~\ref{Sec:CriticalExponentsTI}. This gives the opportunity to study this general phenomenon in an explicit example.

\emph{Parallel drive}: As we demonstrate in Sec.~\ref{Sec:TIParallelDrive}, there is one case evading the ambiguity between microscopic and effective couplings: a drive performed in parallel to the phase boundary. This implies that \emph{only} subleading couplings are driven. This special case therefore offers the unique possibility to study and identify adiabaticity breaking and scaling due to subleading couplings only. In this case, there is just one drive scale according to \Eq{Eq:IntroSubScaling}, which is now competing with the finite ground state correlation length $\xi$. This scenario is very different from the \KZM discussed so far, as we stay at a constant distance to the critical line (see also Refs.~\cite{Divakaran2008,Mondal2009,Divakaran2010,Dutta2015}). The competition of the scales also allows us to restore adiabaticity, once $\xi \ll \xi_j$ (similarly to \cite{Rams2019}). This scenario is fully in line with -- and can be viewed as a special instance of -- the generalized \KZM; our present approach provides the direct link between the scaling / \RG  based and the observable based perspectives.

\emph{Non-universal scales}: Besides the universal scaling exponents from \Eq{Eq:IntroSubScaling},\eq{Eq:IntroKZMScaling}, we extract the non-universal scales (crossover velocities and required angles) for both models, see orange dot in \Fig{Fig:Summary}{b}. To qualitatively understand the effect of non-universal contributions, e.g., from larger momentum modes, we use the minimal model with $z=3$ with an explicit cutoff $\Lambda$, which captures the not further specified non-universal contributions. In particular, we are interested in how extended the new scaling regime in \Fig{Fig:Summary}{b} (full circle) is, depending on $\Lambda$. By varying the cutoff, the range of velocities, which allow one to observe the different scalings (Sec.~\ref{Sec:ModelDependence}), can be enlarged.

\emph{Plan of the paper}: The three main ingredients to understand and complement the (generalized) \RG perspective onto the \KZM are the equilibrium critical exponent spectrum of a  model, the interplay of a drive with this spectrum, and physical observables to extract the scaling. In Secs.~\ref{Sec:EquilibriumXY}, \ref{Sec:EngineeringADrive} and \ref{Sec:Techniques} these first two ingredients are worked out for the transverse XY model in detail to make the analysis self-contained. In Sec.~\ref{Chap:MinimalModel} the scaling of the excitation density is worked out for an exactly solvable minimal model and in Sec.~\ref{Eq:GeneralizedDrivesXY} for the transverse XY model. In Sec.~\ref{Sec:TIParallelDrive} we discuss the case of a purely parallel drive and in Sec.~\ref{Sec:ModelDependence} the role of the cutoff on the observability of the (subleading) scaling.

\section{Transverse XY model \label{Sec:EquilibriumXY}}

The scalings in the \KZM, \Eqs{Eq:IntroKZMScaling} and \eq{Eq:IntroSubScaling}, are based on the equilibrium critical exponents. Therefore, our first step is to identify these exponents and in particular the exponent \emph{spectrum}, including irrelevant exponents for the specific model at hand, the transverse XY model. Furthermore, we need to identify how the time-evolution of the system can be described.

We are mainly interested in the quantum Ising model, but to be able to tune the first subleading coupling independently we need at least two independent couplings, which are indeed present in the transverse XY model. The Hamiltonian for this transverse XY model for $N$ sites and periodic boundary conditions $\sigma_1^{x,y}=\sigma_{N+1}^{x,y}$ reads
\begin{align}
&H=-g\sum_l \sigma_l^z - J_x \sum_l \sigma_l^x \sigma_{l+1}^x-J_y \sum_l \sigma_l^y \sigma_{l+1}^y.
\end{align}
It is described by the microscopic couplings $\{g,J_x,J_y\}$ (where $J_y=0$ for the transverse Ising model) and the lattice spacing $a$. Here we consider $J_x,J_y>0$ implying a ferromagnetic coupling of spins. The equilibrium transverse XY model has two phases: the paramagnetic phase dominated by the transverse field $g\sum_l \sigma_l^z$ with ground state $|\uparrow \uparrow \dots \rangle$ and the ferromagnetic phase dominated by the XY-terms $\sum_l \sigma_l^{x,y} \sigma_{l+1}^{x,y}$. To extract the critical point and critical exponents, the model is mapped to non-interacting fermions by a Jordan-Wigner transformation, which takes for an even number of fermions the form (indicated by the $+$)\cite{Katsura1962,Lieb1961,Pfeuty1970,Barouch1970,Barouch1971,Sachdev2011,Dziarmaga2005}\footnote{We use the conventions of Ref.~\cite{Dziarmaga2005}.} (see Appendix~\ref{App:JordanWigner} for more details):
\begin{align}
\begin{aligned}
 &H^+=-\sum_l \left[ Jc_l^\dagger c_{l+1} +\gamma c_l^\dagger c_{l+1}^\dagger -gc_l^\dagger c_l +\frac{g}{2}+\text{h.c.}\right],\\
 &J:=J_x+J_y, \,\ \gamma:=J_x-J_y.
 \end{aligned}
 \end{align}
This Hamiltonian becomes particularly simple in Fourier space, where we use the convention used in Ref.~\cite{Dziarmaga2005}
\begin{align*}
c_l&=\frac{e^{-i\frac{\pi}{4}}}{\sqrt{N}}\sum_k c_k e^{ik(la)} \,\ , \\
k_j&= \frac{2\pi}{Na}\left[ -\frac{N}{2}+\left(j-\frac12\right) \right], j\in\{1,\dots,N\},
\end{align*}
which results in
\begin{align}
\begin{aligned}
&H^+=\frac12  \sum_k \left(c_k^\dagger \, c_{-k} \right) \, h_k \begin{pmatrix} c_{k} \\ c_{-k}^\dagger \end{pmatrix} +\text{const.},\\
&h_k=\begin{pmatrix} 2\left(g-J\cos(ka)\right) & 2\gamma \sin(ka) \\ 2\gamma \sin(ka) & -2(g-J\cos(ka)) \end{pmatrix}.\\
\label{Eq:MomentumFermions}
\end{aligned}
\end{align}

\begin{figure}[h]
\centering
\includegraphics[width=0.80\textwidth]{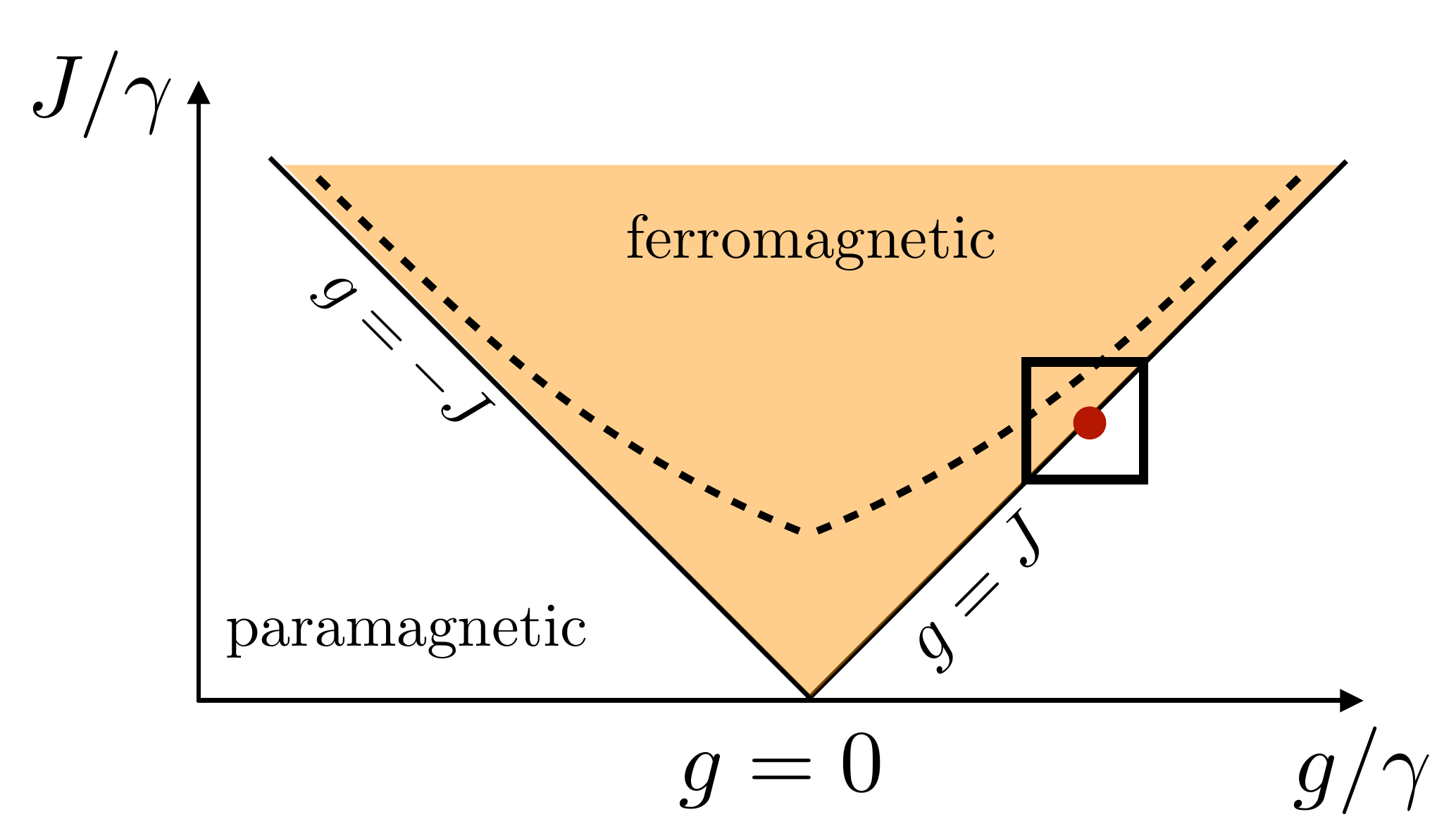}
\caption{(Simplified) phase diagram for the dimensionless spin couplings $J/\gamma$ and $g/\gamma$. The phase boundaries describe second order phase transitions. The dashed line indicates the region of incommensurability, described by $(J/\gamma)^2-1=J|g|/\gamma^2$ (typically written as $1-(\gamma/J)^2=|g/J|$, \cite{Bunder1999}). The red dot and square indicate the part of the critical region, which we will focus on.}
\label{Fig:XYPhaseDiagram}
\end{figure}

To extract the energy spectrum of this non-diagonal Hamiltonian, a canonical Bogoliubov-transformation can be used, which here amounts to diagonalizing the Hamiltonian:
\begin{align}
\begin{aligned}
&\text{eigenstate equation:} &&    h_k | \pm \rangle_k = \pm \epsilon_k |\pm \rangle_k, \\
&\text{diagonalizing unitary:} && U_k^\dagger= \left( \vec{(+)}_{k}, \vec{(-)}_{k} \right),\\
& && U_k h_k U_k^\dagger =\epsilon_k \sigma_z.
\label{Eq:StaticBogoliubov}
\end{aligned}
\end{align}
It is used to define new quasi-particle operators $\chi_k$ according to 
\begin{align}
\begin{pmatrix} c_k \\ c_{-k}^\dagger \end{pmatrix}=U^\dagger_k \begin{pmatrix} \chi_k \\ \chi_{-k}^\dagger \end{pmatrix} .
\end{align}
Here the transformation coefficients can be chosen real and are typically denoted as 
\begin{align}
\begin{aligned}
&| + \rangle_k = (u_k, v_k)^T, &&  | - \rangle_k  = (v_{-k}, u_{-k})^T, \\
&u_k=u_{-k}, && v_k=-v_{-k},\\
&c_k=u_k\chi_k + v_{-k} \chi_{-k}^\dagger,
\label{Eq:BogoliubovDefinitions}
\end{aligned}
\end{align}
where $| \pm \rangle_k$ are normalized to one. Using these operators, the Hamiltonian takes the form 
\begin{align}
\begin{aligned}
& H^+=\sum_k \epsilon_k\left( \chi_k^\dagger \chi_k -\frac12\right), \\
&\epsilon_{k}=2\sqrt{(g-J\cos(ka))^2+\left(\gamma\sin(ka)\right)^2},
\end{aligned}
\end{align}
where $\pm \epsilon_k$ are the eigenenergies of $h_k$. The energy-gap $\Delta(g)=\text{min}[\epsilon_k]$ closes at $g_c=J$ for $|k|=0$ and for $g=-J$ for $|ka|=\pi$. From the gap, the relevant critical exponents $z,\nu$ can be extracted according to
\begin{align}
\begin{aligned}
&\text{energy-gap:} && \Delta \sim |g-g_c|^{z \nu},\\
&\text{correlation length:} && \xi^{-1}\sim |g-g_c|^{\nu}, \\
& &&\Delta \sim \xi^{-z},
\end{aligned}
\end{align}
using finite-size scaling. The largest finite correlation length is $\xi\sim N$ and therefore the values $z=1$ and $\nu=1$ can be read off. At the level of the microscopic parameters the phase diagram is displayed in \Fig{Fig:XYPhaseDiagram}{}. The phase diagram is often plotted for variable-pairs $(\gamma/J,g/J)$ (e.g., Refs.~\cite{Dutta2015, Rams2011}). Here it will turn out to be more useful to use $g/\gamma$ and $J/\gamma$ instead. The reason is that we want to control and drive the terms $\sim c_k^\dagger c_k$ and $\sim k^2 c_k^\dagger c_k$ independently (see Sec.~\ref{Sec:CriticalExponentsTI}), and therefore we keep $\gamma$ fixed.

\subsection{Dynamical Bogoliubov transformation - solving the dynamical system}

In the following, we will consider the non-equilibrium situation, where the coefficients of the transverse Ising model are time dependent. We are interested in how strongly the system gets excited during the time evolution. Therefore, we consider the density of excited quasi-particles at time $t$ \cite{Dziarmaga2005}
\begin{align}
\begin{aligned}
n_E(t)&=\frac{1}{N} \sum_k \langle \Psi(t) | \chi_{k,t}^\dagger \chi_{k,t} | \Psi(t) \rangle =\frac{1}{N} \sum_k p_k .
\end{aligned}
\end{align}
The drives are e.g. of the form $g\to g(t)=v_nt^n+g^*$ (`order-$n$ drive'), where $v_n$ denotes a generalized `velocity'. The explicitly time-dependent evolution under $H^+(t)$ can be solved by making the Ansatz of a \emph{time-dependent} Bogoliubov transformation \cite{Dziarmaga2005,Dziarmaga2010}, where we follow closely the discussion in Refs.~\cite{Dziarmaga2005,Biaonczyk2018}. 
The starting point is the equilibrium case, where the ground state can be written using the Bogoliubov-coefficients
\begin{align}
\begin{aligned}
&|\Psi\rangle = \prod_{k>0} (u_k -v_{k} c_k^\dagger c_{-k}^\dagger )|0 \rangle=|\text{GS} \rangle,
\end{aligned}
\end{align}
which is the vacuum state of the Bogoliubov operators ($|0 \rangle$ is the c-fermion vacuum). The time-dependent state can as well be written in this form \cite{Dziarmaga2005,Biaonczyk2018}
\begin{align}
&|\Psi (t) \rangle = \prod_{k>0} (U_k(t) -V_{k}(t) c_k^\dagger c_{-k}^\dagger ) |0 \rangle.
\label{Eq:DiabaticRepresentation1}
\end{align}
The time-evolution of the coefficients in \Eq{Eq:DiabaticRepresentation1}, starting from the ground state at $t_i$, is given by a Schrödinger equation \cite{Dziarmaga2005} [see again \Eq{Eq:StaticBogoliubov}]
\begin{align}
\begin{aligned}
&|A(t)\rangle_k := \begin{pmatrix} U_k(t) \\ V_k(t) \end{pmatrix}, \,\ |A(t_i)\rangle_k =| + (t_i)\rangle_k \\
&i\hbar \partial_t | A (t) \rangle_k = h_k(t) |A (t) \rangle_k.\\
\label{Eq:DiabaticRepresentation}
\end{aligned}
\end{align}
(in the following we set $\hbar=1$). Therefore, solving the dynamics of the many-body state is reduced to finding the solutions $| A(t)\rangle_k$ to these $N$ two-state systems $h_k(t)$ in \Eq{Eq:MomentumFermions}, similar to Landau-Zener problems \cite{Damski2005,Zener1932,Majorana1932,Stueckelberg1932}. Nevertheless, the state in \Eq{Eq:DiabaticRepresentation1} will not necessarily be a ground state anymore. To make this transparent, we can rewrite this state as
\begin{align}
|\Psi (t) \rangle = \prod_{k>0} (\adiabaticbasisA(t) -\adiabaticbasisB(t) \chi_{k,t}^\dagger \chi_{-k,t}^\dagger)| \text{GS}_t \rangle ,
\label{Eq:AdiabaticRepresentiation1}
\end{align}
with $\adiabaticbasisA$ and $\adiabaticbasisB$ to be defined shortly. We call this the adiabatic representation, as it is referring to the instantaneous ground state at time $t: | \text{GS}_t \rangle$. The coefficients of the adiabatic case can directly be inferred by rewriting \Eq{Eq:DiabaticRepresentation} and using \Eq{Eq:StaticBogoliubov}
\begin{align}
&(U_k(t),V_k(t)) =:\adiabaticbasisA(t) (u_{k,t},v_{k,t}) + \adiabaticbasisB(t) (v_{-k,t},u_{-k,t}), \nonumber \\
&|A(t)\rangle_k= \adiabaticbasisA(t) | + (t)\rangle_k + \adiabaticbasisB(t) | - (t) \rangle_k. \label{Eq:adiabaticBogoliubov} 
\end{align}
Finally, the density of excited quasi-particles (excitation density) can directly be deduced from \Eq{Eq:AdiabaticRepresentiation1}
\begin{align}
\begin{aligned}
n_E(t)&=\frac{1}{N} \sum_k \langle \Psi(t) | \chi_{k,t}^\dagger \chi_{k,t} | \Psi(t) \rangle \\
&=\frac{1}{N} \sum_k p_k = \frac{1}{N}\sum_k |\adiabaticbasisB(t)|^2.
\label{Eq:ExcitationDensityGeneral}
\end{aligned}
\end{align}

\subsection{Critical exponent spectrum \& field theory \label{Sec:CriticalExponentsTI}}
As we have seen, the fermionic representation of the spin model allows us to extract the critical exponents $z$ and $\nu$ directly. They are the input for the standard \KZM once the energy-gap $\Delta$ is driven in time with $\text{dim}[\Delta]=1/\nu$. Nevertheless, as discussed in Sec.~\ref{Sec:Introduction}, we also want to consider drives of subleading/irrelevant couplings. To extract these couplings and their scaling dimensions we analyze the transverse XY model from the (equilibrium) \RG-perspective. 

Close to the critical point only the long-wavelength modes $k\to 0$ play an important role, justifying an expansion in powers of $k$ of the trigonometric functions in $h_k$. The validity of such an expansion is restricted to momenta $k<\Lambda$, where $\Lambda$ is a UV-cutoff. We are interested in the theory close to the phase transition at $g\approx g_c$ and want to extract the scaling dimensions of the couplings close to this transition. Our starting point is the thermodynamic limit $N\to \infty$ of \Eq{Eq:MomentumFermions} with the restriction of the momenta according to the UV-cutoff $\Lambda=1/a$ (see also \Tab{Tab:FermionDimensions} for the relation of the new operators to the old ones):
\begin{align}
&H^+ \approx \int \displaylimits_{-\Lambda}^{\Lambda} \frac{dk}{2\pi}\left[ \Delta \phi_k^\dagger \phi_k + \frac12D_1 k[\phi_k^\dagger \phi_{-k}^\dagger+\phi_{-k}\phi_k] \right. \nonumber \\
& \qquad \qquad \qquad \left. + D_2 k^2 \phi_k^\dagger \phi_k +\dots \right], \label{Eq:thermoXY} \\
&\Delta=2(g-J), \,\ D_1=2\gamma a, \,\ D_2=J a^2, \,\ \dots . \nonumber
\end{align} 
The essence of the \RG-approach is the idea that the coupling constants actually depend on the length scales under consideration \cite{Cardy2015}. This is formalized by the \RG (e.g. momentum-shell \RG), which gives a constructive way to calculate this length-scale dependence of the couplings. Due to the simplicity of the Gaussian model, a dimensional analysis is enough to extract the scaling dimensions of the couplings, which determine the length-scale dependence in the \RG. We still have the freedom to scale out one of the couplings in \Eq{Eq:thermoXY}. By doing so, the corresponding operator stays unchanged under \RG-transformations. The choice of the coupling we scale out determines what kind of phase transition and universality class we are describing. The reason is that by scaling out one coupling the corresponding operator is always present in the theory, even though all other (rescaled) couplings might vanish. To make this explicit: At the critical point $\Delta=0$ in \Eq{Eq:thermoXY} the leading term is the $D_1$-term (in Sec.~\ref{Sec:GammaCoupling} we discuss another choice and how it affects the spectrum). Scaling out $D_1$ by rescaling the time will give us the proper theory for the Ising-transition:
\begin{align}
H'^+ \approx& \int \displaylimits_{-\Lambda}^{\Lambda} \frac{dk}{2\pi}\left[ \Delta' \phi_k^\dagger \phi_k + \frac12 k[\phi_k^\dagger \phi_{-k}^\dagger+\phi_{-k}\phi_k] \right. \nonumber \\
& \left. + D_2' k^2 \phi_k^\dagger \phi_k +\dots \right],
\label{Eq:thermoXYrescaled}
\end{align}
where the couplings are defined in \Tab{Tab:FermionDimensions}. 

\begin{table}[h]
\center
\begin{tabular}{l | l | l | l}
\toprule
\textbf{microscopic} & \textbf{rescaled} & \textbf{dimensionful} & \textbf{dimensionless}\\
\colrule 
$t$ & $t'=2\gamma a t$ & $[t']=[k]^{-1}$ & $\hat{t}=2\gamma t $ \\
$\phi_k=c_k \sqrt{Na}$ & $\phi_k'= \phi_k$ & $[\phi_k']=[k]^{-1/2}$ & $\hat{\phi}_k=c_k \sqrt{N}$ \\ 
$D_1=2\gamma a$ & & & \\ \colrule
$\Delta=2(g-J)$ & $\Delta'=\frac{g-J}{\gamma a}$ & $[\Delta']=[k]$ & $\hat{\Delta}=\frac{2(g-J)}{2\gamma}$ \\
$v_\perp=2(v_g-v_J)$ & & & $\hat{v}_\perp=\frac{2(v_g-v_J)}{(2\gamma)^{n+1}}$\\ \colrule
$D_2=Ja^2$ & $D_2'=\frac{J}{2\gamma} a $ & $[D_2']=[k]^{-1}$ & $\hat{D}_2=\frac{J}{2\gamma} $ \\
$v_\parallel= v_J a^2$ & & & $\hat{v}_\parallel=\frac{v_J}{(2\gamma)^{n+1}}$\\ \colrule
$D_j$=\dots  & $D_j'=\frac{D_j}{2\gamma a}$ & $[D_j']=[k]^{-(j-1)}$ & $\hat{D}_j=D_j' \Lambda^{j-1}$ \\
\botrule
\end{tabular}
\caption{Overview of the operators and couplings in the fermionic theory for the microscopic \Eq{Eq:thermoXY}, rescaled \Eq{Eq:thermoXYrescaled}, and dimensionless version \Eq{Eq:DimensionlessXY}. The corresponding couplings and dimensions ($a$: lattice spacing) are given, where $[k]$ denotes the dimension of momentum. Here we used $\Lambda=1/a$ as the scale to define the dimensionless couplings, especially we have $\hat{k}=ka$. For later use also the velocities for an order-$n$ drive are added.}
\label{Tab:FermionDimensions}
\end{table}

All physical dimensions of the couplings $g_j'$ can be expressed as $[k]^{\text{dim}[g_j]}$, defining the scaling dimension as given in \Tab{Tab:FermionDimensions}. In particular, we have $z=-\text{dim}[t]=1$ and $\text{dim}[\Delta]=1=1/\nu$, as we already have seen. In Sec.~\ref{Chap:MinimalModel}, we discuss a fermionic model with $D_1=D_2=0$ and therefore $1/\nu=z=3$. To see the significance of these scaling dimensions, we consider dimensionless couplings that can be defined by multiplying the couplings with the proper power of the UV-cutoff, \Tab{Tab:FermionDimensions}. The Hamiltonian, using these dimensionless couplings, takes the form 
\begin{align}
\hat{H}^+\approx  & \int_{-1}^{1} \frac{d(ka)}{2\pi}\left[ \hat{\Delta} \hat{\phi}_k^\dagger \hat{\phi}_k + \frac12 (ka)[\hat{\phi}_k^\dagger \hat{\phi}_{-k}^\dagger+\hat{\phi}_{-k}\hat{\phi}_k] \right. \nonumber \\
& \left. + \hat{D}_2 (ka)^2 \hat{\phi}_k^\dagger \hat{\phi}_k +\dots \right].
\label{Eq:DimensionlessXY}
\end{align}
We can now ask how these dimensionless couplings change under an (infinitesimal) change of the cutoff: $\Lambda' \to \Lambda' -d\Lambda'$ (see, e.g., Refs.~\cite{Fradkin2010,Cardy2015}). Formally, we can determine this cutoff-dependence for Gaussian models using
\begin{align}
&\Lambda'\frac{\partial g_j}{\partial \Lambda'}\stackrel{!}{=}0, &&\left. \Lambda'\frac{\partial \hat{g}_j}{\partial \Lambda'}\right|_{g_j}=-\text{dim}[g_j] \hat{g}_j=:\hat{\beta}_j(\hat{g}_j), 
\label{Eq:FlowEquations}
\end{align}
where the couplings at large spatial distances are given by solving the equation towards $\Lambda' \to 0$. A positive scaling dimension implies a growth of the couplings with respect to the fixed point on larger length scales (relevant coupling) and a negative scaling dimension a shrinking (irrelevant coupling). Two examples of scale-dependent couplings as solutions to the flow equations in \Eq{Eq:FlowEquations} are given by (where the initial scale is set by $\Lambda'=\Lambda$)

\begin{align}
\begin{aligned}
\text{relevant/growing:} &&\hat{\Delta}(\Lambda')=\hat{\Delta} \left(\frac{\Lambda}{\Lambda'} \right)^{+1},\\
\text{irrelevant/shrinking:} &&\hat{D}_2(\Lambda')=\hat{D}_2  \left( \frac{\Lambda}{\Lambda'} \right)^{-1}.
\label{Eq:ExampleRGFlow}
\end{aligned}
\end{align}

This set of flow-equations $\hat{\beta}_j$ has a simple fixed point $\vec{C}^*$, here describing the scale-invariant fixed point of the second order phase transition:
\begin{align}
\vec{C} = \begin{pmatrix} \hat{\Delta} \\ \hat{D}_2 \\ \hat{D}_3 \\ \vdots \end{pmatrix}, && \vec{C}^*= \begin{pmatrix} \hat{\Delta}^* \\ \hat{D}_2^* \\ \hat{D}_3^* \\ \vdots \end{pmatrix}=\begin{pmatrix} 0  \\ 0 \\ 0 \\  \vdots \end{pmatrix}.
\end{align}
At such a fixed point, we can make a stability-analysis of the \RG-flow described by $\vec{\hat{\beta}}$ and find the stable/irrelevant and unstable/relevant directions. To this end, we can formally calculate the Jacobian of the $\vec{\hat{\beta}}$-vector field, which here is just a diagonal matrix. The eigenvalues are by definition the (negative) scaling dimensions, and the eigenvectors the stability directions, see \Tab{Tab:DriveOverviewXY} (second and last column). We emphasize that the stability directions are the \emph{essential step} to construct the phase diagram in terms of the (effective) couplings of the universal theory close to criticality. If we want to make the scaling with respect to the coupling $D_j$ observable we need to drive in the proper (eigen)direction in \Tab{Tab:DriveOverviewXY} (last column).

\begin{figure}[t]
\centering
\begin{tikzpicture}
\node[anchor=south west,inner sep=0] at (0,0) {\includegraphics[width=0.95\textwidth]{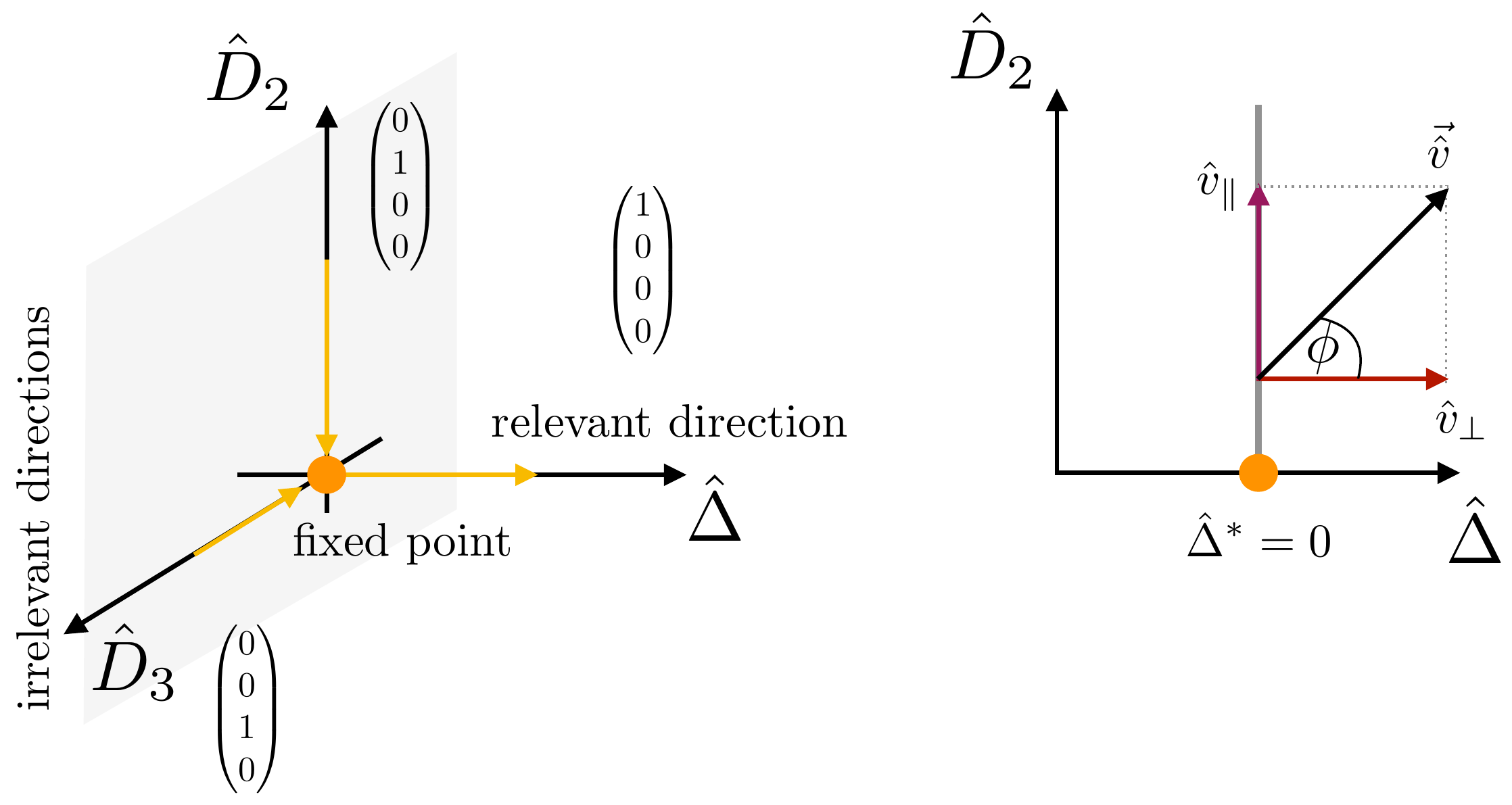}};
\node at (0,4.6) {\textbf{(a)}};
\node at (5,4.6) {\textbf{(b)}};
\end{tikzpicture}
\caption{\textbf{(a)} Geometry of the \RG-flow at the fixed point in the fermionic coupling space (here we ignore the other couplings for illustrational purposes). The direction of the arrows indicate, whether they are irrelevant (flowing into the fixed point) or relevant (flowing out). \textbf{(b)} Longitudinal ($\hat{v}_{\parallel}$) and transversal ($\hat{v}_{\perp}$) drive.}
\label{Fig:FermionCouplingSpace}
\end{figure}

\emph{Phase diagram in fermionic representation}: For a free theory this is simple, as all couplings are independent\footnote{In a more general setup the directions can be inferred as described from the stability matrix of the full set of the \RG $\hat{\beta}$-functions at the critical point, see Ref.~\cite{Mathey2020}.}. A reduced part of the coupling space is shown in \Fig{Fig:FermionCouplingSpace}. In the following we will refer to the $\hat{\Delta}$-direction as the `transversal' direction, as it controls the distance to the critical point. All other directions are  labeled `longitudinal'.

\begin{figure*}[t]
\centering
\includegraphics[width=0.80\textwidth]{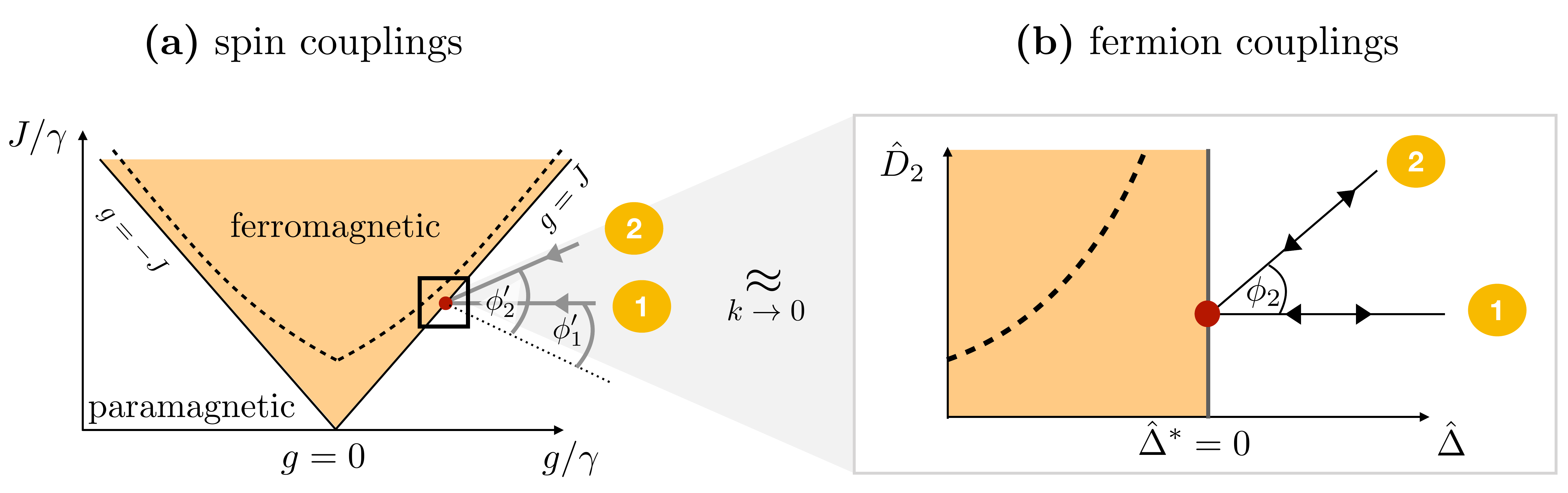}
\caption{(Simplified) phase diagrams for the dimensionless spin \textbf{(a)} and fermion couplings \textbf{(b)}. The geometry has to be inferred from the fermionic phase diagram, for $k\to 0$ the two couplings in (a) and (b) can directly be related, see \Eq{Eq:RelationCouplingsPre}. A purely transversal (1) and a general (2) even drive are shown, which reach a critical point at $t=0$. The angles are $\phi_{1,2}$ ($\phi_1=0$) in the fermionic language and $\phi'_{1,2}$ in the spin language respectively. In particular, $\phi_1=0$ corresponds to $\phi_1' \neq 0$. Here $\hat{\Delta}$ is the relevant coupling and $\hat{D}_2$ the first irrelevant coupling. The bold dashed line indicates the border to incommensurability \cite{Bunder1999}.}
\label{Fig:SpinVsFermion}
\end{figure*}

\emph{Relation to spin models}: In a final step we compare the phase diagrams in terms of the microscopic spin couplings and the effective fermionic ones in \Fig{Fig:SpinVsFermion}{}. When translating fermionic couplings back to spin-couplings, we can first of all make the identification (in the corresponding subspace for $k\to 0$, which allows us to neglect higher powers in $k$):
\begin{align}
\begin{pmatrix} g/\gamma \\ J/\gamma \end{pmatrix} \approx \begin{pmatrix} \hat{\Delta}+2\hat{D}_2 \\ 2\hat{D}_2 \end{pmatrix}.
\label{Eq:RelationCouplingsPre}
\end{align}
One consequence of this mapping between spin-couplings and fermion-couplings is that it \emph{is not angle-preserving}. This is the central point that makes it important to distinguish the microscopic and effective phase diagram, a further discussion is postponed to Sec.~\ref{Sec:OrthogonalityIssue}. In general, the phase diagram of the transverse XY model has more features than just the Ising transitions, as there can also be gap-closings at, e.g., $|ka|=\pi$ and multicritical points, which we will not investigate. Furthermore, there is a region of incommensurability (see \Fig{Fig:XYPhaseDiagram}{}), where the minimal gap of the dispersion is neither located at $ka=0$ nor $|ka|=\pi$ \cite{Dutta2015}. One potential issue of the transverse XY model is apparent: The (naive) fixed point (see \Fig{Fig:FermionCouplingSpace}) of the fermionic theory coalesces with the $\pi$-gap closing at $g=J=0$, which would modify the simple picture given above as not only the $k$-modes close to $0$ are important. Therefore, we will consider the region of finite $J$ and $\hat{D}_2$ as indicated by the red dot in \Fig{Fig:SpinVsFermion}{}.

\section{Constructing a drive \label{Sec:EngineeringADrive}}

Having extracted the fixed point $\vec{C}^*$, scaling dimensions, and stability directions at the fermionic level we can directly apply the \RG-results from Ref.~\cite{Mathey2020}. They allow us to construct a drive of, e.g., the relevant and one irrelevant coupling, such that for intermediate velocities adiabaticity will be broken due to the subleading drive and at very low velocities due to the leading one, see \Fig{Fig:Summary}{} again. Therefore, we will construct a drive like the one shown in \Fig{Fig:FermionCouplingSpace}{b}, which consists of a drive along the proper directions in the effective (fermionic) phase diagram. 
According to Ref.~\cite{Mathey2020}, driving any coupling with $\hat{t}^n$ close to the fixed point $\vec{C}^*$: $\hat{g}_j \to \hat{g}_j(\hat{t})=\hat{v}_j \hat{t}^n+\hat{g}_j^*$ will give rise to a finite length scale, which scales with the velocity $\xi_{j} \sim \hat{v}_j^{-1/(nz + \text{dim}[g_j])}$ and becomes observable once $\text{dim}[v]=nz+\text{dim}[g_j]>0$. Therefore, to make a certain subleading scaling observable, we need to pick a large enough $n$. An overview for different drivings and the possibly extractable scalings is given in \Tab{Tab:DriveOverviewXY} for $z=\nu=1$ (see also Ref.~\cite{Mondal2009} for the nonlinear cases).

\begin{table}[h]
 \center
 \begin{tabular}{c | c | ccc | c}
 \toprule
\multicolumn{6}{c}{\underline{shifted critical exponent spectrum (transverse XY)}} \\
 & equilibrium & \multicolumn{3}{c|}{driven} & direction \\
 $g_j$ & $\text{dim}[g_j]$ &  \multicolumn{3}{c|}{$\text{dim}[v]=nz + \text{dim}[g_j]$} & \\
 & & $n=1$ & $n=2$ & $n=4$ & \\
 $\hat{\Delta}$& \colorbox{lightred1}{$+1$} & \colorbox{lightred1}{$+2$} & \colorbox{lightred1}{$+3$} & \colorbox{lightred1}{$+5$} & $(1,0,0,0,\dots)$ \\
 $\hat{D}_2$ & $-1$ & $0$  & \colorbox{lightred2}{$+1$} & \colorbox{lightred2}{$+3$} & $(0,1,0,0,\dots)$ \\
 $\hat{D}_3$ & $-2$ & $-1$ & $0$ & \colorbox{lightred3}{$+2$} & $(0,0,1,0,\dots)$ \\
 $\hat{D}_4$ & $-3$ & $-2$ & $-1$ & \colorbox{lightred4}{$+1$} & $(0,0,0,1,\dots)$ \\
\vdots & \vdots & \vdots & \vdots & \vdots & \vdots \\ \botrule
\end{tabular}
\caption{Equilibrium critical exponent spectrum (second column) and shifted versions for different drives of order $n$ for the quantum Ising universality class in $d=1$, which will be considered. The colored exponents are relevant and in principle are associated with different observable length scales. The color code indicates that higher velocities (brighter) or lower velocities (darker) relative to each other are needed to make the corresponding (length) scale observable.}
\label{Tab:DriveOverviewXY}
\end{table}
As we want to consider driving multiple couplings, which could have in principle different dimensions, a first preparation step is to construct proper dimensionless couplings. To make the connection between the field theory in the critical region and the spin model, we consider driving the dimensionless combinations

\begin{align}
\hat{h}_k(\hat{t})=&\big[\underbrace{ \hat{\Delta}^0+\hat{v}_\perp \hat{t}^n}_{=(g(t)-J(t))/\gamma} + \underbrace{2(\hat{D}_2^0+\hat{v}_\parallel \hat{t}^n)}_{=J(t)/\gamma}(1-\cos(ka)) \big] \sigma_z \nonumber \\
 &+ \sin(ka) \sigma_x .
 \label{Eq:XYall}
\end{align}
Therefore, by driving the physical couplings $J(t)$ and $g(t)$ we can approximately drive the proper dimensionless couplings of the critical theory for $k\to0$:\footnote{To be precise, by tuning $g(t)$ and $J(t)$ in the given way, we are actually tuning all $k^{2m}$-terms. Nevertheless, for a quadratic drive (the case of interest) the driven higher-order $k$-terms do not lead to observable scaling according to \Tab{Tab:DriveOverviewXY}.}

\begin{align}
\begin{aligned}
&\hat{\Delta}(\hat{t})\approx \hat{\Delta}^0+\hat{v}_\perp \hat{t}^n, && \hat{v}_\perp = \frac{2(v_g-v_J)}{(2\gamma)^{n+1}},\\
&\hat{D}_2(\hat{t})\approx \hat{D}_2^0 + \hat{v}_\parallel \hat{t}^n, && \hat{v}_\parallel=\frac{v_J}{(2\gamma)^{n+1}}.
\end{aligned}
\end{align}

\begin{figure}[h]
\centering

\begin{tikzpicture}

\node[anchor=south west,inner sep=0] at (0,0) {\includegraphics[width=0.9\textwidth]{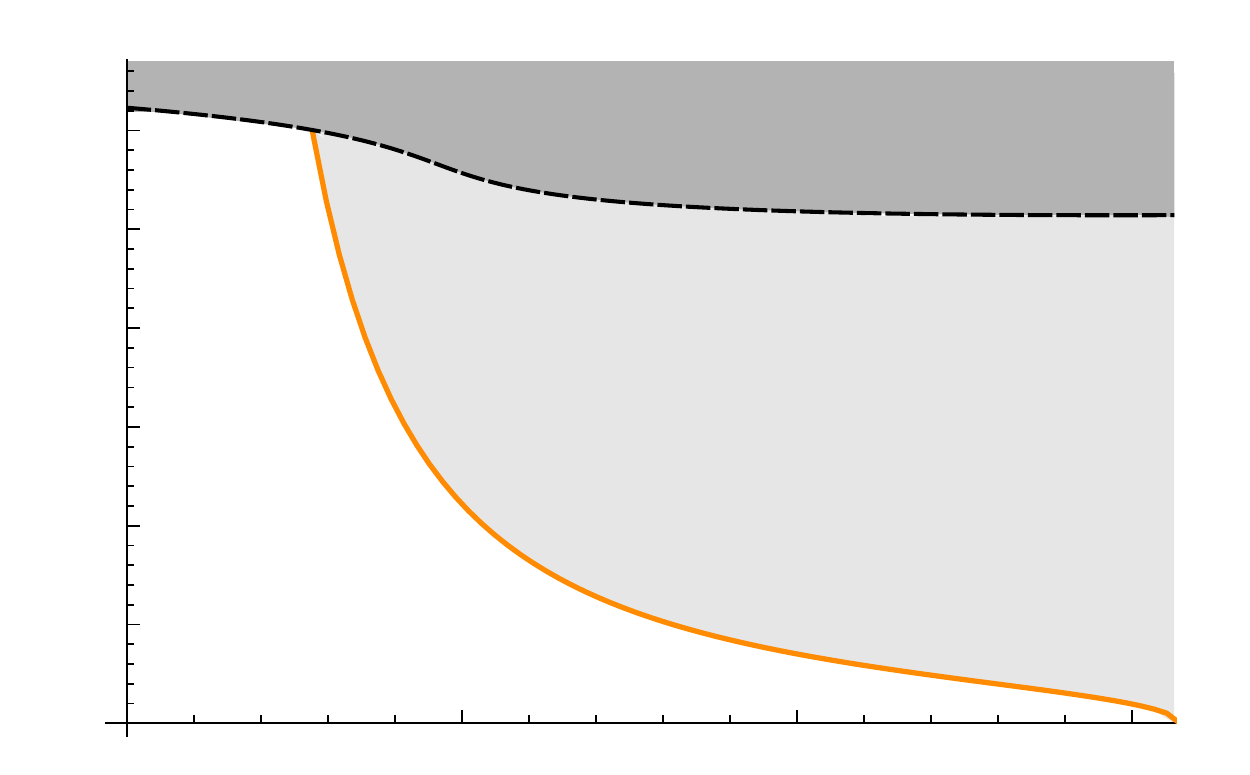}};

\node at (0.5, 4.5) {\large $\hat{v}$};
\node at (0.5, 4.1) {$\hat{v}_{\text{cut}}$};

\node at (7.6, 0.2) {\large $\phi$};
\node at (7.3, 0.1) {$\frac{\pi}{2}$};

\node at (2, 2) {\KZM};
\node at (5.1,3) {subleading scaling};
\node at (5.1,4) {non-universal};

\node at (2,1.3) {$n_E \sim \hat{v}_\perp^{\frac{1}{nz+1}}$};
\node at (5.1,2.3) {$n_E \sim \hat{v}_\parallel^{\frac{1}{nz-1}}$};

\end{tikzpicture}
\caption{Schematic crossover velocity $\hat{v}^*(\phi)$ (solid orange line), which separates the \KZM scaling at lower velocities and the subleading scaling above the orange solid line, see \Fig{Fig:MinimalModelFirstSub}{} for an explicit example for some fixed $\phi$. The dashed black line ($\hat{v}_{\text{cut}}$) represents the crossover to the non-universal regime at larger velocities and depends strongly on $\Lambda$. An explicit example is given in \Fig{Fig:MinimalModelCrossover}{c}.}
\label{Fig:SchematicCrossover}
\end{figure}
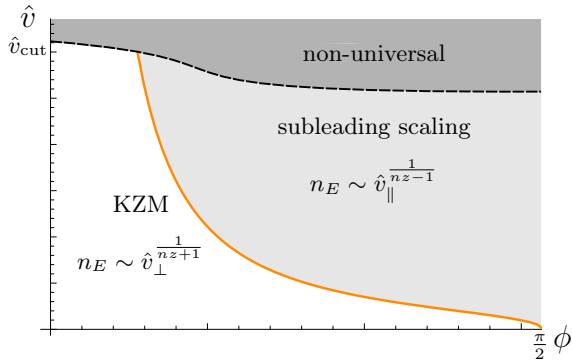

The corresponding directions in the (fermionic) dimensionless coupling space are orthogonal\footnote{It is important to note that orthogonality in the fermionic coupling-space does not imply orthogonality in spin-coupling space and vice versa, see Sec.~\ref{Sec:OrthogonalityIssue} and \Fig{Fig:SpinVsFermion}{}.} [see \Fig{Fig:FermionCouplingSpace}{b}], and therefore we can write the drive in the reduced coupling space as
\begin{align}
\hat{\vec{v}} =\hat{v}_{\perp} \mathbf{e}_{\hat{\Delta}} +\hat{v}_\parallel\mathbf{e}_{\hat{D}_2}=\hat{v} \begin{pmatrix} \cos(\phi) \\ \sin(\phi) \end{pmatrix},
\end{align}
where $\pi/2-\phi$ describes the angle enclosed with the subleading direction ($\hat{D}_2$-direction), see also \Fig{Fig:FermionCouplingSpace}{b}. The definition is chosen such that it fits to the notation in Ref.~\cite{Mathey2020}. Such a drive will lead to the two scales, as already discussed:
\begin{align}
\xi_{\perp}\sim \hat{v}_\perp^{-\frac{1}{nz+1}}, && \xi_{\parallel}\sim \hat{v}_\parallel^{-\frac{1}{nz-1}}.
\end{align}
The smaller of these scales will be observable. We see that a drive of order $\geq 2$ is needed to make the subleading scale relevant, see again \Tab{Tab:DriveOverviewXY}. For very low velocities, $\xi_{\perp}$ will be smaller and thereby observable, by increasing the velocity up to some crossover scale $\hat{v}^*(\phi)$ the scale $\xi_{\parallel}$ will become smaller and thereby observable. This is presented schematically in \Fig{Fig:SchematicCrossover}{} for the related scaling of the excitation density $n_E$ as a function of $\phi$. In practice, there will also be a velocity $\hat{v}_{\text{cut}}$, which separates the universal scaling regime from a non-universal regime at high velocities.

\subsection{Phase diagrams \& orthogonality issue \label{Sec:OrthogonalityIssue}}
As already mentioned, the mapping between the spin-coupling space and fermionic coupling space is not angle-preserving (for $k\to 0$). This becomes an important issue once we want to define the notion of a transversal and parallel drive. In the fermionic case we know the geometry of the coupling space, which was inferred from the \RG-analysis. A naive use of the notions `transversal' and `longitudinal' in the spin-coupling space can be misleading. To make this transparent, consider the purely transversal drive $\phi_1=0$ in the fermionic leading coupling $\hat{\Delta}$. In spin-coupling space the drive takes rather the form of path $(1)$ with $\phi_1'\neq 0$ in \Fig{Fig:SpinVsFermion}{}, such that naively we would think of this drive as not being purely transversal in the spin-coupling space. The resolution is that what \emph{defines} transversal and longitudinal needs to be inferred from the \RG-analysis (here the exactly solvable fermionic version) and cannot in general be done at the level of the \emph{microscopic} coupling phase diagram. Therefore, we compare the spin-coupling phase diagram of the XY model with the fermionic version in \Fig{Fig:SpinVsFermion}. To quantify the deviations in the angles, we define the transformation matrix~$\mathcal{M}$
\begin{align}
\mathcal{M} \begin{pmatrix} g/\gamma\\ J/\gamma \end{pmatrix} \approx \begin{pmatrix} \hat{\Delta} \\ \hat{D}_2  \end{pmatrix}, && \mathcal{M}= \begin{pmatrix} 1 & -1 \\ 0 & \frac12  \end{pmatrix},
\label{Eq:RelationCouplings}
\end{align}
which gives us the corresponding fermionic couplings (again valid for $k \to 0$ in the vicinity of the critical point). $\mathcal{M}$ describes a non-orthogonal transformation between spin- and fermion-couplings: $\mathcal{M}^{-1} \neq \mathcal{M}^T$. The correct fermionic angle $\pi/2-\phi$ relative to the subleading direction/phase boundary and the (misleading) spin-coupling angle $\pi/2-\phi'$ enclosed with the respective phase boundaries are defined as
\begin{align}
\begin{aligned}
\sin(\phi)&=\frac{\left\langle \hat{\vec{v}},\mathbf{e}_{\hat{D}_2} \right\rangle}{\left|\hat{\vec{v}}\right|\left| \mathbf{e}_{\hat{D}_2} \right|}, \\
\sin(\phi')&=\frac{\left\langle \mathcal{M}^{-1}\hat{\vec{v}},\mathcal{M}^{-1}\mathbf{e}_{\hat{D}_2} \right\rangle}{\left|\mathcal{M}^{-1}\hat{\vec{v}}\right| \left| \mathcal{M}^{-1}\mathbf{e}_{\hat{D}_2} \right|},
\label{Eq:AngleDefinition}
\end{aligned}
\end{align}
where $\mathbf{e}_j$ is the unit-vector in coupling-direction $j$. In principle, $\phi$ and $\phi'$ are not the same. Although being transversal depends on the choice of the coordinate system, `longitudinal' is an invariant property independently of the choice of $\mathcal{M}$ [by construction of \Eq{Eq:AngleDefinition}]. Therefore, a purely longitudinal drive is longitudinal in all coordinate systems in close vicinity to the critical point. We analyze such a drive in Sec.~\ref{Sec:TIParallelDrive}.

\section{Physical observables - exact \& approximate approach \label{Sec:Techniques}}
In the previous sections, we discussed the universal properties of the transverse XY model, as well as the construction of a proper drive to make the subleading scaling observable, which correspond to the first two steps of the general agenda in Sec.~\ref{Sec:KeyResults}. The last and final step is to relate the breaking of adiabaticity to observables. A direct measure of adiabaticity breaking is given by the density of excited quasi-particles in \Eq{Eq:ExcitationDensityGeneral}. Nevertheless, the physical model is the spin model, and therefore we have to identify the meaning of the excitation density in the spin-representation. Following the logic of Ref.~\cite{Dziarmaga2005}, we identify the excitation density as the density of spin flips, once we are deep in the paramagnetic phase. Formally, we need to translate \Eq{Eq:ExcitationDensityGeneral} into the spin-language and identify the fermionic quasi-particle operators $\chi_k$ with the corresponding spin-operators at the end of the drive \cite{Dziarmaga2005}.

For a drive ending (deep) in the paramagnetic phase, we get for $g-J\gg \gamma$ or $J(g-J) \gg \gamma^2$: $\chi_k \approx c_k$. Therefore, the excitation density $n_E$ takes the form\footnote{The spin Hamiltonian becomes a transverse XX chain for a generalized drive and $|t| \to \infty$: 
\begin{align*}
\hat{H}_{\text{end}} \propto& -\left(\frac{\hat{v}_\perp}{2}+\hat{v}_\parallel \right)\sum_l \sigma^z_l -\frac{\hat{v}_\parallel}{2} \sum_l \left( \sigma^x_l\sigma_{l+1}^x+\sigma_l^y\sigma_{l+1}^y \right).
\end{align*}
This Hamiltonian is particularly simple, as it is diagonal in momentum-space without any further (Bogoliubov-) transformation and therefore $\chi_k=c_k$. The ground state is $| \uparrow \uparrow \hdots \rangle$.}
\begin{align}
\begin{aligned}
n_E&= \frac{1}{N}\left\langle \sum_k \chi_k^\dagger \chi_k \right\rangle \approx \frac{1}{N}\left\langle \sum_k c_k^\dagger c_k \right\rangle \\
 &=\frac{1}{N} \left\langle \sum_l c_l^\dagger c_l \right\rangle = \frac{1}{N}\left\langle \sum_l \frac{1}{2}\left(1-\sigma_l^z\right) \right\rangle,
\end{aligned}
\end{align}
such that it results from single spin-flips in accord with the paramagnetic phase \cite{Dziarmaga2005}. Therefore, the deviation from perfect magnetization in $z$-direction, $1-\langle \sigma_z \rangle$, at the end of the drive can be used as a direct observable. For $N\to \infty$ the excitation density reads
\begin{align}
n_E=\frac{1}{N} \sum_k p_k, && \lim_{N\to \infty} n_E = \int_{-\pi}^{\pi} p_k \frac{d(ka)}{2\pi}.
\label{Eq:ExcitationDensity}
\end{align}
We can furthermore separate the universal and non-universal parts of the Hamiltonian, based on the UV cutoff $\Lambda$ (which can differ from $1/a$ by a prefactor) as
\begin{align}
n_E&=\frac{a\Lambda}{\pi}\int_{0}^{\pi/a\Lambda^{-1}} p_k \, d\hat{k}  \label{Eq:SeparationExcitaitonDensity} \\
&= \underbrace{\frac{a\Lambda}{\pi}\int_{0}^{1} p_k \, d\hat{k}}_{\text{universal}}+ \underbrace{\frac{a\Lambda}{\pi}\int_{1}^{\pi/a\Lambda^{-1}} p_k \, d\hat{k}}_{\text{non-universal}}, \nonumber
\end{align}
assuming $p_k=p_{-k}$. In the next Sec.~\ref{Sec:DimensionalConsideration}, we identify the only two parameters, controlling the dynamics and ultimately the behaviour of $p_k$, which we approximate in the Secs.~\ref{Sec:AdiabaticApproximations} and \ref{Sec:BasicsAI} from different perspectives. The role of different $\Lambda$'s are discussed in Sec.~\ref{Sec:ModelDependence}.

\subsection{Dimensional considerations \label{Sec:DimensionalConsideration}}
We identify the two dimensionless parameters $\adparameter$ and $\hat{\mu}_k$, which also govern the analytically exact and approximate solutions in the next subsections. The prototypical, `universal' two-level Hamiltonian valid for $k \leq \Lambda$ for the  transverse XY and similar models reads:
\begin{align}
\begin{aligned}
&h_k(t)=(v_kt^n+M_k)\sigma_z+\Omega_k \sigma_x, \\
&v_k:= v_\perp + v_\parallel k^\subexponent,\\
&M_k:=\Delta^0 + D_\subexponent^0 k^\subexponent,\\
&\Omega_k:= D_z k^z .
\label{Eq:PrototypicalModel}
\end{aligned}
\end{align}
This form is valid close to the critical point, reached at $t=0$ [see again \Eq{Eq:thermoXY}: $z=1$ and $\subexponent=2$ for the transverse XY model]. To identify the two dimensionless parameters for such a model, we can directly rescale the model \Eq{Eq:PrototypicalModel}, such that the off-diagonal terms become $1$ (similar in spirit to the adiabatic-impulse approximation \cite{Damski2005,Damski2006}), see also \Tab{Tab:PrototypicalCouplings}:
\begin{align}
&\text{rescaled:}&&\bar{h}_k= \left( \adparameter \hat{\adtime}_k^n +\hat{\mu}_k\right) \sigma_z + \sigma_x, \nonumber  \\
&\text{eigenvalues:}&&\mathcal{E}(k,\hat{\adtime})=\pm \sqrt{\left( \adparameter \hat{\adtime}_k^n +\hat{\mu}_k\right)^2 + 1}, \label{Eq:DimensionlessTwoLevel}\\
&&&\hat{\adtime}_k:=\Omega_k t, \,\ \adparameter: = \frac{v_k}{\Omega_k^{n+1}}, \,\ \hat{\mu}_k:=\frac{M_k}{\Omega_k}. \nonumber 
\end{align}
The only parameters left are $\adparameter$ and $\hat{\mu}_k$ (see also Refs.~\cite{Vitanov1999a,Suominen1992} for the linear and quadratic case). As we will show in the following, $\adparameter$, a generalized `velocity', controls adiabaticity and we call it the adiabaticity parameter. The simple but decisive relation $\adparameterstar \approx 1$ indicates the breaking of adiabaticity (as long as $\hat{\mu}_k$ is negligible), which we will investigate in the next subsections. For the given model $\adparameter$ and $\hat{\mu}_k$ read:
\begin{align}
\begin{aligned}
&\adparameter=\hat{v}_\perp\left( \frac{\Lambda}{k}\right)^{nz+1/\nu}+\hat{v}_\parallel \left( \frac{\Lambda}{k}\right)^{nz+\text{dim}[D_\subexponent]},\\
&\hat{\mu}_k=\hat{\Delta}^0\left( \frac{\Lambda}{k}\right)^{1/\nu}+\hat{D}_\subexponent^0 \left( \frac{\Lambda}{k}\right)^{\text{dim}[D_\subexponent]},
\label{Eq:ExampleEpsilonMu}
\end{aligned}
\end{align}
where the two terms in $\hat{\mu}_k$ are reminiscent of the scale-dependent couplings from \Eq{Eq:ExampleRGFlow}, especially $\text{dim}[D_\subexponent]<0$ being an irrelevant exponent. Similarly, $\adparameter$ entails the two $k$-dependent velocity couplings. This is the main result of the dimensional analysis.

\begin{table}
\begin{tabular}{l | l | l}
\toprule
\multicolumn{3}{c}{Couplings in the prototypical model}  \\ \colrule
$t$ & $\hat{\tau}=D_z (k/\Lambda)^z t \Lambda^{z}$ &  \\
$\Delta^0$ & $\hat{\Delta}^0=\frac{\Delta^0}{D_z}\Lambda^{-z}$ & $\text{dim}[\Delta^0]=z=1/\nu$\\
$D_\subexponent^0$ & $\hat{D}_\subexponent^0 = \frac{D_\subexponent^0}{D_z} \Lambda^{\subexponent-z}$ & $\text{dim}[D_\subexponent^0]=-(\subexponent-z)$ \\
$v_\perp$ &  $\hat{v}_\perp =\frac{v_\perp}{D_z^{n+1}} \Lambda^{-(nz+z)}$ & $\text{dim}[v_\perp]=nz+\text{dim}[\Delta^0]$\\
$v_\parallel$ & $\hat{v}_\parallel = \frac{v_\parallel}{D_z^{n+1}} \Lambda^{-(nz-(\subexponent-z))}$ & $\text{dim}[v_\parallel]=nz+\text{dim}[D_\subexponent^0]$ \\ \botrule
\end{tabular}
\caption{Couplings and scaling dimensions of the prototypical model.}
\label{Tab:PrototypicalCouplings}
\end{table}

\subsection{Approximation schemes}

We are interested in the evolution of the $k$-resolved excitation density $p_k$, especially for nonlinear, polynomial drives, to be able to evaluate \Eq{Eq:ExcitationDensity}. For an arbitrary nonlinear drive even the two-level evolution is not analytically solvable. The necessity for the nonlinear cases results from $z=1$ for the transverse Ising model, which implies that a linear drive does not allow one to make subleading scalings observable (see \Tab{Tab:DriveOverviewXY}). Nevertheless, it allows us to consider (analytically exact) a minimal fermionic model with $z=3$ (see Sec.~\ref{Chap:MinimalModel}). We will consider drives starting in one of the phases at $|t_i|=\infty$ and either ending at the transition at $t_f=0$ (used in Sec.~\ref{Eq:GeneralizedDrivesXY}) or at $|t_f|=\infty$ (used in Sec.~\ref{Chap:MinimalModel}).

Due to the lack of exact solutions (for nonlinear drives), we approximate $p_k$ and clarify the meaning of $\adparameter$ and $\hat{\mu}_k$ on the breaking or restoring of adiabaticity. To this end, we take two different perspectives. The first is the adiabatic perspective.
This includes the leading order of the adiabatic perturbation theory \cite{DeGrandi2010,DeGrandi2010b,Gritsev2010,Deng2009b}, 
where the initial state is one of the eigenstates and we assume a weak occupation of the other (eigen)states and are interested in how $\adparameter$ and $\hat{\mu}_k$ control this assumption. The role of $\hat{\mu}_k$ becomes prominent for parallel drives, see Sec.~\ref{Sec:TIParallelDrive}. Furthermore, for drives from $t_i=-\infty$ to $t_f=\infty$ non-analytic contributions are dominant, which includes the (adiabatic) \DDP approximation \cite{Davis1976,Joye1993,Vitanov1999} (Appendix~\ref{App:AdiabaitcApproach}) and the analytically exact asymptotic Landau-Zener solution as a special case. The analytically exact solution will be our starting point in Sec.~\ref{Chap:MinimalModel}. The second perspective is the adiabatic-impulse approximation \cite{Damski2005,Damski2006} (see Sec.~\ref{Sec:BasicsAI}), rather based on the physical intuition of the \KZM. It complements the adiabatic perspective by working accurately in the limit of $\adparameter \gg 1$ and strong occupation.\footnote{The ground state of the full model corresponds to the excited states of the two-level systems, see again \Eq{Eq:DiabaticRepresentation}. Therefore, we are interested in the (de)excitation probability for the two-level systems. Nevertheless, this is not changing any of the arguments and we keep referring to initial states as the ground states.}

\subsubsection{Adiabatic approximations \label{Sec:AdiabaticApproximations}}

\emph{First order adiabatic perturbation theory}. The starting point is the adiabatic representation as in \Eq{Eq:AdiabaticRepresentiation1} and \Eq{Eq:adiabaticBogoliubov} of the state [but for the rescaled model \Eq{Eq:DimensionlessTwoLevel}]. Our quantity of interest is $p_k=|\adiabaticbasisB(\hat{\adtime}_k)|^2$, which can first of all be approximated by a perturbative expansion. Assuming $\adiabaticbasisA(\hat{\adtime}_{k,i})=1$ and a weak occupation of the excited state, the leading contribution in powers of $\adparameter$ at $\hat{\adtime}_{k,f}=0$ for $\hat{\adtime}_{k,i}=-\infty$ is given by \cite{DeGrandi2010b,DeGrandi2010}
\begin{align}
\begin{aligned}
p_k&\approx  \left(\frac{n!}{2^{n+1}}\right)^2  \adparameter^2 \frac{1}{(\mathcal{E}_k(0))^{2(n+2)}}\\
&\sim  \begin{cases} \left(\adparameter \right)^2 & :\quad \hat{\mu}_k \ll 1 \\ \left(\frac{\adparameter}{\hat{\mu}_k^{n+2}}\right)^2& :\quad \hat{\mu}_k \gg 1\end{cases}.
\label{Eq:AdiabaticPerturbationTheory}
\end{aligned}
\end{align}
As long as $\adparameter \ll 1$ this (first-order) approximation is self-consistent, such that $\hat{\epsilon}_{k^*}\approx 1$ gives an estimate of its breakdown (see also Ref.~\cite{Dziarmaga2010})\footnote{For a more refined treatment beyond this first order see Refs.\cite{DeGrandi2010,DeGrandi2010b,Gritsev2010}.} . Nevertheless, it also encodes that once $\hat{\mu}_k \gg 1$ adiabaticity can be restored, as we will see in Sec.~\ref{Sec:TIParallelDrive}.

\emph{\DDP approximation \& Landau-Zener}: In the limit $\hat{\adtime}_{k,f} \to + \infty$ the leading contribution in the limit $\adparameter\to 0$ stems from a non-analytic contribution, which we discuss in Appendix~\ref{App:AdiabaitcApproach} and refer to as the \DDP approximation. For the linear case, the approximation actually yields the \emph{exact} asymptotic Landau-Zener-Majorana-Stückelberg \cite{Zener1932,Majorana1932,Stueckelberg1932} result
\begin{align}
\begin{aligned}
p_k& =\exp\left(-\pi \adparameter^{-1}\right).
\label{Eq:ExactSolutionLinearDrive}
\end{aligned}
\end{align}
The formula nicely shows the emergence of the adiabaticity parameter as identified in \Eq{Eq:DimensionlessTwoLevel}. Once $\pi^{-1} \adparameter\gtrsim 1$ the $k$-resolved excitation density is $\mathcal{O}(1)$ in agreement with the adiabaticity breaking. The formula guides the discussion in Sec.~\ref{Chap:MinimalModel}.

\subsubsection{Adiabatic-impulse approximation \label{Sec:BasicsAI}} 
Based on the intuition of the \KZM, the $k$-resolved excitation densities $p_k$ can be approximated by separating the evolution into an adiabatic part and a (frozen) impulse part (adiabatic-impulse (AI) approximation \cite{Damski2005,Damski2006}) for each two-level system, which was already successfully applied to the transverse Ising model \cite{Damski2006}.
We will mainly use the AI approximation for drives in the transverse XY model, which start at $t_i=\infty$ deep in the paramagnetic phase and end at $t_f=0$ at the transition (or vice versa). Therefore, each single $k$-mode evolution has its `minimal' gap at $t=0$. Switching to the rescaled model, this implies $\adparameter \hat{\adtime}_k^n \ge 0$ [see again \Eq{Eq:DimensionlessTwoLevel}]. The basic intuition is that, starting from the ground state, the evolution is initially adiabatic up to time $\hat{\adtime}_k^*$, where it essentially freezes. The excitation density is accordingly approximated as 
\begin{align}
p_k\approx |\langle - (0,k)| + (\hat{\adtime}_k^*,k)\rangle|^2=|\langle - (\hat{\adtime}_k^*,k)| + (0,k)\rangle|^2
\label{Eq:AIProbability}
\end{align}
(in both cases we stay on the paramagnetic side). The only ingredients are the known eigenstates of the Hamiltonian as well as the adiabaticity-breaking time $\hat{\adtime}^*_k$. To estimate the time of adiabaticity breaking, we ask, if, at a given time $t$, the necessary time to reach the `minimal' gap is larger or smaller than the characteristic time scale (inverse gap). Adiabaticity is estimated to be broken, once the necessary time gets as small as the characteristic time scale:
\begin{align}
\begin{aligned}
&\frac{1}{2\mathcal{E}(k,\hat{\adtime}_k^*)}=\frac{1}{2\sqrt{\left( \adparameter \hat{\adtime}^{* \,n}_k +\hat{\mu}_k\right)^2 + 1}}= \alpha_n \hat{\adtime}^*_k .
\label{Eq:BreakingTimeDefinition}
\end{aligned}
\end{align}
There are two cases of interest: once $\hat{\mu}_k$ is negligible, $\hat{t}_k=0$ corresponds to the anti-`crossing' center and $\alpha_n$ can be fixed by comparing to a diabatic expansion (see \cite{Damski2006}). From this point of view, the approximation is complementary to the adiabatic expansion. This is reasonable, as it is expected that the overall excitation density is dominated by the `fast' modes with $\adparameter \gg 1$ \cite{Dziarmaga2010}. This is the relevant scenario for the generalized drives, as $\hat{\mu}_k \to0$ in the critical region for $k\to0$ [see \Eq{Eq:ExampleEpsilonMu} for $\hat{\Delta}^0=0$]. The other, extreme, case corresponds to a purely parallel drive with a fixed distance $\hat{\Delta}^0>0$ to the critical line. Here $\hat{\mu}_k$ grows for $k\to 0$ [see \Eq{Eq:ExampleEpsilonMu} for $\hat{\Delta}^0>0$]. Combining \Eqs{Eq:AIProbability} and \eq{Eq:BreakingTimeDefinition} we get that excitations can be suppressed in this case, similarly to the adiabatic expansion in \Eq{Eq:AdiabaticPerturbationTheory}. Nevertheless, the occupation is overestimated in the AI approximation. We discuss this special case in Sec.~\ref{Sec:TIParallelDrive}.

\section{Analytical solution: fermionic minimal model \label{Chap:MinimalModel}}

To demonstrate the emergence of two competing scales in a model driven transversally as well as longitudinally close to a second order phase transition, we first of all consider an analytically solvable case (similar in spirit to the exactly solvable model discussed in Ref.~\cite{Mathey2020}). We show that similarly to \Eq{Eq:IntroTwoScales}, also the excitation density is composed out of two scales, which can both be observable by tuning the velocity:
\begin{align}
\xi^* \sim \text{Min}[\xi_\perp,\xi_\parallel] &&   \Leftrightarrow &&  n_E\sim  \text{Max}[n_{\perp},n_{\parallel}] .
\end{align}
To extract these scales, we analyze the dominant contributions of the $k$-resolved excitations $p_k$ in the (integrated) excitation density $n_E$.

\subsection{Minimal model and generalized drive}
The starting point is a minimal fermionic model, which has a structure similar to transverse XY model but is in a different universality class ($z=3,\nu=1/3$). Then a linear drive is enough to make the subleading scaling observable [see \Eq{Eq:IntroSubScaling}], and our mechanism can be shown to emerge within an analytical analysis (see Sec.~\ref{Sec:AdiabaticApproximations}). At this (exact) level we can outline the general strategy, which will then be used for the transverse XY model. The long wavelength model is defined as (for $N\to \infty$)

\begin{align}
\begin{aligned}
&H=\int_{-\Lambda}^{\Lambda} \frac{dk}{2\pi}\left[ (\Delta^0+{v}_\perp {t}) {\phi}^\dagger_k {\phi}_k \right. \\
&\qquad \qquad + \frac12 D_3 k^3\left[{\phi}^\dagger_{k}{\phi}^\dagger_{-k} +{\phi}_{-k} {\phi}_k\right] \\
& \left. \qquad \qquad +({D}_4^0+{v}_\parallel {t}) k^4 \hat{\phi}^\dagger_k \hat{\phi}_k \right] .
\label{Eq:DimensionfullMinimalModel}
\end{aligned}
\end{align}

It can be thought of as the expansion of a spin model represented in Jordan-Wigner fermions, valid only up to some UV-cutoff $\Lambda$, which we chose to be $1/a$ for all plots. Actually, a comparable spin model (`extended quantum XY chain') was recently used in Ref.~\cite{Sadhukhan2019}. In rescaled, dimensionless couplings the Hamiltonian of the two-level model is written as ($\hat{k}:=k/\Lambda$), see also \Tab{Tab:CouplingsMinimalModel}:
\begin{align}
\begin{aligned}
&\hat{h}_k= \begin{pmatrix}\hat{\Delta}(t)+  \hat{D}_4(t)\hat{k}^4 &  \hat{k}^3 \\ \hat{k}^3 & -(\hat{\Delta}(t)+ \hat{D}_4(t)\hat{k}^4) \end{pmatrix} . \\
\label{Eq:MinimalModel}
\end{aligned}
\end{align}

\begin{table}[h]
\begin{tabular}{l | l | l}
\toprule
$\Delta^0$ & $\hat{\Delta}^0=\frac{\Delta^0}{D_3}\Lambda^{-3}$ & $\text{dim}[\Delta^0]=3$ \\
$v_\perp$ & $\hat{v}_\perp=\frac{v_\perp}{D_3^2}\Lambda^{-6}$ & $\text{dim}[v_\perp]=6$ \\
$D_4^0$ & $\hat{D}_4^0=\frac{D_4^0}{D_3}\Lambda^1$ & $\text{dim}[D_4^0]=-1$ \\
$v_\parallel$ & $\hat{v}_\parallel= \frac{v_\parallel}{D_3^2}\Lambda^{-2}$ & $\text{dim}[v_\parallel]=2$ \\ \botrule
\end{tabular}
\caption{Couplings and scaling dimensions for the minimal model.}
\label{Tab:CouplingsMinimalModel}
\end{table}

\begin{figure}[H]
\centering
\begin{tikzpicture}
\node[anchor=south west,inner sep=0] at (0,0) {\includegraphics[width=0.95\textwidth]{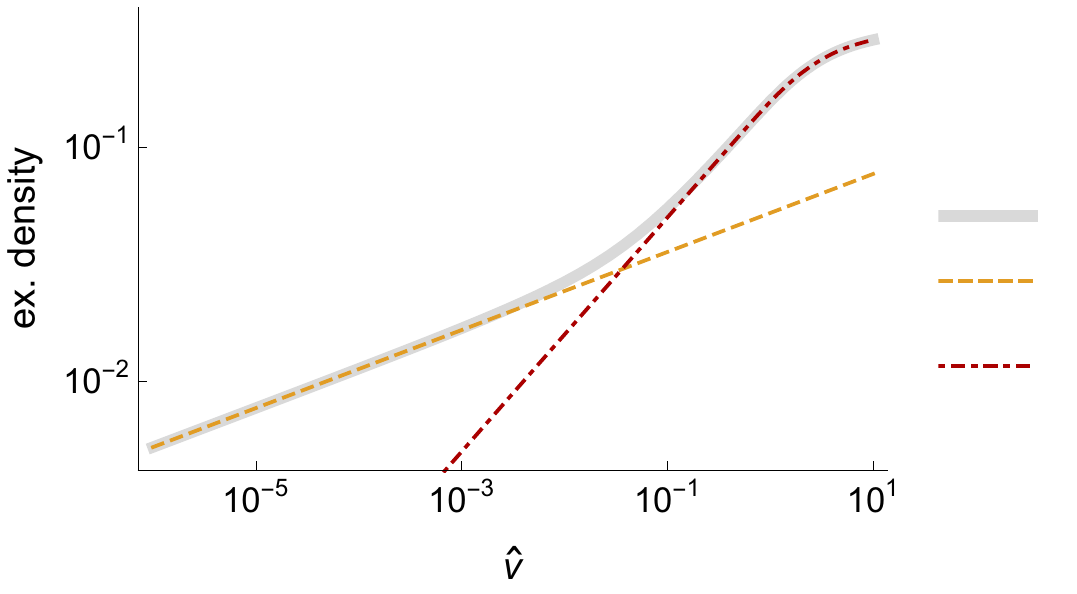}};
\filldraw[color=orange!40, fill=orange!40, ultra thick, opacity=0.2] (1.1,1) rectangle (3.5,3);
\filldraw[color=red!40, fill=red!40, ultra thick, opacity=0.2] (5,1) rectangle (6,4.5);
\node at (2.2,2.2) {$\sim \hat{v}^{\frac16}$};
\node at (5.5,4.2) {$\sim \hat{v}^{\frac12}$};
\node[anchor=south west] at (7.1,2.9) {\large $n_E(\hat{v},\phi)$};
\node[anchor=south west] at (7.1,2.315) {\large $n_{\perp}(\hat{v}_\perp)$};
\node[anchor=south west] at (7.1,1.7) {\large $n_{\parallel}(\hat{v}_\parallel)$};

\node [rotate=90] at (0.23,3.8) {$n_E$};

\end{tikzpicture}
\caption{Plotted (log-log) are the excitation density $n_E(\hat{v},\phi)$ and the approximations in the different scaling regimes, given in \Eq{Eq:TwoApproximations}, for $\phi=\pi/2-10^{-4}$. Two different scaling-regimes can be identified with exponents $\frac16$ and $\frac12$, corresponding to the KZM scaling and the subleading scaling. The extent of both scaling regimes is indicated by the colored boxes and depends on $\phi$.}
\label{Fig:MinimalModelFirstSub}
\end{figure}
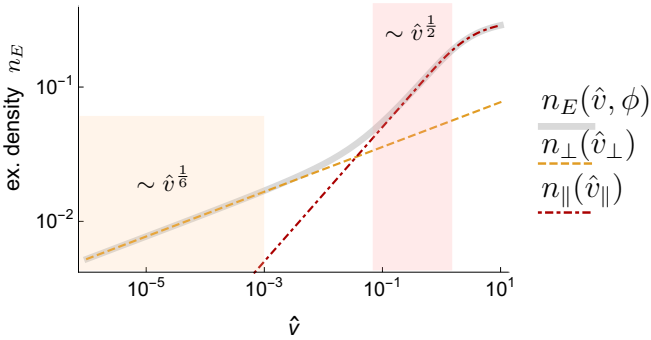

First of all, this model has a dynamical exponent of $z=3$, and a leading critical exponent given by $\nu=1/3$. This allows us to make the scaling dimension of the first subleading term, $D_4k^4$ with $\text{dim}[D_4]$, observable. To this end, we consider a drive of the relevant $\hat{\Delta}$-parameter and the subleading $k^4$-term, which starts at $t_i=-\infty$ and ends at $t_{f}=+\infty$:
\begin{align}
\left.
\begin{aligned}
&\hat{\Delta}(\hat{t})=\hat{v}_\perp \hat{t} \\
&\hat{D}_4(\hat{t})=\hat{v}_\parallel \hat{t}\\
\end{aligned} \right\} \,\   \vec{\hat{v}}=\begin{pmatrix} \hat{v}_\perp \\ \hat{v}_\parallel \end{pmatrix} = \hat{v}\begin{pmatrix} \cos(\phi) \\ \sin(\phi) \end{pmatrix}.
\end{align}
The \RG prediction for the excitation density $n_E(\hat{v},\phi)$ for such a drive-protocol yields [\Eqs{Eq:IntroSubScaling} and \eq{Eq:IntroExvsLength}]:
\begin{align}
\begin{aligned}
&\text{\KZM scaling:} && n_E(\hat{v}_\perp \gg \hat{v}_\parallel) &\sim \hat{v}_\perp^{\frac{1}{6}},\\
&\text{subleading scaling:} &&n_E(\hat{v}_\parallel \gg \hat{v}_\perp ) &\sim \hat{v}_\parallel^{\frac{1}{2}}.
\label{Eq:ExpectationsMinimalModel}
\end{aligned}
\end{align}
The strategy now is to extract these scales from the exactly known $p(\hat{k},\vec{\hat{v}})$ from \Eq{Eq:ExactSolutionLinearDrive}, and thereby $n_E(\hat{v},\phi)$, from \Eq{Eq:ExcitationDensity}. The exact $k$-resolved excitation density and (integrated) excitation density read:
\begin{align}
\begin{aligned}
&p(\hat{k},\vec{\hat{v}})=\exp \left(-\pi \frac{\hat{k}^6}{\hat{v}_\perp+\hat{v}_\parallel \hat{k}^4}\right)=\exp(-\pi \adparameter^{-1}),\\
&n_E(\hat{v},\phi)=\frac{a\Lambda}{\pi}\int_{0}^{1}p(\hat{k},\vec{v})d\hat{k}.
\end{aligned}
\end{align}
A remark on the integration range: Since by construction the model is only valid up to $k=\Lambda \Leftrightarrow \hat{k}=1$ the $k$-integration is restricted as well. Differently put, we only consider the universal content and only use the first part in \Eq{Eq:SeparationExcitaitonDensity}. This is reasonable in the range, where we expect the \KZM to apply, especially towards smaller velocities. Nevertheless, it limits the validity of the model towards larger velocities, in particular once the density starts to saturate. An example of the excitation density $n_E$ for different velocities $\hat{v}$ for a general drive $\vec{\hat{v}}$ is given in \Fig{Fig:MinimalModelFirstSub}{}, saturating at $n_E \to 1/\pi$. 

To gain more insight into the different (scaling) regimes of $n_E$, we approximate the full expression in multiple steps. First of all, $p(\hat{k},\vec{\hat{v}})$ has two different regimes for $\hat{k}\to 0$ and for larger $\hat{k}$, determined by the asymptotic forms of the adiabaticity parameter $\adparameter$
\begin{align}
\begin{aligned}
&p\left(\hat{k} \ll \hat{\kappa} ,\vec{\hat{v}}\right) \sim \exp \left(-\pi \frac{\hat{k}^6}{\hat{v}_\perp }\right), \\ 
&p\left(\hat{k} \gg \hat{\kappa} ,\vec{\hat{v}}\right) \sim \exp \left( -\pi \frac{\hat{k}^2}{\hat{v}_\parallel} \right),  
\label{Eq:ProbabilityDistributionsMinimalModel}
\end{aligned}
\end{align}
where we introduced the crossover scale ${\hat{\kappa}:=(\hat{v}_\perp/\hat{v}_\parallel)^{1/4}}$. Which of these forms will be observable in $n_E$ depends strongly on which contribution will dominate. For $\hat{v}_\parallel=0$, we get the expected \KZM scaling, for $\hat{v}_\perp\to 0$ (and therefore $\hat{\kappa}\to 0$) we get the subleading scaling, see \Eq{Eq:ExpectationsMinimalModel}.

\begin{figure*}[t]
\centering

\begin{tikzpicture}

\node[anchor=south east,inner sep=0] at (-3,0.4) {\includegraphics[width=0.35\textwidth]{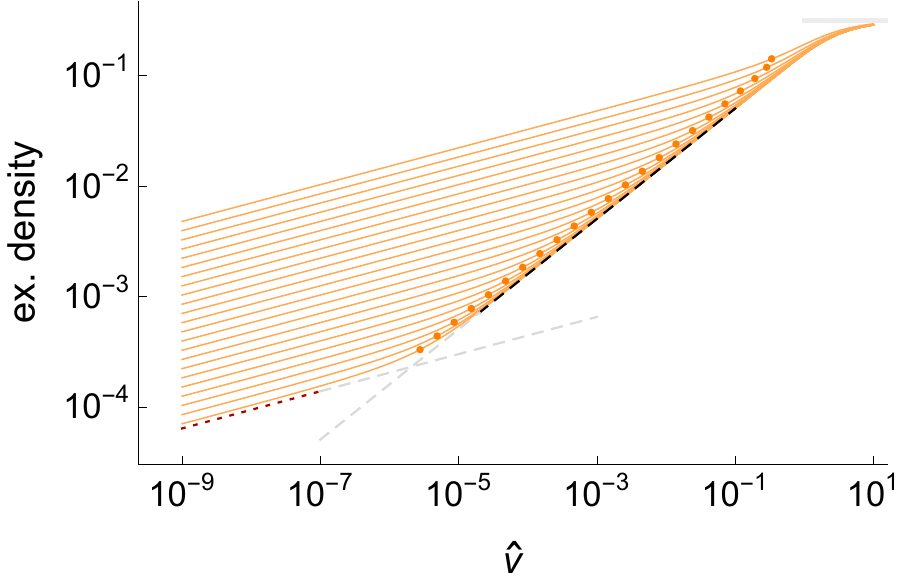}};
\node[anchor=south west,inner sep=0] at (-0.5,0.4) {\includegraphics[width=0.475\textwidth]{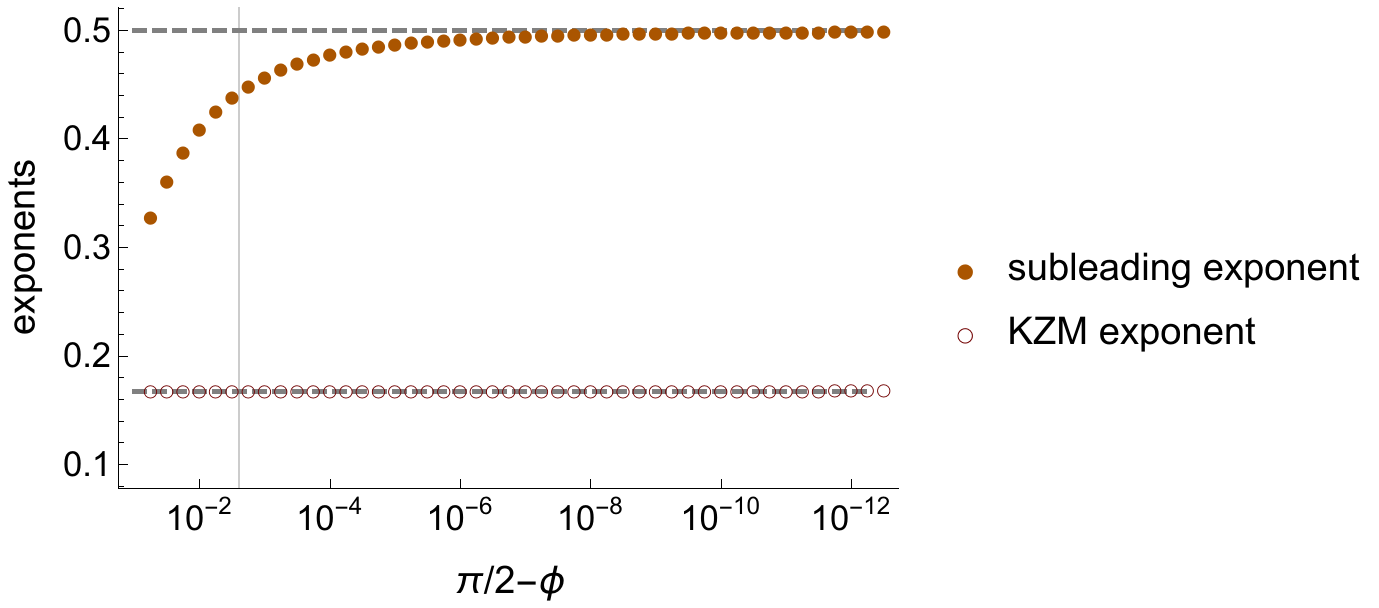}};
\node[anchor=north east,inner sep=0] at (-3,0) {\includegraphics[width=0.35\textwidth]{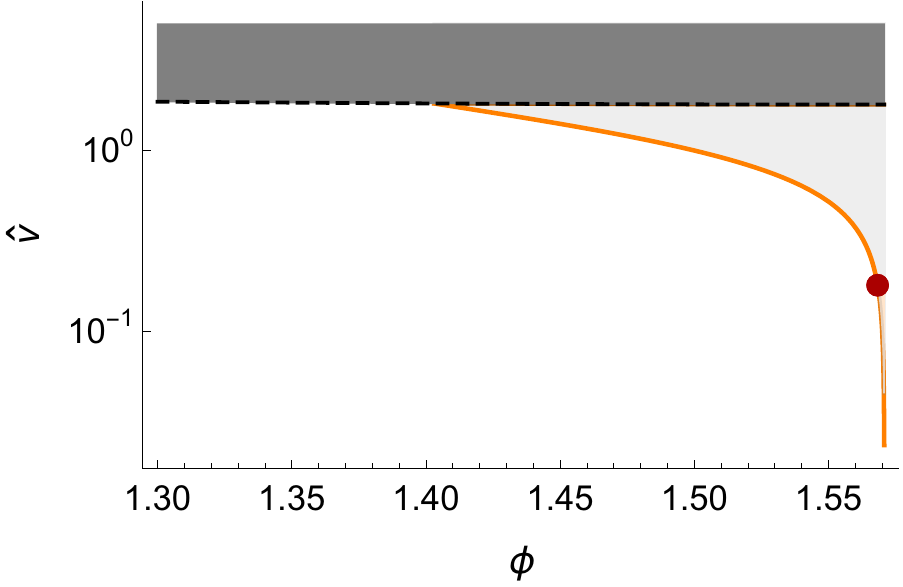}};
\node[anchor=north west,inner sep=0] at (-0.5,0) {\includegraphics[width=0.5\textwidth]{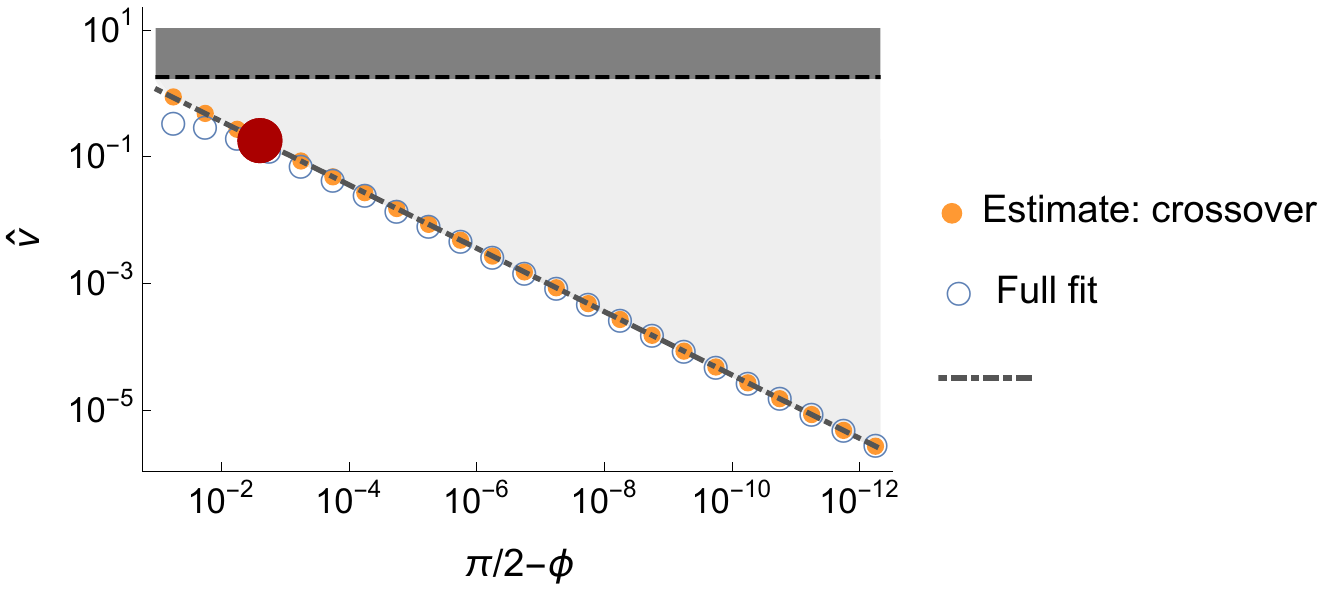}};
\node at (-7.5,4) {\textbf{(a)}};
\node at (8,4) {\textbf{(b)}};
\node at (-7.5,-2.8) {\textbf{(c)}};
\node at (8,-2.8) {\textbf{(d)}};

\node at (7.1,-2.55) {\Eq{Eq:AnalyticalEstimateCrossover}};

\node [rotate=90] at (-9.1,3.9) {$n_E$};

\node at (-5.5,-0.4) {\textcolor{black}{non-universal}};
\node at (-5,-2.5) {KZM};
\node at (3.5,-1.2) {subleading};
\node at (3.5,-0.36) {\textcolor{black}{non-universal}};
\draw[color=gray!60, fill=gray!60, opacity=0.2] (-3,-2.75) -- (-0.5,0) -- (-0.5,-4) -- cycle;
\draw[gray, thick] (-3.5,-3.25) rectangle (-3,-1.75);
\draw[color=gray!60] (-0.5,-4) rectangle (5.6,0);
\end{tikzpicture}

\caption{\textbf{(a)} Set of curves $n_E(\hat{v},\phi)$, as in \Fig{Fig:MinimalModelFirstSub}{}, for the minimal fermionic model, and for different angles $\phi$ on a log-log scale. For low velocities the \KZM scaling results and the subleading scaling emerges for larger velocities up to some non-universal regime. The orange dots indicate the crossover velocities. \textbf{(b)} Scaling exponents extracted from the full fit of the curves in (a) (up to the non-universal regime) on a log-linear scale and the predicted exponents from the \RG (dashed lines). The vertical line indicates $\phi_{\text{min}}$. \textbf{(c)} Overview of the different regimes (log-scale) for a wider range of $\phi$. The new scaling is only well-extractable for $\phi\ge \phi_{\text{min}}$ (denoted by the red dot). The non-universal regime, indicated by the dashed line ($\hat{v}_{\text{cut}}$), is estimated from the saturation of $n_E$. \textbf{(d)} Universal scaling of the crossover velocity on a log-log scale for $\phi \to \pi/2$ that is estimated using  \Eq{Eq:CrossoverEstimateMinimalModel} (orange dots) and a full fit (blue circles) of the curves in \text{(a)}. Both sets are well approximated by the analytical expression \Eq{Eq:AnalyticalEstimateCrossover}{} (gray dash-dotted line) in agreement with the \RG-predictions.}
\label{Fig:MinimalModelCrossover}
\end{figure*}
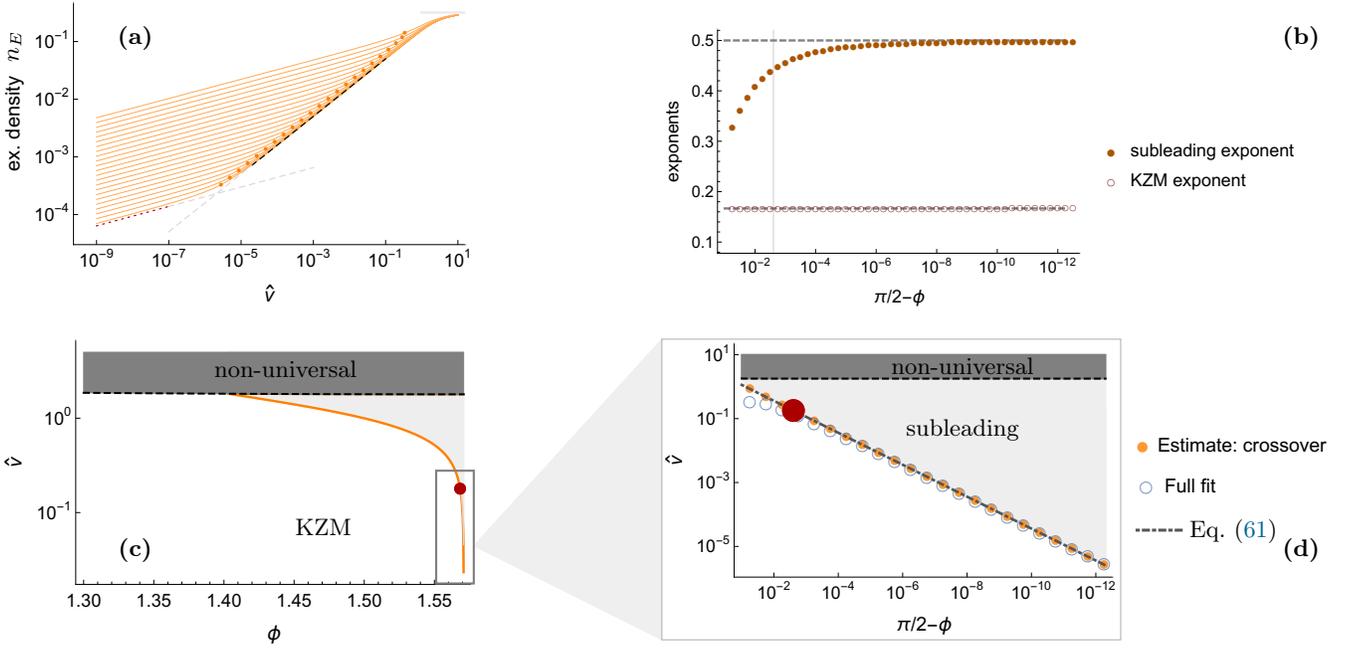

We see from \Fig{Fig:MinimalModelFirstSub}{} that the \KZM scaling emerges for low velocities, and that, for an intermediate range of velocities, the subleading scaling becomes observable. For higher velocities a non-universal regime is entered due to the saturation of the excitation density. To better understand the general case, we decompose the excitation density as [using that $p(k,\vec{\hat{v}})$ is symmetric in $k$]:
\begin{align}
n_E(\hat{v},\phi)&=\frac{a\Lambda}{\pi} \int_{0}^{\hat{\kappa}}p\left(k,\vec{\hat{v}}\right)d\hat{k} + \frac{a\Lambda}{\pi}\int_{\hat{\kappa}}^{1} p\left(k,\vec{\hat{v}}\right)d\hat{k} \nonumber \\
&\approx n_{\perp}(\hat{v}_\perp)+n_{\parallel}(\hat{v}_\parallel).
\end{align}
A rough approximation involves using \Eq{Eq:ProbabilityDistributionsMinimalModel}, setting all integration boundaries back to the full width $[0,1]$:
\begin{align}
\begin{aligned}
 n_{\perp}(\hat{v}_\perp)&\approx \frac{a\Lambda}{\pi} \int_{0}^{1} \exp \left( -\pi \frac{\hat{k}^6}{\hat{v}_\perp }\right) d\hat{k}, \\
 n_{\parallel}(\hat{v}_\parallel) &\approx \frac{a\Lambda}{\pi} \int_{0}^{1} \exp \left( -\pi \frac{\hat{k}^2}{\hat{v}_\parallel }\right) d\hat{k},
\label{Eq:TwoApproximations}
\end{aligned}
\end{align}
and approximating $n_E(\hat{v},\phi)\approx \text{Max}[n_{\perp}(\hat{v}_\perp),n_{\parallel}(\hat{v}_\parallel)]$. This becomes exact in the extreme cases $ \hat{\kappa} \to 0$ or $\hat{\kappa} \to 1$. These approximations are also shown in \Fig{Fig:MinimalModelFirstSub}{}, where we can see that the full excitation density has essentially two regimes, one described by $n_{\perp}(\hat{v}_\perp)$ at very low velocities and $n_{\parallel}(\hat{v}_\parallel)$ at higher velocities. Once the widths of the two $k$-resolved excitation densities are much smaller than $1$, we can roughly write 
\begin{align}
\begin{aligned}
& n_{\perp}(\hat{v}_\perp)  \approx b \frac{a\Lambda}{\pi} \cos(\phi)^{\frac16} \cdot \hat{v}^{\frac{1}{6}}, \qquad  b:= \frac{\Gamma(7/6)}{(\pi)^{1/6}},\\
& n_{\parallel}(\hat{v}_\parallel) \approx \frac{1}{2} \frac{a\Lambda}{\pi} \sin(\phi)^{\frac12} \cdot \hat{v}^{\frac{1}{2}}.
\label{Eq:AnalyticalExcitationDensities}
\end{aligned}
\end{align}
Therefore, the first term generates the \KZM scaling and the second the subleading scaling. The two identified scaling regimes are separated by a crossover velocity $\hat{v}^*$, which indicates the crossing over from the \KZM scaling at $\hat{v}<\hat{v}^*$ and the subleading scaling at velocities $\hat{v}>\hat{v}^*$. This scale also depends on the critical exponents and was estimated in Ref.~\cite{Mathey2020} for a drive of the leading coupling and one subleading coupling $\hat{g}_j$, where it is shown that:
\begin{align}
\begin{aligned}
&& |\hat{v}^* \cos(\phi) |^{\frac{1}{z+1/\nu}}&\sim |\hat{v}^* \sin(\phi)|^{\frac{1}{z+\text{dim}[g_j]}}, \\
\label{Eq:CrossoverRG}
\phi \to \pi/2: && \hat{v}^* &\sim |\pi/2-\phi|^{\frac{z+\text{dim}[g_j]}{1/\nu-\text{dim}[g_j]}}.
\end{aligned}
\end{align}
Here we identify $\hat{g}_j=\hat{D}_4$ with $\text{dim}[D_4]=-1$ and we expect from the \RG prediction a scaling of the form ${\hat{v}^*\sim |\pi/2-\phi|^{1/2}}$. Therefore, we are interested in extracting the two scaling exponents (\KZM and subleading) as well as the crossover scale $\hat{v}^*$ as a function of $\phi$. In a first step, we extract the crossover scaling analytically: At the level of the explicit model at hand [\Eq{Eq:MinimalModel}], this velocity scale can be estimated from (once we are in the scaling regime, cf. also \Fig{Fig:MinimalModelFirstSub}{}):
\begin{align}
n_{\perp}(\hat{v}_\perp^*) \approx n_{\parallel}(\hat{v}_\parallel^*).
\label{Eq:CrossoverEstimateMinimalModel}
\end{align}
For $\phi \to \pi/2$ this expression can be evaluated analytically based on \Eq{Eq:AnalyticalExcitationDensities} and gives 
\begin{align}
\hat{v}^*(\phi) \approx  (2 b)^3 |\pi/2-\phi|^{1/2},
\label{Eq:AnalyticalEstimateCrossover}
\end{align}
which is in full agreement with the \RG-predicted scaling. Besides the analytical estimate given above, we can also extract $\hat{v}^*(\phi)$ by directly (numerically) fitting the full curve $n_E(\hat{v},\phi)$ for fixed $\phi$, which is briefly described in Appendix~\ref{App:FittingDetails}. A typical set of curves for different $\phi$ is shown in \Fig{Fig:MinimalModelCrossover}{a}; the corresponding crossover velocity is plotted as a function of $|\pi/2-\phi|$ in \Fig{Fig:MinimalModelCrossover}{d}. The direct fit shows good agreement of the estimate \Eqs{Eq:CrossoverEstimateMinimalModel} and \eq{Eq:AnalyticalEstimateCrossover} and the RG-predicted scaling. The crossover relation from the full fit approaches the \RG-value for $\phi \to \pi/2$ and also fits well to the simple estimate. Nevertheless, as we can already anticipate from \Fig{Fig:MinimalModelCrossover}{b-d}, the velocity regime displaying subleading scaling gets very narrow for intermediate to small $\phi$, which makes it hard to extract a sensible exponent, see especially \Fig{Fig:MinimalModelCrossover}{c}. We quantify this by the value $\phi_{\text{min}}$ [red dot in \Fig{Fig:MinimalModelCrossover}{c,d}], which we define as the angle for which the subleading regime spans roughly one order of magnitude (to allow for a sensible measurement of the exponent).

\begin{figure}[h]

\centering

\begin{tikzpicture}

\node[anchor=south west,inner sep=0] at (0,0) {\includegraphics[width=0.9\textwidth]{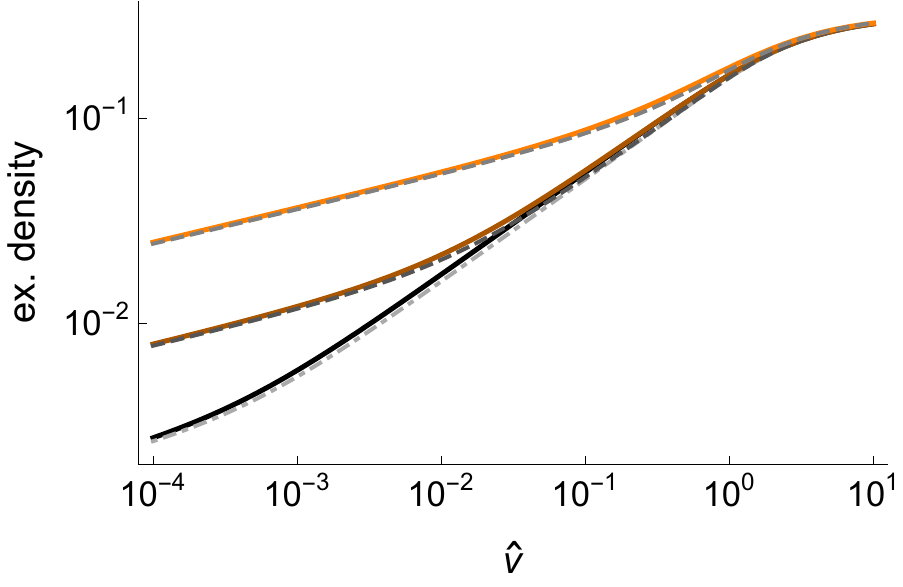}};
\node [rotate=90] at (0.25,4.2) {$n_E$};

\end{tikzpicture}

\caption{Comparison of the adiabatic-impulse approximation (full lines) of the excitation density and the exact result (dashed lines) for the minimal model for angles $\phi=\pi/2-10^{-2}$, $\phi=\pi/2-10^{-5}$ and $\phi=\pi/2-10^{-8}$ (from top to bottom) on a log-log scale.}
\label{Fig:AIvsExactMinimalModel}
\end{figure}

To finalize the discussion of the generalized drive at the level of the minimal model, we compare the exact result to the adiabatic-impulse approximation again for the case $\hat{t}_i=-\infty$ and $\hat{t}_{f}=+\infty$, \Fig{Fig:AIvsExactMinimalModel}{}. In this case, the AI approximation takes the form $p_k=|\langle -(\hat{\adtime}_k^*,k)| + (-\hat{\adtime}_k^*,k) \rangle|^2$, with $\hat{\adtime}_k^*$ from \Eq{Eq:BreakingTimeDefinition} for $\hat{\mu}_k=0$. The agreement between the two results is quite good, which is especially interesting as in the generalized setting we have two competing (length and time) scales. This opens the possibility to understand the \RG results from this more intuitive perspective. 

\subsection{Recovering the RG result}

Due to the exact solvability and knowledge of $p (\hat{k},\vec{\hat{v}})$ we can recover the \RG crossover-scaling result also from another simple argument. Considering the minimal model above, we already saw that $\hat{\kappa}=(\hat{v}_\perp/\hat{v}_\parallel)^{1/4}$ separates the two regimes in $p(\hat{k},\vec{\hat{v}})$. The subleading contribution $n_{\parallel}$ actually has the form
\begin{align}
\begin{aligned}
n_{\parallel} &\approx \int^{1}_{\hat{\kappa}} p\left(\hat{k},\hat{v}_\parallel\right) d\hat{k}\\
&=\hat{v}_\parallel^{\frac12} \int^{1/\hat{v}_\parallel^{\frac12}}_{\hat{\kappa} /\hat{v}_\parallel^{\frac12}} p\left(\bar{k},1\right) d\bar{k}, \quad \bar{k}:=\hat{k}/\hat{v}_\parallel^{\frac12} .
\end{aligned}
\end{align}
The expression becomes proportional to $\hat{v}_\parallel^{1/2}$ once the integral is constant to a good approximation, requiring 
\begin{align}
\begin{aligned}
\hat{\kappa}/\hat{v}_\parallel^{\frac12} \ll 1, && 1/\hat{v}_\parallel^{\frac12} \gg 1.
\end{aligned}
\end{align}
The first condition indicates the separation from the leading \KZM scaling and the second condition is the requirement not to be in the non-universal regime. Setting the first inequality to an equality recovers the predicted \RG-scaling: $\hat{v}^*(\phi) \sim |\pi/2-\phi|^{1/2}$. From the second condition we get that scaling is visible for $\hat{v}_\parallel \ll \hat{v}_{\text{cut}}\approx 1$. 

All these steps can be repeated for a more general setup with some dynamical critical exponent $z$ and scaling dimensions $\text{dim}[g_j]$ valid for variants of the Gaussian model discussed here. First of all we have:
 \begin{align}
\hat{\kappa}\sim (\hat{v}_\perp/\hat{v}_\parallel)^{1/(1/\nu-\text{dim}[g_j])},
\end{align}
and furthermore the conditions read
\begin{align}
\begin{aligned}
\hat{\kappa}/\hat{v}_\parallel^{1/(z+\text{dim}[g_j])} \ll 1, && 1/\hat{v}_\parallel^{1/(z+\text{dim}[g_j])} \gg 1.
\end{aligned}
\end{align}
The crossover velocity therefore is estimated as 
\begin{align}
\phi \to \pi/2: && \hat{v}^* &\sim |\pi/2-\phi|^{\frac{z+\text{dim}[g_j]}{1/\nu-\text{dim}[g_j]}}.
\end{align}

\subsection{Purely parallel drive}
An alternative to extract the subleading scaling is to consider a purely parallel drive, where we fix $\hat{\Delta}(\hat{t})=\hat{\Delta}^0$ and only drive along the subleading direction $\hat{D}_4(\hat{t})=\hat{v}_\parallel \hat{t}$. The excitation density reads for a linear drive with $\hat{t}_i=-\infty$ and $\hat{t}_{f}=+\infty$: 
\begin{align}
\begin{aligned}
&p(\hat{k},\hat{v}_\parallel)= \exp \left(-\pi \frac{\hat{k}^6}{\hat{v}_\parallel \hat{k}^4} \right), \\
&n_E(\hat{v}_\parallel \ll 1) \approx \frac{a\Lambda}{2\pi} \sqrt{\hat{v}_\parallel},
\end{aligned}
\end{align}
which is independent of $\hat{\Delta}^0$. It allows us directly to extract the predicted subleading scaling for low enough velocities, which therefore makes this protocol a useful tool to extract the subleading scaling. Nevertheless, this consideration is oversimplified, as we can always shift out $\hat{\Delta}^0$, such that it plays no role in the asymptotic case of $\hat{t}_i=-\infty$ to $\hat{t}_f=+\infty$. We resort to a more refined discussion of parallel drives in Sec.~\ref{Sec:TIParallelDrive}. In particular, such drives include the case of driving \emph{along the gapless line}, discussions of this topic can be found in Refs.~\cite{Divakaran2008,Mondal2009,Divakaran2010,Dutta2015} (in Sec.~\ref{Sec:GammaCoupling} we derive the scaling law found in Ref.~\cite{Divakaran2008} from the \RG-perspective).

\section{Generalized drives in the transverse XY model \label{Eq:GeneralizedDrivesXY}}
Due to $z=\nu=1$ in the transverse XY model, a linear drive only allows us to observe the leading \KZM scaling. We need at least a drive of order $2$ to make the scaling of driven subleading couplings observable, therefore we will consider drives of order $n=1,2$ (see also \cite{Sen2008} for the non-linear transversal case). In these cases, we have the following \RG-predictions listed in \Tab{Tab:RGResultsTI}{}, where the empty entries correspond to negative, meaning non-observable, exponents without fine tuning.
In the next subsections we verify these universal predictions for the scaling exponents of models for $n=1,2$ in the Ising universality class and determine the (non-universal) values $\hat{v}^*$ for the transverse XY model, mainly based on the AI approximation. To analyze scaling from further subleading couplings, like $D_4$, a generalized XY model and a quartic drive ($n=4$) can be used by adding additional terms to the spin model (see also Refs.~\cite{Suzuki1971,Suzuki1971a,Sadhukhan2019}) like\footnote{The fermionic couplings (for $k\to0$) are related to the spin-couplings according to 
\begin{align}
\mathcal{M} \begin{pmatrix} g/\gamma\\ J/\gamma \\ J_2/\gamma \end{pmatrix} \approx \begin{pmatrix} \hat{\Delta} \\ \hat{D}_2 \\ \hat{D}_4 \end{pmatrix}, && \mathcal{M}= \begin{pmatrix} 1 & -1 & -1 \\ 0 & \frac12 & 2 \\ 0 & -\frac{1}{24} & -\frac{2}{3} \end{pmatrix}.
\end{align}}
\begin{align}
\Delta H&=-\frac{J_2}{2}\sum_l (\sigma_l^x\sigma_{l+2}^{x}+\sigma_l^y\sigma_{l+2}^y)\sigma_{l+1}^z  \label{Eq:DeltaH}\\
&=-J_2\sum_l (c_l^\dagger c_{l+2}+\text{h.c.}) =-2J_2\sum_k \cos(2ka)c_k^\dagger c_k. \nonumber
 \label{Eq:DeltaH}
\end{align}

\begin{table}[h]
\begin{minipage}{0.45\textwidth}
\begin{tikzpicture}[x=0.95cm,y=.6cm]
    \draw (0,0) grid [step=1] (4,4);
    \node at (0.5,0.5) {3rd};
    \node at (0.5,1.5) {1st};
    \node at (0.5,2.5) {KZM};
    \node at (1.5,3.5) {1};
    \node at (2.5,3.5) {2};
    \node at (3.5,3.5) {4};
    
     \node at (1.5,2.5) {$\frac12$};
    \node at (1.5,1.5) {};
    \node at (1.5,0.5) {};

        \node at (2.5,2.5) {$\frac13$};
    \node at (2.5,1.5) {$1$};
    \node at (2.5,0.5) {};
    
      \node at (3.5,2.5) {$\frac15$};
    \node at (3.5,1.5) {$\frac13$};
    \node at (3.5,0.5) {$1$};

    \draw (0,4) -- (1,3);
    \node at (0.9,3.9) [below left,inner sep=1pt] {$n$};
    \node at (0.2,3.1) [above right,inner sep=1pt] {$\alpha$};
\end{tikzpicture}
\end{minipage}
\begin{minipage}{0.45\textwidth}

\begin{tikzpicture}[x=0.95cm,y=.6cm]
    \draw (0,0) grid [step=1] (3,3);
    \node at (0.5,0.5) {3rd};
    \node at (0.5,1.5) {1st };

    \node at (1.5,2.5) {2};
    \node at (2.5,2.5) {4};
    
     \node at (1.5,2.5) {};
    \node at (1.5,1.5) {};
    \node at (1.5,0.5) {};

        \node at (1.5,2.5) {};
    \node at (1.5,1.5) {$\frac12$};
    \node at (1.5,0.5) {};
    
      \node at (2.5,2.5) {};
    \node at (2.5,1.5) {$\frac32$};
    \node at (2.5,0.5) {$\frac14$};

    \draw (0,3) -- (1,2);
    \node at (0.9,2.9) [below left,inner sep=1pt] {$n$};
    \node at (0.1,2.0) [above right,inner sep=1pt] {$\beta$};
\end{tikzpicture}
\end{minipage}
\caption{Overview of the \RG predictions for the Ising-transition in the transverse XY model, where the exponents refer to $n_E\sim \hat{v}^{\alpha}$ and to $\hat{v}^*\sim |\pi/2-\phi|^{\beta}$ based on \Eq{Eq:IntroSubScaling} and \Eq{Eq:CrossoverRG} for the \KZM case and the 1st and 3rd subleading couplings.}
\label{Tab:RGResultsTI}
\end{table}

\subsection{Transverse XY: linear drive}
\begin{figure}[h]
\center

\begin{tikzpicture}

\node[anchor=south west,inner sep=0] at (0,0) {\includegraphics[width=0.7\textwidth]{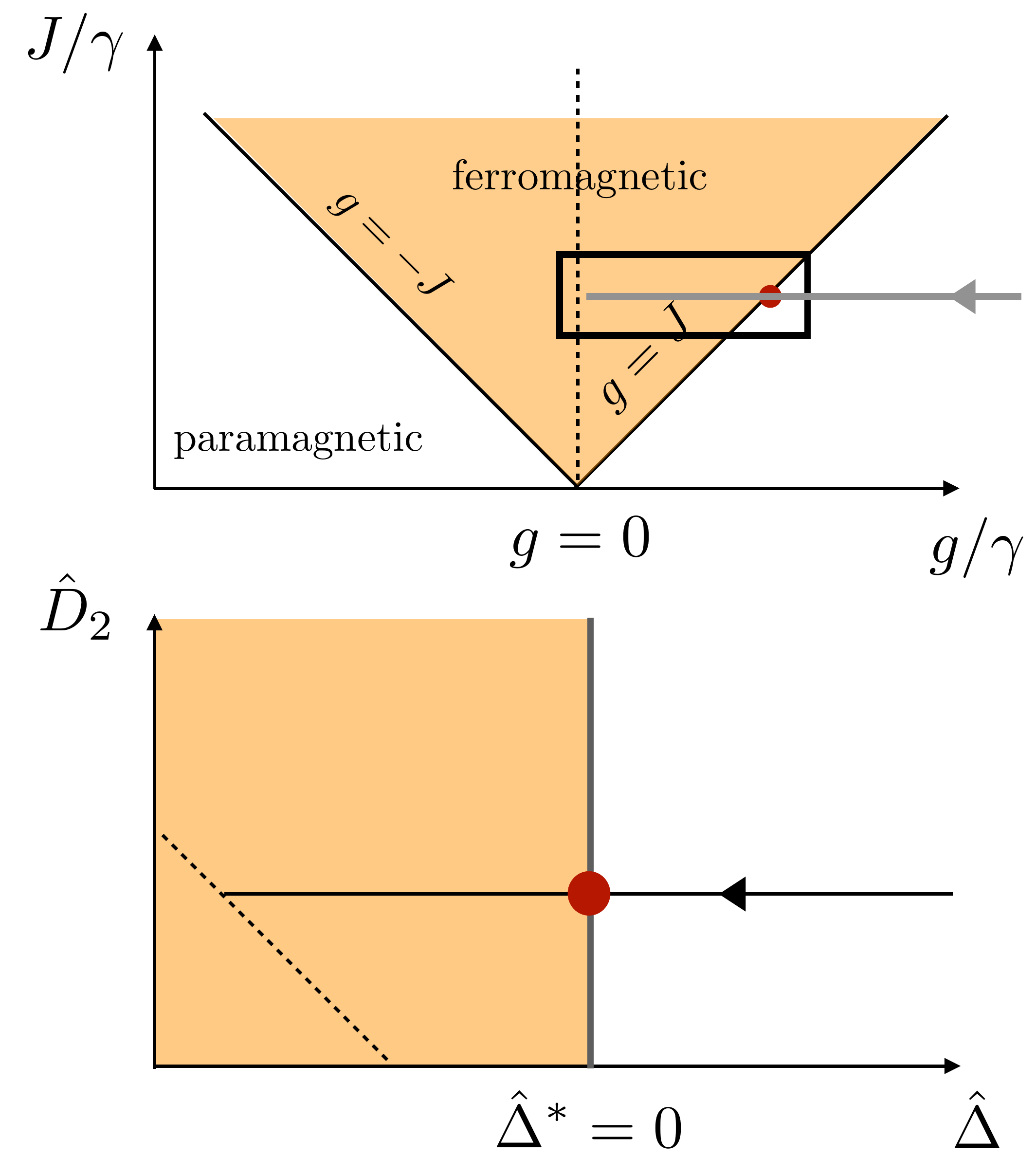}};

\node at (7,5.5) {\textbf{(a)}};
\node at (7,1.5) {\textbf{(b)}};

\end{tikzpicture}

\caption{Transversal drive in the transverse XY/Ising model considered in \cite{Dziarmaga2005}, \textbf{(a)} in the spin-coupling space and \textbf{(b)} the fermionic coupling space.}
\label{Fig:TILinearDrive}
\end{figure}

Since the transverse XY model \Eq{Eq:MomentumFermions} at the Ising transition has $z=1,\nu=1$ a linear drive only allows us to make the transversal scaling (standard \KZM) visible with an exponent $n_E\sim \hat{v}_{\perp}^{1/(z+1/\nu)}=\hat{v}_{\perp}^{1/2}$, see Tabs.~\ref{Tab:DriveOverviewXY} and \ref{Tab:RGResultsTI}. This setup was solved analytically exact by Dziarmaga \cite{Dziarmaga2005} (transverse Ising), where a linear drive $g(t)/J=-\hat{v}_g \hat{t}$ was considered, starting from the ground state at $t_i=-\infty$ up to $t_{f}=0$, see \Fig{Fig:TILinearDrive}{}. From our perspective, this corresponds to $\hat{\Delta}=-\hat{v}_\perp \hat{t}$ [with: $\hat{D}_2(\hat{t})=\hat{D}_2^0$], up to $\hat{\Delta}(\hat{t}_{f})=-2\hat{D}_2^0$, defining $\hat{t}_{f}$. The velocities are related by $\hat{v}_g=2\hat{v}_\perp$ (see \Tab{Tab:FermionDimensions} for $\gamma=1$). Strictly speaking, the Landau-Zener formula as given in \Eq{Eq:ExactSolutionLinearDrive} is not directly applicable when $\hat{t}_{f} \neq \infty$. Nevertheless, for low momenta and velocities $\hat{t}_{f}\gg \hat{t}^*$, where $\hat{t}^*$ is the time of adiabaticity-breaking, estimated by requiring $\hat{v}_{k^*}\approx 1$ or using the AI approximation. Therefore, $\hat{t}_{f}\to \infty$ will not change the result, for more details see Ref.~\cite{Dziarmaga2005}.

The asymptotic probability reads for small $k$ in the universal regime:
\begin{align}
\begin{aligned}
&p(k,\hat{v}_\perp)\approx \exp \left[-\pi \frac{(ka)^2}{\hat{v}_\perp}\right], \\
&n_E(\hat{v}_\perp)=\frac{1}{\pi} \int_{0}^{1} d(ka) p(k,\hat{v}_\perp) \sim  \frac{1}{2\pi} \hat{v}_\perp^{\frac12},
\end{aligned}
\end{align}
which is valid for low velocities, and directly allows us to read off the expected scaling from \Tab{Tab:RGResultsTI}.

\subsection{Transverse XY: higher-order drives}

Following the same strategy as in the minimal model and using the AI approximation as well as numerical integrations of \Eq{Eq:DiabaticRepresentation}, we determine the different scaling exponents, the crossover velocities, and the overall scaling regimes for a quadratic drive. In this section, we consider a drive starting deep in the paramagnetic phase and ending at the transition (see also Ref.~\cite{Biaonczyk2018}). Alternatively, one can start from the transition and drive into the phase, yielding the same result (see, e.g., Refs.~\cite{Damski2006,Dziarmaga2010}), in which case the final excitations are spin flips (see Sec.~\ref{Sec:Techniques} and Ref.~\cite{Dziarmaga2005}). The main reason for this choice is that it enables us to apply analytic approximations while avoiding the ferromagnetic phase that has some complicating features for the transverse XY model (see, e.g., Ref.~\cite{Deng2009} for a discussion how `non-critical' modes can play a role). 

Formally, we consider the drive of the leading coupling $\hat{\Delta}$ and a subleading coupling $\hat{D}_j$, where once again for the simplest case:
\begin{align}
\begin{aligned}
&\hat{\Delta}(\hat{t})\approx \frac{2(g(t)-J(t))}{2\gamma}=\hat{v}_\perp \hat{t}^n, \\
&\hat{D}_2(\hat{t}) \approx \frac{J(t)}{2\gamma}=\hat{D}_2^0 +\hat{v}_\parallel \hat{t}^n.
\end{aligned}
\end{align}
One major difference compared to the minimal model is that we need the constant $\hat{D}_2^0 \neq 0$, as already indicated in \Fig{Fig:SpinVsFermion}{} by the red dot. The reason is that we want to circumvent the region in coupling space, where the $|ka|=\pi$ and $k=0$-gap closing coalesce. We set $\hat{D}_2^0=1$ (corresponding to $J/\gamma=2$ at the transition) in the following.

As discussed in Sec.~\ref{Sec:AdiabaticApproximations} and Sec.~\ref{Sec:BasicsAI} the approximations are closely connected to the adiabaticity parameter $\adparameter$. It was the simple structure of this parameter $\adparameter$ that allowed us, in the minimal model, to decompose the excitation density as $n_E\sim \text{Max}[n_{\perp}(\hat{v}_\perp),n_{\parallel}(\hat{v}_\parallel)]$. A similar logic applies for the transverse XY model and an order-$n$ drive. Here $\adparameter$ reads (for an order-$n$ drive, see \Tab{Tab:FermionDimensions}):
\begin{align}
\begin{aligned}
\adparameter &= \frac{ \hat{v}_\perp+ 2\hat{v}_\parallel(1-\cos(ka))}{\sin(ka)^{n+1}} \\
&\approx \hat{v}_\perp (ka)^{-(n+1)}+\hat{v}_\parallel (ka)^{-(n-1)}.
\label{Eq:AdiabaticityParameterXY}
\end{aligned}
\end{align}

We can again introduce a crossover scale in $k$-space: ${\hat{\kappa}:=(\hat{v}_\perp/\hat{v}_\parallel)^{1/2}}$ to separate the regimes in \Eq{Eq:AdiabaticityParameterXY}, where we identify the \KZM and subleading scaling regime. A qualitative estimate of adiabaticity breaking and corresponding scaling is given by
\begin{align}
\adparameterstar \stackrel{!}{\approx} 1: \begin{cases} \hat{k}^* \ll \hat{\kappa} \,\ (\text{KZM}):&  \hat{k}^* \sim \hat{v}_\perp^{\frac{1}{n+1}}, \\ \hat{\kappa} \ll \hat{k}^* \ll 1 \,\ (\text{sub}):&  \hat{k}^* \sim \hat{v}_\parallel^{\frac{1}{n-1}} .\end{cases}
\end{align}

\begin{figure*}[t]
\centering
\begin{tikzpicture}

\node[anchor=north east,inner sep=0] at (-1,0.5) {\includegraphics[width=0.45\textwidth]{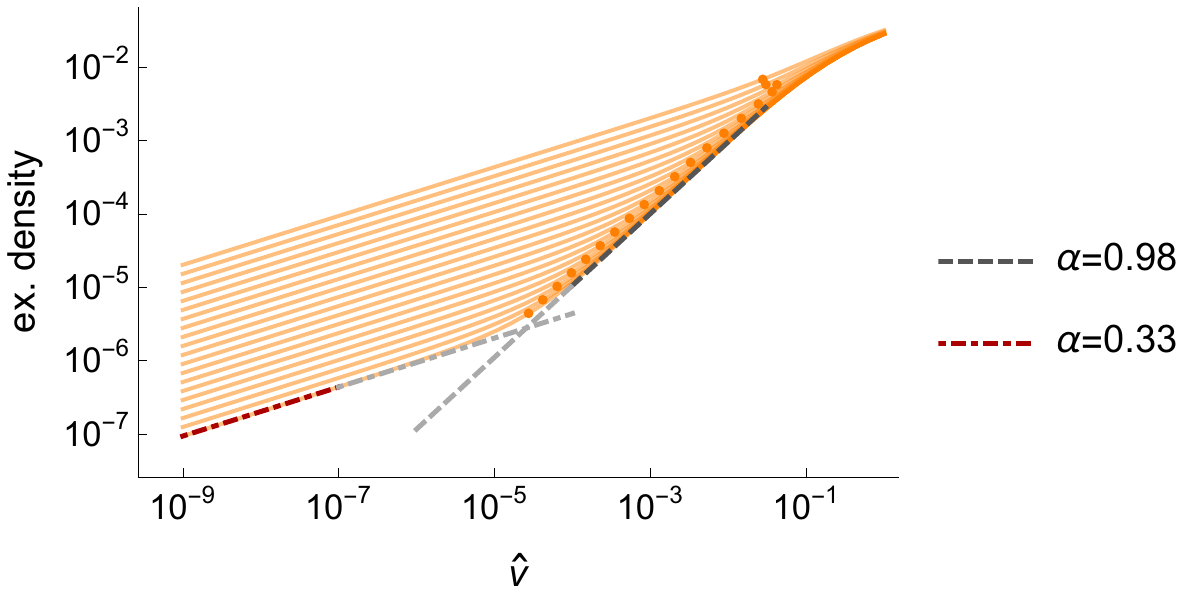}};
\node[anchor=south west,inner sep=0] at (-0.5,1) {\includegraphics[width=0.5\textwidth]{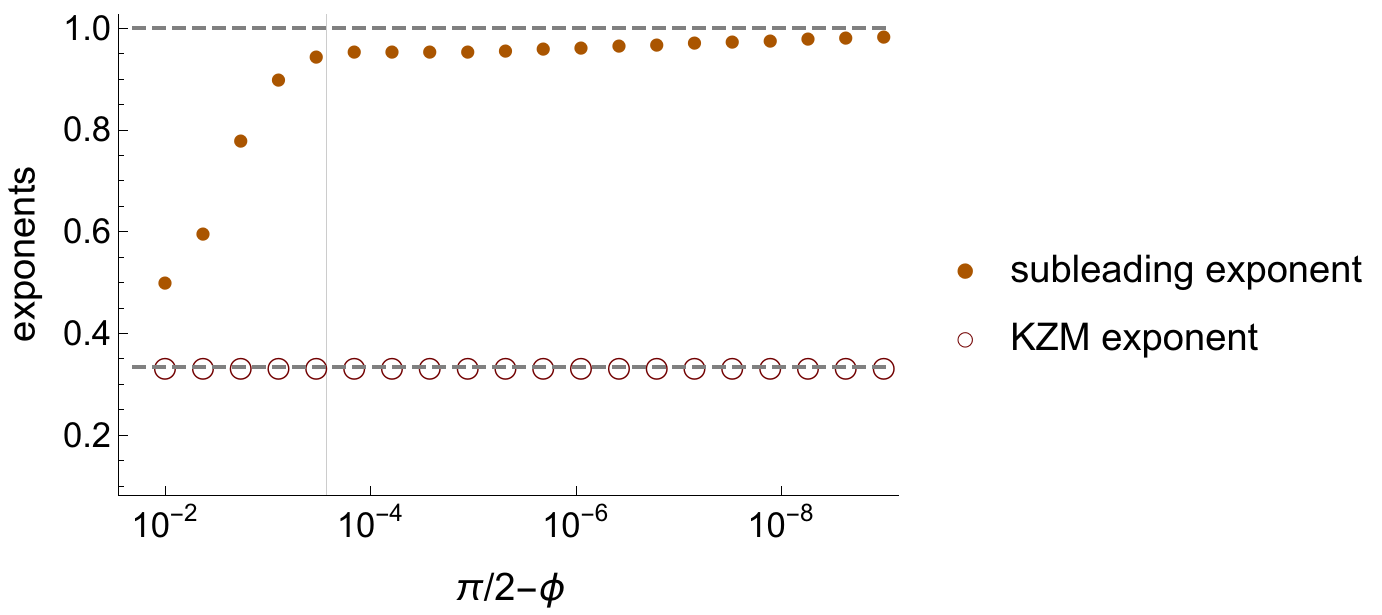}};
\node[anchor=south east,inner sep=0] at (-2.75,1) {\includegraphics[width=0.35\textwidth]{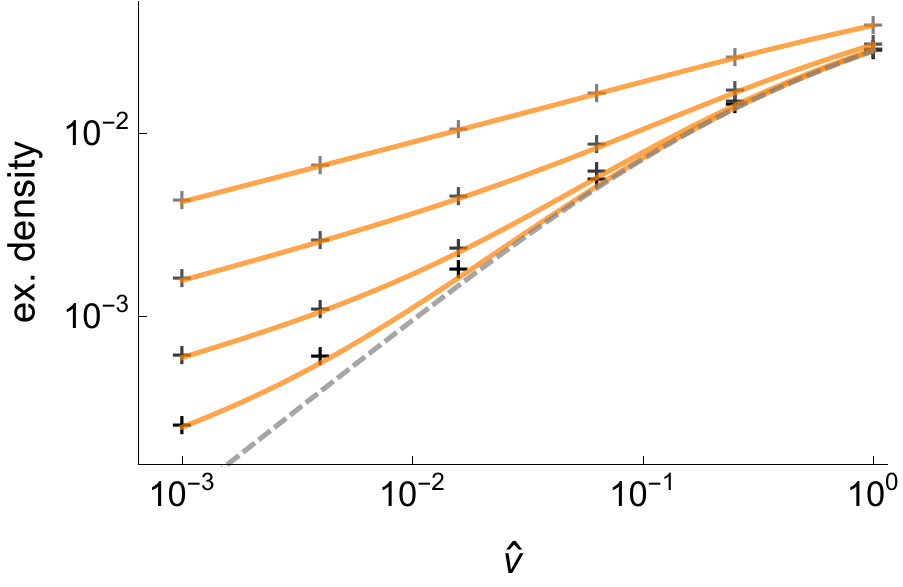}};
\node[anchor=north west,inner sep=0] at (-0.5,0.5) {\includegraphics[width=0.5\textwidth]{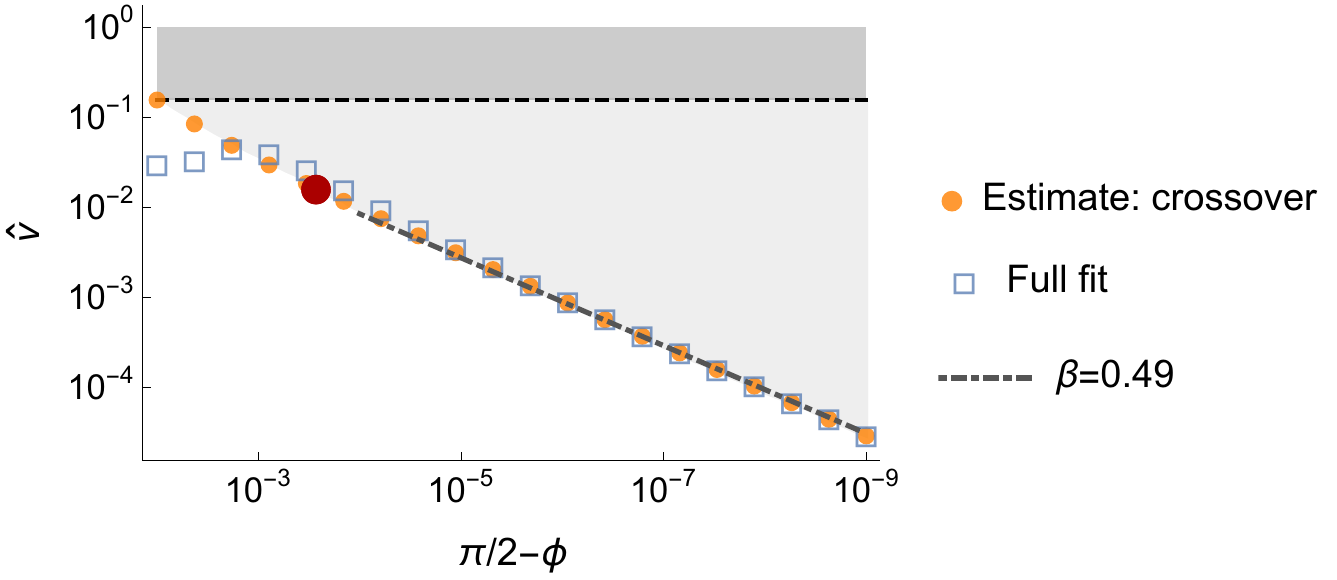}};
\node at (-7.5,5) {\textbf{(a)}};
\node at (8,5) {\textbf{(c)}};
\node at (-7.5,-0) {\textbf{(b)}};
\node at (8,-0) {\textbf{(d)}};

\node [rotate=90] at (-8.85,4.5) {$n_E$};
\node [rotate=90] at (-8.87,-0.1) {$n_E$};

\node at (3.3,0.1) {non-universal};
\node at (2,-2.2) {KZM};
\node at (4.2,-0.8) {subleading};
\draw[color=gray!60, fill=gray!60, opacity=0.2] (-3.5,0.5) -- (-9.2,0.9) -- (-2,0.9) -- cycle;
\draw[gray, thick] (-4.7,-0.5) rectangle (-3,0.5);
\draw[color=gray!60] (-9.2,0.9) rectangle (-2,5.5);

\end{tikzpicture}

\caption{\textbf{(a)} Set of curves $n_E(\hat{v},\phi)$ (log-log scale) for the transverse XY model (quadratic drive ($n=2$), starting deep in the paramagnetic phase and stopping at the transition or vice versa), and for different angles $\pi/2-\phi=10^{-1},5\times 10^{-3},2\times 10^{-4},10^{-5}$ (from top to bottom) from numerical simulations ($+$, starting at the transition) and the AI approximation (orange lines; dashed line describes $\phi\to\pi/2$). \textbf{(b)} Extended set of curves using the AI approximation. For low velocities the KZM-scaling results and the subleading scaling emerges for larger velocities up to some non-universal regime. The orange dots indicate the crossover velocities. \textbf{(c)} Scaling exponents extracted from the full fit of the curves in (b) (up to the non-universal regime) and the predicted exponents from the \RG (dashed lines). The vertical line indicates $\phi_{\text{min}}$. \textbf{(d)} Universal scaling of the crossover velocity for $\phi \to \pi/2$ that is estimated using again \Eq{Eq:CrossoverEstimateMinimalModel} (orange dots) and a full fit (blue squares) of the curves in (b). The extracted exponent $\beta=0.49$ fits well to the \RG-prediction of $1/2$.}
\label{Fig:quadraticTICrossover}
\end{figure*}
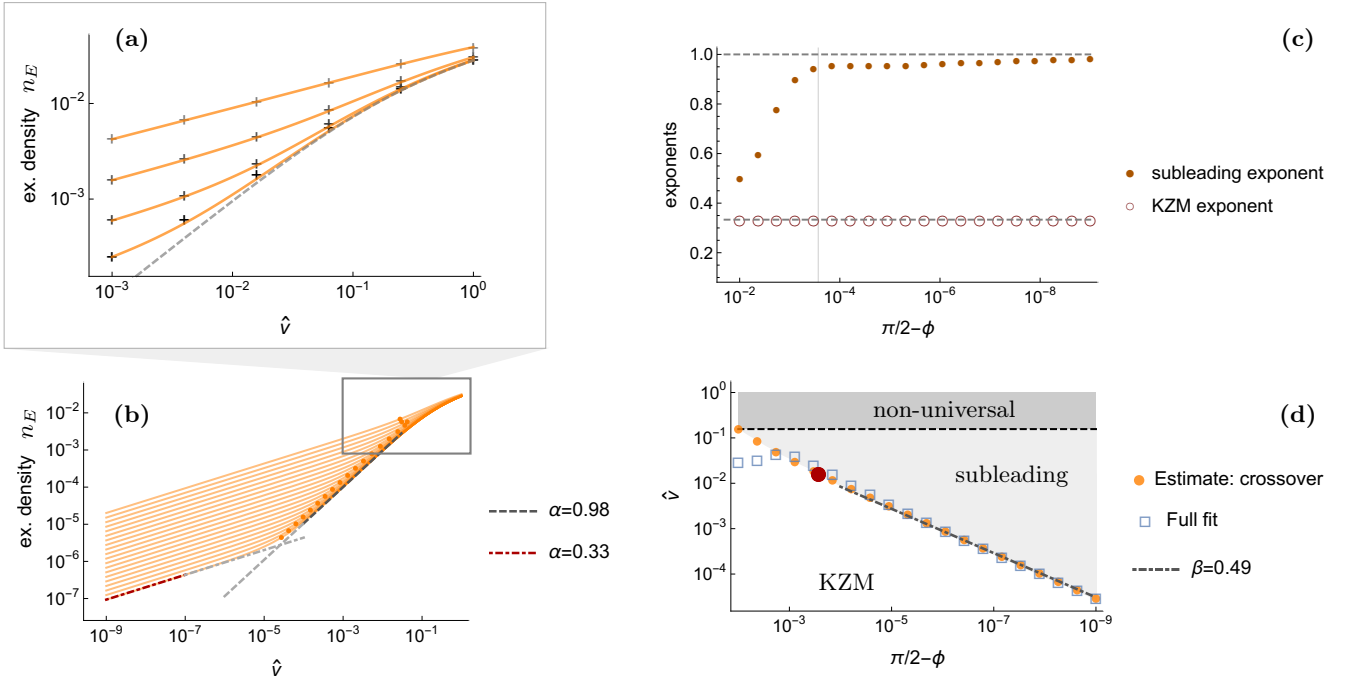
In a first step, we use the AI approximation to determine the excitation density for small velocities and compare it to numerical integrations of \Eq{Eq:XYall} (for a finite number of lattice sites $N$) for a few cases\footnote{We use the adiabatic basis to solve the dynamics numerically. The system is initially prepared at the critical point and is stopped at $J(t_f)/(2\gamma) \approx 600$.}. The results are in fair agreement, as shown in \Fig{Fig:quadraticTICrossover}{a}. The system size is chosen such that the length scale $\xi^* \sim n_E^{-1}$ is smaller than the system size $N$. Otherwise we expect finite-size effects to play a dominant role.

In a second step, we determine the crossover scale $\hat{v}^*(\phi)$, which indicates the crossover from the subleading scaling at higher velocities to \KZM scaling at lower velocities. The results are summarized in \Fig{Fig:quadraticTICrossover}{}. We see a similar emerging picture compared to the minimal model, as expected from the scaling of the adiabaticity parameter \Eq{Eq:AdiabaticityParameterXY}: For $\phi \to \pi/2$ the subleading scaling regime becomes prominent over a few orders of magnitude [\Fig{Fig:quadraticTICrossover}{b}] and allows us to extract the expected scaling exponents [\Fig{Fig:quadraticTICrossover}{c}], as well as the predicted scaling of the crossover velocities $\hat{v}^*$, \Fig{Fig:quadraticTICrossover}{d}. Also here intermediate angles smaller than $\phi_{\text{min}}$ [red dot in \Fig{Fig:quadraticTICrossover}{d}] will not allow one to extract a sensible subleading scaling exponent.

\subsection{Transverse XY: purely parallel drive \label{Sec:TIParallelDrive}}
\begin{figure}[H]
\centering
\begin{tikzpicture}

\node[anchor=south west,inner sep=0] at (0,0) {\includegraphics[width=0.7\textwidth]{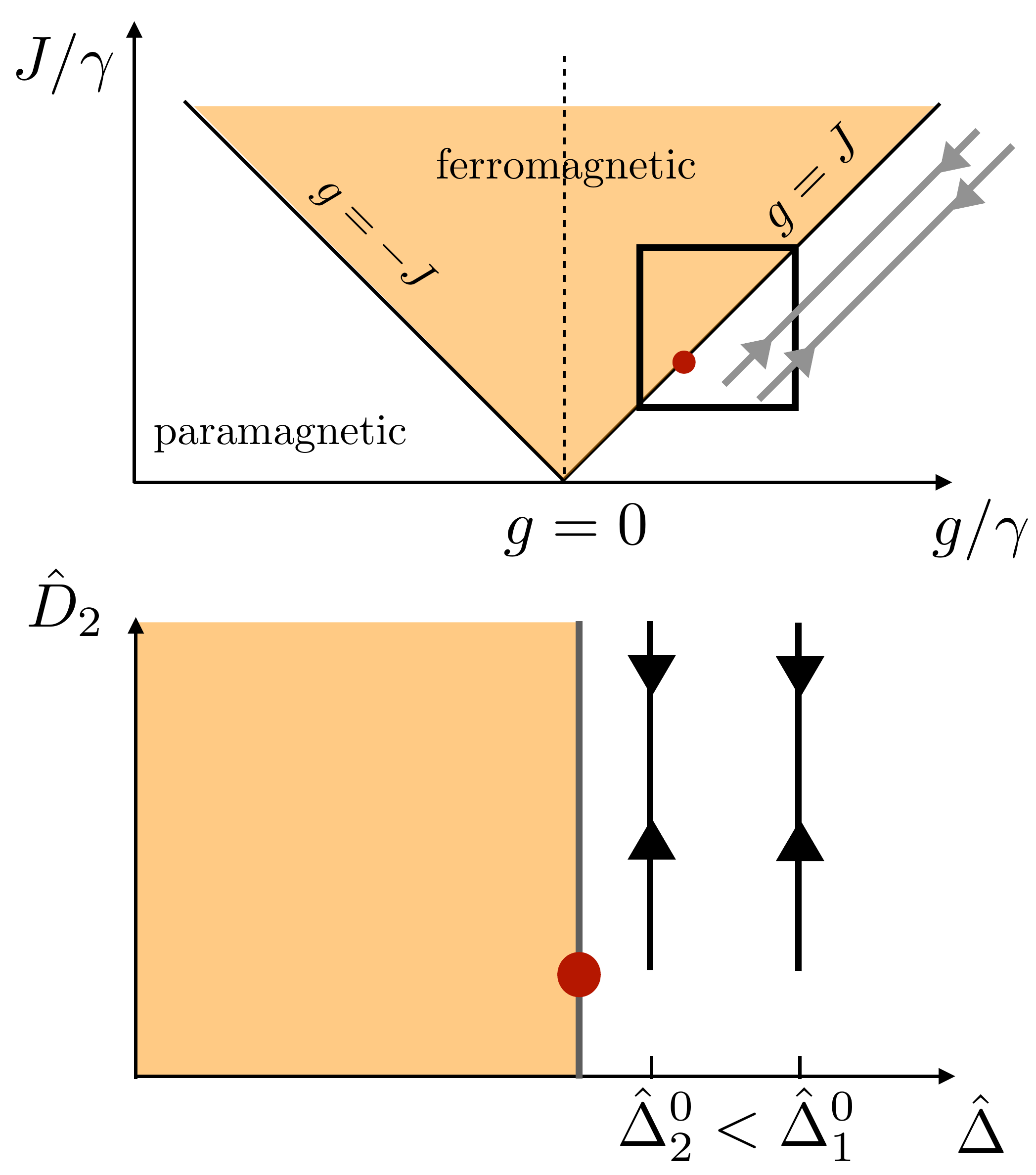}};

\node at (7,5.5) {\textbf{(a)}};
\node at (7,1.5) {\textbf{(b)}};

\end{tikzpicture}

\caption{Parallel drive in the transverse XY model in \textbf{(a)} the spin-coupling space and \textbf{(b)} the fermionic coupling space for different constant distances to the critical line.}
\label{Fig:TransverseXYparallel}
\end{figure}

So far we have analyzed drive protocols, where the critical point was reached during the drive. Here we consider the situation of a purely parallel drive for a fixed distance to the critical line $\hat{\Delta}^0 \ge 0$, where only the subleading coupling is driven as shown in \Fig{Fig:TransverseXYparallel}{}. This situation is very different from the standard \KZM scenario, nevertheless the \RG picture suggests that the subleading scaling could be made observable also for such a drive. In more physical terms, it implies that excitations are created by any drive, which leads to adiabaticity breaking. In particular, this intuition is valid for any direction of drive, parallel or longitudinal to the phase boundary, as long as the criterion $nz+\text{dim}[g_j]>0$ is fulfilled.

\begin{figure}[h]
\centering

\begin{tikzpicture}

\node[anchor=south west,inner sep=0] at (0,0) {\includegraphics[width=0.9\textwidth]{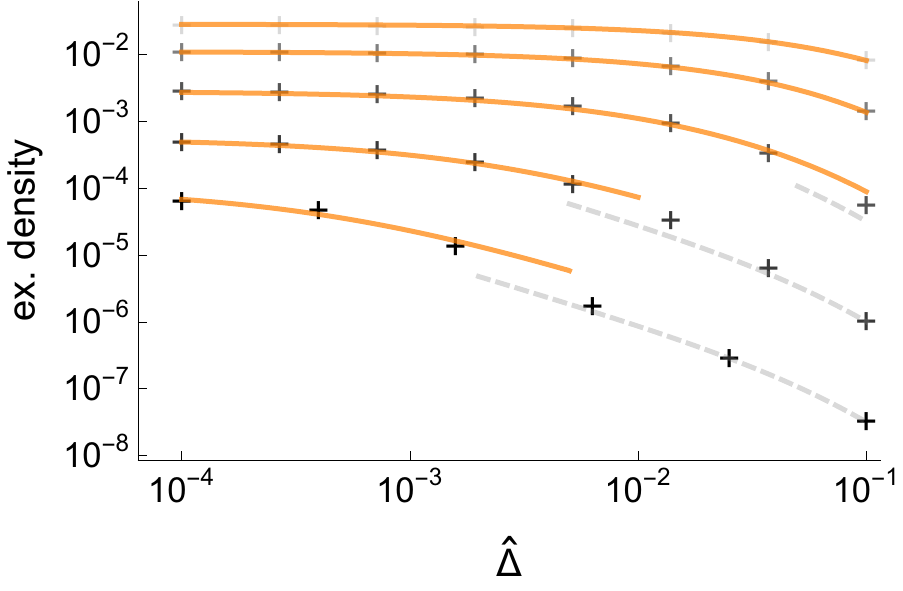}};
\node [rotate=90] at (0.25,4.25) {$n_E$};

\end{tikzpicture}

\caption{Excitation density (numerical: $+$; AI: full lines) for a purely parallel drive ($n=2$) as a function of the gap $\hat{\Delta}$ on a log-log scale for $\hat{v}=1.0\times 10^{0},1.8\times 10^{-1},3.2\times 10^{-2},5.6\times 10^{-3},1.0 \times 10^{-3}$ from top to bottom ($N=2\times 10^{3}$; for the smallest velocity $N=10^{4}$). For $\hat{\Delta} \to 0$ a constant value is reached, which can be used to determine the subleading scaling again, see \Fig{Fig:TransverseXYGapless}{}. The dashed lines correspond to the density obtained from using $p_k$ from the first-order adiabatic perturbation result \Eq{Eq:AdiabaticPerturbationTheory}.}
\label{Fig:TransverseXYSmallGaps}
\end{figure}

\begin{figure}[h]
\centering

\begin{tikzpicture}

\node[anchor=south west,inner sep=0] at (0,0) {\includegraphics[width=0.9\textwidth]{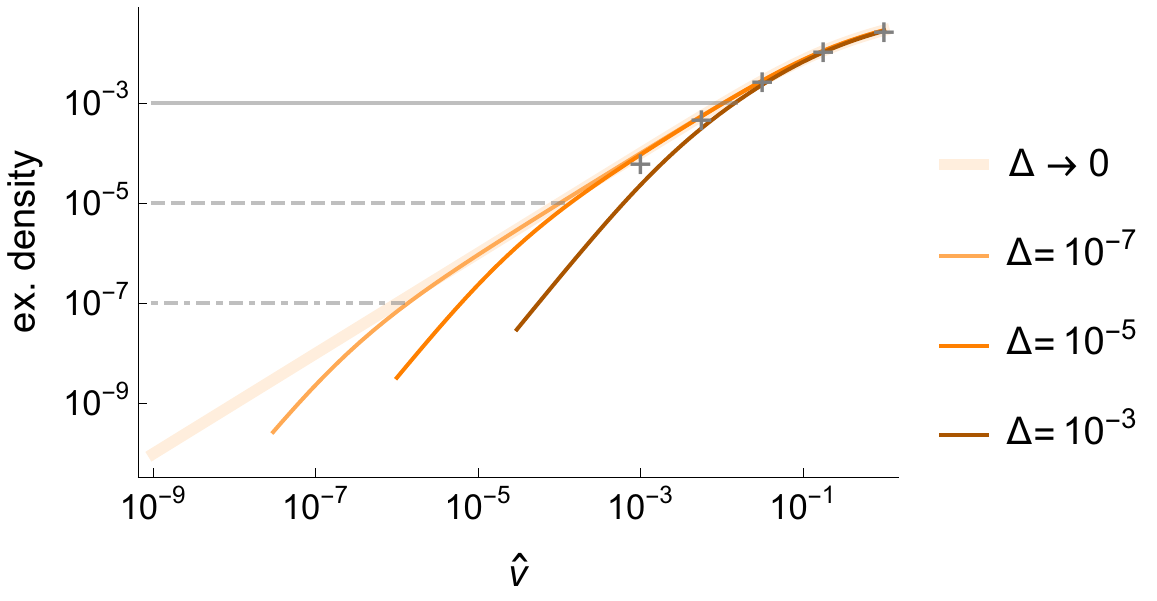}};
\node [rotate=90] at (0.23,3.35) {$n_E$};

\end{tikzpicture}

\caption{AI approximation for a purely parallel quadratic drive for small fixed gaps $\hat{\Delta}=\hat{\Delta}^0$ on a log-log scale, the crosses are the numerical values from \Fig{Fig:TransverseXYSmallGaps}{} for the smallest gaps. The smaller the gap, the more extended the subleading scaling regime ($n_E \sim \hat{v}_{\parallel}^{1}$) becomes up to a scale $\sim \hat{\Delta}^0$ indicated by the horizontal lines.}
\label{Fig:TransverseXYGapless}
\end{figure}

A first, very basic intuition is that once $\hat{\Delta}^0$ is large, the system will stay adiabatic for the whole drive. Only once the $\hat{\Delta}^0$ becomes small enough (in a sense we clarify in the following) adiabaticity can be broken, signaled by a finite excitation density $n_E$. Therefore, one approach to extract the subleading scaling is to fix a drive-velocity $\hat{v}_\parallel$ and perform drives for different $\hat{\Delta}^0$ (\Fig{Fig:TransverseXYparallel}{}). Once adiabaticity is broken, we expect the excitation density to reach a constant finite value. For $\hat{\Delta}^0\to, 0$ this saturation of $n_E$ is directly observable in \Fig{Fig:TransverseXYSmallGaps}{}. The pair $(\hat{v} ,n_E(\hat{\Delta}^0\to 0))$ can again be used to extract the subleading scaling exponent similarly to the discussions before, see \Fig{Fig:TransverseXYGapless}.

Guided by the above picture, we compare a numerical integration\footnote{We use the adiabatic basis to solve the dynamics numerically, which is stopped at $J(t_f)/(2\gamma)\approx 600$.} for a drive ($n=2$) starting at $\hat{D}_2(t_i)\approx J(t_i)/(2\gamma)=1$ and $\hat{\Delta}(t)=\hat{\Delta}^0$ with the AI approximation \Eqs{Eq:AIProbability} and \eq{Eq:BreakingTimeDefinition}, see \Fig{Fig:TransverseXYSmallGaps}{}. We expect a reasonable fit in the non-adiabatic (saturated) regime, but not in the adiabatic one. In the latter regime the (perturbative) adiabatic approximation of $p_k$, \Eq{Eq:AdiabaticPerturbationTheory}, works better (shown as the dashed lines), supporting the idea of a crossover to an adiabatic evolution.

More in line of the discussion of the generalized drives, we can also vary the velocity for a fixed gap. This allows us to extract the subleading scaling, seen in \Fig{Fig:TransverseXYGapless}{}. Nevertheless, the window of clean algebraic scaling due to the subleading term is limited to intermediate velocities. The size of the finite gap sets a velocity scale, below which the behaviour changes (where adiabaticity is restored). The limiting case would be the drive along the gapless line. Similar drives were already used in the transverse XY model. An overview is given in \cite{Divakaran2008,Divakaran2010,Dutta2015}. We discuss this case in Sec.~\ref{Sec:GammaCoupling}.

As we have seen in \Fig{Fig:TransverseXYSmallGaps}{} for small enough gaps $\hat{\Delta}^0$ the excitation density crosses over to a constant value for $\hat{\Delta}^0 < \hat{\Delta}^{0*}$, implying that adiabaticity is broken for the given velocity. To connect this observation with the guiding idea of competing length scales, we first of all associate the saturated regime with the length scale induced by the drive: $\xi_{\parallel}\sim n_E^{-1}$. The second scale is the equilibrium correlation length $\xi \sim (\hat{\Delta}^{0})^{-\nu}$ (for the transverse XY model the correlation length is analytically known \cite{Barouch1971,Bunder1999}). 

\begin{figure}[h]
\centering

\begin{tikzpicture}

\node[anchor=south west,inner sep=0] at (0,0) {\includegraphics[width=0.9\textwidth]{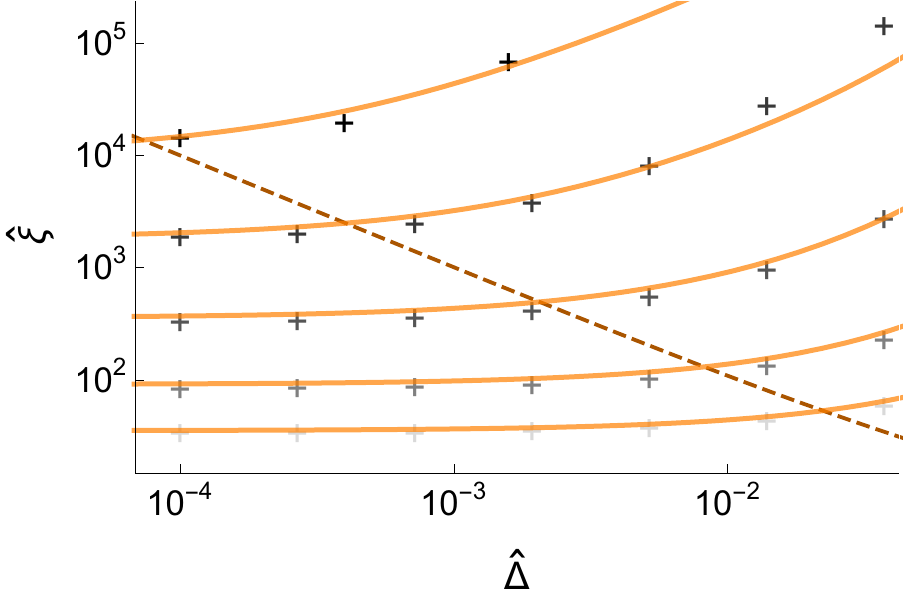}};
\node at (3,2.2) {\large non-adiabatic};
\draw[color=orange!60, fill=orange!60, opacity=0.4] (8,1.3) -- (8,5.2) -- (1.2,5.2)--(1.2,4) -- cycle;
\node at (6,3.3) {\large adiabatic};

\end{tikzpicture}

\caption{Comparison of the ground state correlation length $\xi$ (red dashed) in the transverse XY model (for a fixed $J$ close to $\hat{D}_2^0=1$) and the `excitation' length scale $\xi_\parallel$ on a log-log scale, defined by the inverse excitation density (solid lines; same as in \Fig{Fig:TransverseXYSmallGaps}{}; $+$: (inverted) numerical data from \Fig{Fig:TransverseXYSmallGaps}{}).}
\label{Fig:TransverseXYParallelLengthScales}
\end{figure}
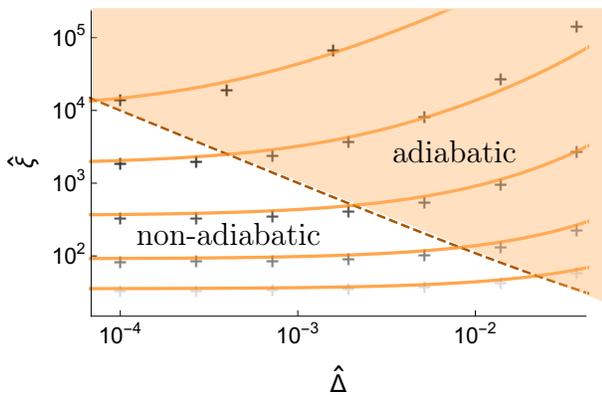

Following the idea that only the \emph{smaller} scale is observable, the crossing of the two curves should give an estimate of $\hat{\Delta}^{0*}$, separating the adiabatic from the non-adiabatic region (shown in \Fig{Fig:TransverseXYParallelLengthScales}{}). A related scenario regarding the competition of length scales was discussed in the case of a transversal drive with a finite symmetry breaking bias (with an additional term $-g_\parallel \sum_l \sigma_l^x$ in the Hamiltonian) \cite{Rams2019}, which also allows one to restore adiabaticity (see also \cite{Rysti2019} for an experimental investigation of the \KZM with a symmetry breaking bias). Here the field $g_\parallel$ is a second equilibrium relevant coupling, such that a finite value induces a finite length scale, similar to $\hat{\Delta}^0$.

This competition of (length) scales is reflected in the competition of the two parameters $\adparameter$ and $\hat{\mu}_k$: from the (perturbative) adiabatic side, adiabaticity breaking is suppressed once $\hat{\mu}_{k} \gg 1$ [see \Eq{Eq:AdiabaticPerturbationTheory}]. The scale induced by $\hat{\mu}_{k_\mu}\approx 1$ has to be compared to $\hat{v}_{k_v}\approx1$. Only if $k_v \gg k_\mu$ adiabaticity breaking is possible. The equality gives a condition on the gap size $\hat{\Delta}^{0*}$
\begin{align}
\hat{\Delta}^{0*} \sim \hat{v}_\parallel^{\frac{1/\nu}{nz+\text{dim}[D_\subexponent]}}\stackrel{\text{here}}{=} \hat{v}_\parallel^1,
\end{align}
such that only $\hat{\Delta}_0 <\hat{\Delta}^{0*}$ allow the evolution to be non-adiabatic. For larger gaps, the physical length scale should be given by the ground state correlation length $\xi$ as in \Fig{Fig:TransverseXYParallelLengthScales}{}. Only once the velocity is large enough or the gap is small enough, a non-adiabatic regime is entered and a direct extraction of the subleading scaling becomes possible, as in \Fig{Fig:TransverseXYGapless}{}.

\subsubsection{Transverse XY: driving the $\gamma$-coupling \label{Sec:GammaCoupling}}

The situations we have discussed so far have been the drive of one leading (relevant in equilibrium) and one subleading (irrelevant in equilibrium) coupling. To extract the corresponding scaling dimensions, we chose to scale out $D_1$, which fixes the fixed point theory. This is only possible once $D_1$, as the leading derivative, is \emph{not} driven. 
In Ref.~\cite{Divakaran2008} the coupling $\gamma(t)$ was driven, which is effectively the same as driving $D_1$. The corresponding scaling of $n_E(v)$ can also be quite easily explained from the generalized \KZM perspective. In this case, we actually deal with a different fixed point, which is determined by the lowest non-driven derivative term (assuming no further fine tuning). To this end, we scale out $D_2$, leading to different scaling dimensions and critical exponents $z'=2$ and $\nu'=1/2$. In particular, the $D_1$-direction is now a relevant direction with a positive scaling dimension $\text{dim}[D_1]=+1$. Therefore, a linear drive along the gapless line $\hat{\Delta}^0=0$ leads to a scaling according to \Eqs{Eq:IntroSubScaling} and \eq{Eq:IntroExvsLength}
\begin{align}
n_E(\hat{v}_\parallel) \sim \hat{v}_\parallel^{\frac{1}{2+1}},
\end{align}
which is the exponent found in Ref.~\cite{Divakaran2008}. The generalized expression for a model with an original dynamical exponent $z$, which is driven along the otherwise scaled out direction is given as follows. Before scaling out any of the couplings, we have the lowest $k$-term being $\propto D_z k^z$ and the next subleading one $\propto D_\subexponent k^\subexponent$ with $\subexponent>z$. By driving the $k^z$-term, we have to scale out $D_\subexponent$, which results in the proper dynamical exponent and scaling dimension for an order-$n$ drive:
\begin{align}
z'=\subexponent, && \text{dim}[D_z]=\subexponent-z: && n_E(\hat{v}_\parallel) \sim \hat{v}_\parallel^{\frac{1}{(n+1)\subexponent-z}}.
\end{align}
This is exactly the generalized expression given in Refs.~\cite{Divakaran2008,Mondal2009}, which therefore can be understood as well from the \RG-perspective.

\section{Model dependence of non-universal scales \label{Sec:ModelDependence}}

The minimal model and the transverse Ising model are explicit examples of different universality classes, which are defined by the dimensionality and symmetries only. So far, we have demonstrated the emergence of the predicted new scaling regimes for a generalized drive of leading and subleading couplings, which should in principle be observable in all models of a given universality class. 
This is indeed the case, but the extent of the different scaling regions is not universal and will lead to a different phenomenology for different models in the same universality class. In particular, the dependence of the crossover velocity $\hat{v}^*$ on $\phi$ can vary widely.

\begin{figure}[t]
\centering
\includegraphics[width=0.95\textwidth]{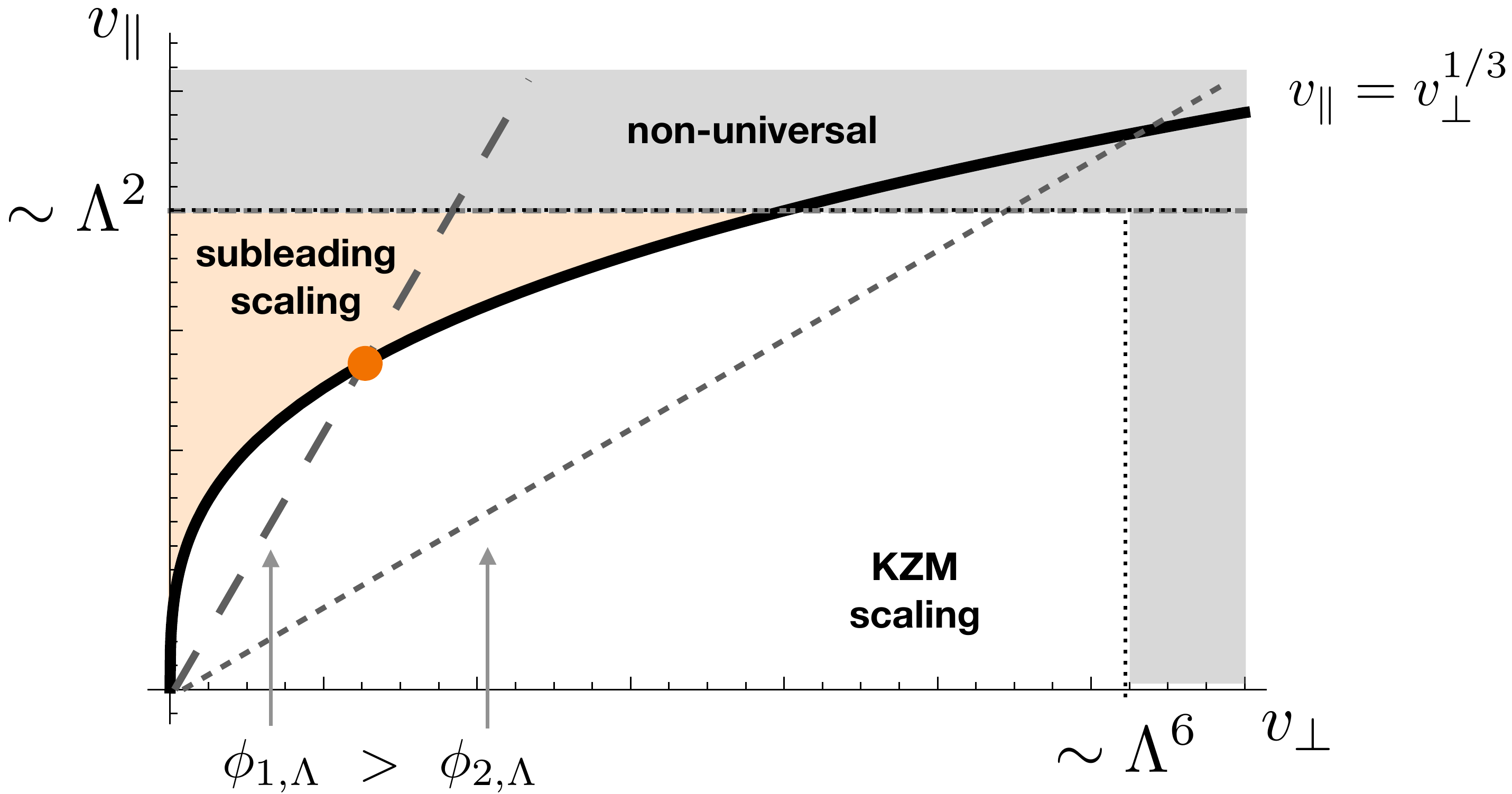}
\caption{Schematic scaling regimes (\KZM and subleading) expressed with respect to the dimensionful velocities; the thick black line separates the \KZM scaling regime (below) from the subleading one (above). The estimated non-universal regime is controlled by the cutoff $\Lambda$. The two dashed lines indicate two trajectories for two different but fixed $\phi_{1|2,\Lambda}$ and varying $\hat{v}$. Only for the $\phi_{1,\Lambda}$ both scalings are observable, as, e.g., in \Fig{Fig:MinimalModelFirstSub}{}, with the dot indicating the crossover. Only if the angle is large enough, the subleading regime can be accessed.}
\label{Fig:NonuniversalRegimes}
\end{figure}

Concretely, in the models considered so far, we find that we need $\pi/2-\phi$ very small to see the subleading scaling, but this is non-universal and model-dependent. We illustrate this model dependence with different versions of the minimal model, which differ only by different cutoffs $\Lambda$. In the critical region, all models of a universality class are described by a long-wavelength field theory, which for example for the minimal model takes the Hamiltonian form in \Eq{Eq:DimensionfullMinimalModel}. Given the dimensionful, microscopic velocities $v_\perp$ and $v_\parallel$, we identified a crossover condition $\hat{\kappa}/\hat{v}_\parallel^{* 1/2}\approx 1$, separating the \KZM and the subleading scaling regime. Expressed in terms of dimensionful velocities it reads:
\begin{align}
v_\parallel^* \approx v_\perp^{* 1/3},
\end{align}
represented in \Fig{Fig:NonuniversalRegimes}{} by the thick line. In particular, the crossover condition does not depend on the cutoff $\Lambda$. Nevertheless, to have a clear scaling regime, we also have a condition to obey on the largest possible velocities, fixed by the requirement to avoid the non-universal regime:
\begin{align}
\begin{aligned}
&\hat{v}_\parallel \ll 1, && v_\parallel \ll \Lambda^2,
\end{aligned}
\end{align}
which is indeed cutoff dependent. The non-universal regimes are indicated by the gray areas in \Fig{Fig:NonuniversalRegimes}{} above the dotted lines. Therefore, increasing the cutoff allows one to enlarge the subleading scaling regime further (a similar logic applies to the \KZM regime). To relate this figure with the discussions in the last sections, consider a protocol, where we fix $\phi_\Lambda$ and change $\hat{v}$. In \Fig{Fig:NonuniversalRegimes}{} this corresponds to one of the dashed lines. The subleading scaling regime in $n_E$ becomes only observable once the corresponding (orange) region in \Fig{Fig:NonuniversalRegimes}{} is passed (e.g., for $\phi_{1,\Lambda}$). We can directly see that the possibility to observe subleading scaling is strongly cutoff dependent.

The discussion so far clearly separated the dimensionful, microscopic velocities from the effect of the cutoff $\Lambda$. A subtlety arises at the level of the angles: By changing the cutoff also the angles change, even though we keep the dimensionful velocities $v_\parallel$ and $v_\perp$ the same (we stay at the same dashed lines in \Fig{Fig:NonuniversalRegimes}{}). The reason is that the angles $\phi_\Lambda$ and the dimensionless velocities $\hat{v}_\parallel$ and $\hat{v}_\perp$ are defined with respect to $\Lambda$. To see this point, consider again the dimensionless velocities 
\begin{align}
\begin{aligned}
&\hat{v}_\perp(\Lambda)=\left(v_\perp/D_3^2\right) \Lambda^{-6}, \\
&\hat{v}_\parallel(\Lambda)=\left( v_\parallel/D_3^2 \right) \Lambda^{-2}, \\
&\tan(\phi_\Lambda)=\hat{v}_\parallel(\Lambda)/\hat{v}_\perp(\Lambda) .
\end{aligned}
\end{align}
Therefore, the angles for different cutoffs are related by:
\begin{align}
\tan(\phi_{\Lambda'})=\tan(\phi_\Lambda)(\Lambda/\Lambda')^{-4}. 
\label{Eq:AngleTransformation}
\end{align}
For angles $\phi_\Lambda$ close to $\pi/2$ this can be approximated by:
\begin{align}
&\phi_\Lambda \to \frac{\pi}{2}: \quad \phi_{\Lambda'} \approx \frac{\pi}{2}-\left(\Lambda'/\Lambda \right)^{-4} \left(\frac{\pi}{2}-\phi_\Lambda\right).
\label{Eq:AngleTransformationApproximation}
\end{align}
This means that a larger cutoff $\Lambda' > \Lambda$ will lead to larger angles. At the level of \Fig{Fig:NonuniversalRegimes}{}, changing $\Lambda$ to a larger $\Lambda'$ therefore has two effects: First of all, the non-universal regime is shifted to larger velocities [according to $\Lambda^2(\Lambda^6) \to \Lambda'^{2} (\Lambda'^{6})$ for the $y(x)$ axis] and second, the labels of the (dashed) trajectories are changed, e.g. $\phi_{1,\Lambda} \to \phi_{1,\Lambda'}$, where the transformation is given by \Eqs{Eq:AngleTransformation} and \eq{Eq:AngleTransformationApproximation}.

\section{Conclusion and outlook}

In this work, we have established the observable phenomenology of the generalized KZM scenario in an exactly solvable and experimentally relevant model.

The generalization includes driving equilibrium \emph{irrelevant} couplings, which can be turned relevant and thus observable. Once the ferromagnetic coupling and the transverse field (for the transverse XY model) can be tuned independently, extracting subleading scaling due to equilibrium irrelevant operators is feasible and becomes measurable at the level of the excitation density. In the limit of large transversal fields, deep in the paramagnetic phase, this density corresponds to the density of spin flips. Therefore, the generalized \KZM also fits into the \emph{observable} perspective of the traditional \KZM. 

We have outlined how a program can look like to construct a proper drive and reveal the new scaling due to irrelevant couplings for the transverse XY model. Besides its analytical appeal, the transverse XY model is in reach of experimental investigations like trapped ion experiments as in Refs.~\cite{Cui2016,Cui2020}, where the transverse Ising model was already analyzed, or compressed quantum simulations \cite{Kraus2011,Boyajian2013,Li2014,Boyajian2015}. In the case of single trapped ions, driving multiple couplings should be possible in the very same framework. The reason is that the XY model as well as the transverse Ising model, once mapped to fermions, are described by a set of two-level Hamiltonians, which can be cast into a Landau-Zener like form for each momentum mode $k$ (see Sec.~\ref{Sec:DimensionalConsideration}). For the general mechanism we have explored here, two ingredients are crucial. The first is the nonlinear character of the drive. Second, it must be possible to control and keep the distance to the critical point, measured by $\hat{\mu}_k$, fixed. The second point becomes important for drives parallel to the phase boundary, as we discussed in Sec.~\ref{Sec:TIParallelDrive}.

There are three further directions to explore this generalized \KZM. A first one concerns the interplay of the generalized \KZM phenomenology and its relation to the adiabatic \RG. The \RG picture applies quite generically to interacting and non-integrable theories with a Wilson-Fisher fixed point, and more complicated critical exponent spectra. Following a similar path as described here, different scalings and crossover scales could be detected by numerical simulations of models, where the fixed point geometry is well-understood by other means. An additional interesting aspect arises in more complicated, interacting theories from the dynamics and possible decay of the quasi-particles or defects beyond the time scale of adiabaticity breaking.

The second is in the direct vicinity of the XY model. So far, we have focused on a particular corner of the phase diagram, the paramagnetic phase, to cleanly study the physics of the Ising critical point. Nevertheless, there are additional features of the phase diagram like the incommensurate region in the ferromagnetic phase and multicritical points, which lead to an even richer phenomenology. The multicritical points give another arena to apply the \RG-perspective. Such drives have already been analyzed \cite{Divakaran2009,Divakaran2010,Deng2009,Mukherjee2010}, and indeed different scalings can be observed, depending on the direction of approach to the critical point. This is in line with the \RG perspective, as the multicritical point is characterized by multiple relevant couplings in equilibrium. Therefore, a generic drive will lead to driving many of these couplings.

The third direction is to turn to more complex models, from the experimental as well as theoretical side. The perspective of driving in different directions along a phase boundary also turns condensed matter systems with a curved phase boundary (e.g., Ref.~\cite{Grams2019}) into promising candidates to explore this generalized \KZM. For such systems, the critical line can be approached with very different angles by varying a single parameter (e.g., temperature or pressure) and choosing different fixed values for the other (even though these angles should not be confused with the angles defined for the universal field theory, see discussion in Sec.~\ref{Sec:OrthogonalityIssue}). This also allows for a locally parallel drive.

\section{Acknowledgments}

We thank M. Białończyk, C. Grams, M. Scherer, and J. Hemberger for useful and inspiring discussions. We acknowledge support by the funding from the European Research Council (ERC) under the Horizon 2020 research and innovation program, Grant Agreement No. 647434 (DOQS), and by the DFG Collaborative Research Center (CRC) 1238 Project No. 277146847 - project C04. This research was supported in part by the National Science Foundation under Grant No. NSF PHY-1748958.

\appendix

\section{Transformations and symmetries of the XY model \label{App:JordanWigner}}

\subsection{Jordan-Wigner transformation}
We introduce the Jordan-Wigner transformation, relating spin and fermion operators, which is the essential step to exactly solve the spin model. Early works on this topic include Refs.~\cite{Kaufman1949,Nambu1950}. The Jordan-Wigner transformation \cite{Jordan1928} maps spin-operators $\sigma_l^x,\sigma_l^y,\sigma_l^z$ (or $\sigma_l^\pm, \sigma_l^z$) to fermionic creation and annihilation operators $c_l,c_l^\dagger$ (see also Refs.~\cite{Katsura1962,Lieb1961,Pfeuty1970,Suzuki1971,Suzuki1971a,Bunder1999,Fradkin2010,Sachdev2011,Dziarmaga2005,Perk2017} and references therein). Here $\sigma_l^\pm$ are defined as 
\begin{align}
\begin{aligned}
&\sigma_l^+=\frac12 (\sigma_l^x+i\sigma_l^y),\\
&\sigma_l^-=\frac12 (\sigma_l^x -i \sigma_l^y), 
\end{aligned}
\end{align}
and the operators fulfill the commutation-relations
\begin{align}
\begin{aligned}
&[\sigma_l^\alpha,\sigma_m^\beta]=2i\epsilon_{\alpha \beta \gamma}\sigma_l^\gamma \delta_{lm},\\
&[\sigma_l^\pm,\sigma_m^z]=\mp 2 \sigma_l^\pm \delta_{lm}.
\end{aligned}
\end{align}
The fermionic creation and annihilation operators fulfill the anti-commutation relations:
\begin{align}
\begin{aligned}
&\{c_l,c_m^\dagger\}=\delta_{lm},\\
&\{c_l,c_m\}=\{c_l^\dagger,c_m^\dagger \}=0.
\end{aligned}
\end{align}
The transformation (or different representation), keeping the commutation-relations intact can be written as:
\begin{align}
\begin{aligned}
&\sigma_n^z=1-2c_n^\dagger c_n, \\
&\sigma_n^x=(c_n^\dagger+c_n) \prod_{m<n} (1-2c_m^\dagger c_m),\\
&\sigma_n^y=i(c_n^\dagger-c_n) \prod_{m<n} (1-2c_m^\dagger c_m),\\
&\sigma_n^+=c_n \prod_{m<n} (1-2c_m^\dagger c_m),\\
&\sigma_n^-=c_n^\dagger \prod_{m<n} (1-2c_m^\dagger c_m).\\
\end{aligned}
\end{align}
The backwards-transformation reads 
\begin{align}
\begin{aligned}
&c_l^\dagger =\sigma_l^-\prod_{m<l}\sigma_m^z, \\
&c_l=\sigma_l^+\prod_{m<l}\sigma_m^z.
\end{aligned}
\end{align}
Some important relations are (using $\sigma_j^\pm \sigma_j^z=\mp \sigma_j^{\pm}$)
\begin{align}
\begin{aligned}
&c_j^\dagger c_{j+1} + c_{j+1}^\dagger c_j= \frac12 \left[ \sigma_j^x \sigma_{j+1}^x + \sigma_j^y \sigma_{j+1}^y \right],\\
&c_j^\dagger c_{j+1}^\dagger + c_{j+1}c_j=\frac12 \left[\sigma_j^x \sigma_{j+1}^x -\sigma_j^y \sigma_{j+1}^y \right].
\end{aligned}
\end{align}
These relations show how controlling the ferromagnetic couplings $J_x$ and $J_y$ in the transverse XY model allows one to control the diagonal and off-diagonal sector of the fermionic theory. Similarly, we get
\begin{align}
c_j^\dagger c_{j+2} +h.c. = \frac12\left[ \sigma_j^x \sigma_{j+2}^x+\sigma_j^y \sigma_{j+2}^y\right]\sigma_{j+1}^z.
\end{align}

\subsection{Symmetries of the Hamiltonian}
The transverse XY model exhibits a $\mathbb{Z}_2$-symmetry (see, e.g., Ref.~\cite{Fradkin2010}), meaning that the Hamiltonian commutes with (tensor)product of all $\sigma^z$-operators
\begin{align}
\left[H, \prod_{l=1}^N \sigma_l^z\right]=0.
\end{align}
In the paramagnetic phase, the ground state of the Hamiltonian shares the $\mathbb{Z}_2$-symmetry, nevertheless in the symmetry-broken (ferromagnetic) phase the ground state does not.
The resulting fermionic Hamiltonian is quadratic in the $c$-operators, implying that the parity-operators
\begin{align}
P^\pm =\frac12 \left[ 1\pm \prod_{l=1}^N \sigma_l^z \right]
\end{align}
commute with the Hamiltonian as well (as the fermion-number is changed by either $0$ or $2$). Therefore, parity is a good quantum number and the Hamiltonian can be split into the two subspaces of even and odd parity \cite{Katsura1962,Dziarmaga2005}:
\begin{align}
H=P^+H^+P^+ +P^-H^-P^- ,
\end{align}
with boundary conditions in $H^-$: $c_1=c_{N+1}$ and in $H^+$: $c_1=-c_{N+1}$.

\section{Adiabatic nonperturbative contribution  \label{App:AdiabaitcApproach}}
In the limit $\hat{\adtime}_{k,f} \to + \infty$ the leading contribution in the limit $\adparameter\to 0$ stems from a non-analytic contribution from the $n$ complex zeros of the energy-difference in the upper half-plane:
\begin{align}
\mathcal{E}(\hat{t}_k^{c,l},k)=0, && \left(\hat{t}_k^{c,l}\right)^n=\left(\frac{-\hat{\mu}_k\pm i}{\adparameter}\right).
\end{align}
The excitation density for each $k$ is approximated as \cite{Davis1976,Joye1993}:
\begin{align}
\begin{aligned}
&p_k \approx \left|\sum_{l=1}^{n} \sigma_l \exp\left(i\mathcal{D}(\hat{\adtime}_k^{c,l})\right)\right|^2,\\
&\mathcal{D}(\hat{\adtime}_k^{c,l}):=2\int_{0}^{\hat{\adtime}_k^{c,l}} \mathcal{E}(\tau') d \tau ' , \\
&\sigma_l=4i \lim_{\hat{\adtime}_k \to \hat{\adtime}_k^{c,l}}(\hat{\adtime}_k-\hat{\adtime}_k^{c,l})\gamma(\hat{\adtime}_k)=\pm 1,
\end{aligned}
\end{align}
which we refer to as the \DDP approximation. Rescaling $\adparameter\hat{\adtime}_k^n=y^n$ we get 
\begin{align}
\mathcal{D}(\hat{\adtime}_k^{c,l})=\adparameter^{-\frac{1}{n}} 2\int_0^{y(\hat{\adtime}_k^{c,l})} \mathcal{E}(y)dy =: \adparameter^{-\frac{1}{n}} \mathcal{I}_l(\hat{\mu}_k),
\end{align}
where the second term in the last equation is an integral, which only depends on $\hat{\mu}_k$. Therefore, $\adparameter$ is the general adiabaticity parameter (see also Ref.~\cite{Suominen1992} for the quadratic case). For a linear drive only one pole is relevant and the excitation density for mode $k$ reads
\begin{align}
\begin{aligned}
p_k &\approx  \exp\left(-2\text{Im}\mathcal{D}(\hat{\adtime}_k^c)\right)\\
&=\exp\left(-\pi \adparameter^{-1}\right),
\end{aligned}
\end{align}
which is actually the \emph{exact} asymptotic Landau-Zener-Majorana-Stückelberg \cite{Zener1932,Majorana1932,Stueckelberg1932} result valid for $\hat{\adtime}_{k,i}=-\infty$ and $\hat{\adtime}_{k,f}=+\infty$.

\section{Fitting the crossover \label{App:FittingDetails}}

 \begin{figure}[t]
\centering

\begin{tikzpicture}

\node[anchor=south west,inner sep=0] at (0,6) {\includegraphics[width=0.9\textwidth]{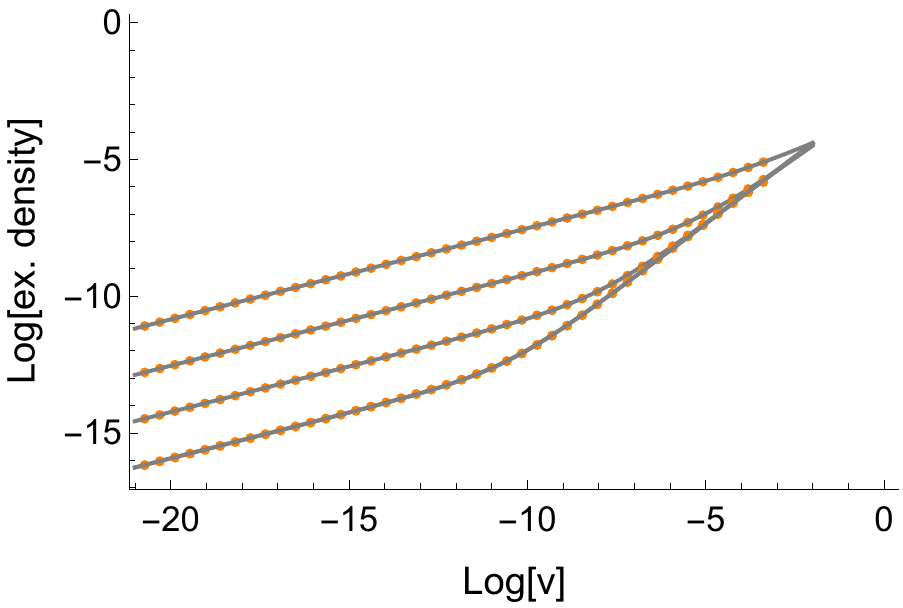}};
\node[anchor=south west,inner sep=0] at (0,0) {\includegraphics[width=0.9\textwidth]{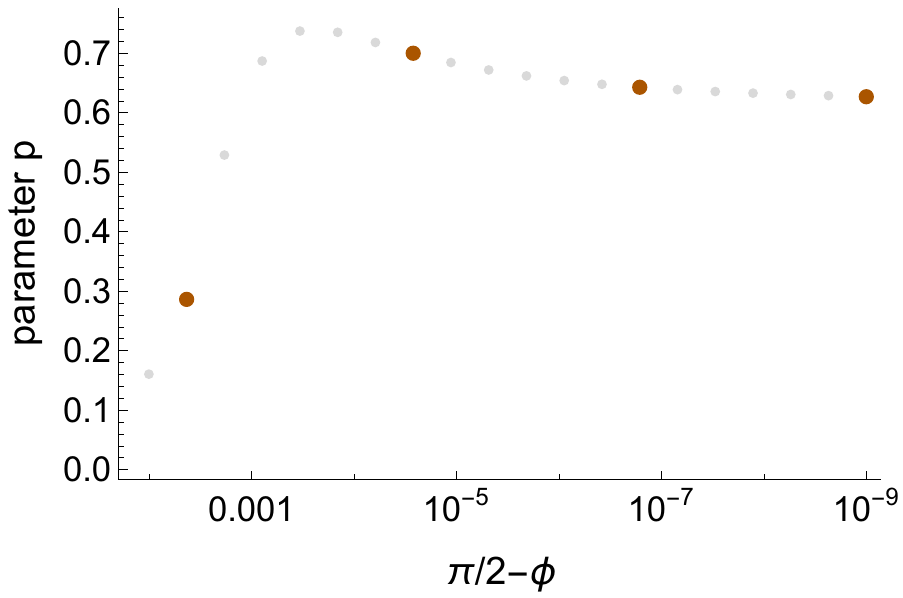}};
\node at (7.5,11) {\textbf{(a)}};
\node at (7.5,5) {\textbf{(b)}};
\end{tikzpicture}
\caption{Example of the numerical fits of $n_E(v,\phi)$ from the quadratic drive in the transverse XY on a (natural) log-log scale. \textbf{(a)} Plot of $n_E(v,\phi)$ (from the AI approximation) for four different angles (orange dots) and the fits according to \Eq{Eq:FittingFunction} (gray lines). \textbf{(b)} Parameter $p$ for the different fittings in (a) [bigger dots; most left is the upper most curve in (a) etc.] and all other angles used in the main text as well.}
\label{Fig:ExampleFit}
\end{figure}

The mechanism we are investigating, different scaling behaviours in the excitation density with the velocity, makes it necessary to identify the crossover velocity $v^*$. To be able to fit the crossover and the different scaling regimes in the $v$-scaling of $n_E(v)$, we use an Ansatz of the form:
\begin{align}
f(v)=A\left(\left(1+\left(\frac{v}{v^*}\right)^{\frac{a-b}{p}}\right)v^{\frac{b}{p}}\right)^{p},
\label{Eq:FittingFunction}
\end{align}
where $v^*$ is an estimate for the crossover scale. The exponents $a$ and $b$ are directly related to the exponents
\begin{align}
\begin{aligned}
&b=\frac{1}{nz+1/\nu},\\
&a=\frac{1}{nz+\text{dim}[g_j]}.
\end{aligned}
\end{align}
 The fitting procedure consists out of extracting the \KZM-exponent $b$ for very small velocities (anticipating $v\ll v^*$) and using this exponent to extract the subleading exponent as well as the crossover velocity according to the function in \Eq{Eq:FittingFunction}. An example is given in \Fig{Fig:ExampleFit}{}, which shows fair agreement. We keep the additional parameter $p$, as the crossover seems to be fitted better by allowing $p$ to be variable. A fixed $p$ can also be used, which might result in a `better' result of the exponents but a worse one for the crossover scale, where we prioritize the better fitting of the crossover underlining that a clear scaling only emerges for rather steep angles close to $\pi/2$. This choice is also more consistent with the estimated crossover scales.


%

\end{document}